\documentclass[preprint]{aastex}
\pdfoutput=1 
\usepackage{natbib}
\usepackage{graphicx}
\usepackage{amssymb}
\newcommand{\Beff}{B_{\rm eff}}
\newcommand{\amin}{\alpha_{\rm min}}
\newcommand{\amax}{\alpha_{\rm max}}

\def\beff{B_{\rm eff}}

\def\taud{\tau_{\rm dyn}}

\slugcomment{June 11th, 2010}
\shorttitle{Saturn's rings opposition effect}
\shortauthors{Salo and French}

\begin{document}

\title{The opposition
  and tilt effects of Saturn's rings from HST observations}

\author{Heikki Salo}
\affil{Dept. of Physics, Astronomy Division, Univ. of Oulu, PO Box 3000, FI-90014 Finland}
    \email{heikki.salo@oulu.fi}
\and
\author{Richard G. French}
\affil{Astronomy Department, Wellesley College,
    Wellesley, MA 02481}

%HUOM: 
%c6v4: alpha=0.461 east --> spokes
%c6v2: alpha=1.925 east --> spokes

\begin{abstract}
 
\vskip 0.5cm The two major factors contributing to the opposition
brightening of Saturn's rings are i) the {\em intrinsic brightening} of
particles due to coherent backscattering and/or shadow-hiding on their
surfaces, and ii) the {\em reduced interparticle shadowing} when the
solar phase angle $\alpha \rightarrow 0^\circ$.  We utilize the extensive
set of Hubble Space Telescope observations (Cuzzi et al.~2002, Icarus
158, 199-223) for different elevation angles $B$ and wavelengths
$\lambda$ to disentangle these contributions. We assume that the
intrinsic contribution is independent of $B$, so that any $B$
dependence of the phase curves is due to interparticle shadowing,
which must also act similarly for all $\lambda$'s. Our study
complements that of Poulet et al.~(2002, Icarus 158, 224), who used a subset
of data for a single $B \sim 10^\circ$, and the French et al.~(2007b,
PASP 119, 623-642) study for the $B \sim 23^\circ$ data set that
included exact opposition.  
We construct a grid of dynamical/photometric simulation models, with
the method of Salo and Karjalainen (2003, Icarus 164, 428-460), and
use these simulations to fit the elevation-dependent part of
opposition brightening. 
Eliminating the modeled interparticle component
yields the intrinsic contribution to the opposition effect: 
for the B and A rings it is almost entirely due to coherent
backscattering; for the C ring, an intraparticle shadow hiding
contribution may also be present.

Based on our simulations, the width of the interparticle shadowing
effect is roughly proportional to $B$. This follows from the
observation that as $B$ decreases, the scattering is primarily from
the rarefied low filling factor upper ring layers, whereas at larger
$B$'s the dense inner parts are visible. Vertical segregation of
particle sizes further enhances this effect.  The elevation angle
dependence of interparticle shadowing also explains most of the B ring
tilt effect (the increase of brightness with elevation). From
comparison of the magnitude of the tilt effect at different filters,
we show that multiple scattering can account for at most a $10\%$
brightness increase as $B \rightarrow 26^\circ$, whereas the remaining
$20\%$ brightening is due to a variable degree of interparticle
shadowing. The negative tilt effect of the middle A ring is well
explained by the the same self-gravity wake models that account for
the observed A ring azimuthal brightness asymmetry (Salo et al.~2004,
Icarus 170, 70-90; French et al.~2007a, Icarus 189, 493-522).


\end{abstract}

\section{Introduction}
Saturn's rings, like most atmosphereless objects in the solar system,
exhibit an opposition effect: a rapid increase in the brightness
when the Sun-observer phase angle $\alpha \rightarrow 0^\circ$.  Most
strikingly, this has been demonstrated directly by the zero-phase Cassini images
\citep{deau2009},
%{\bf [How do we get the proper reference style for Icarus? -Dick]}
showing a bright localized spot on the ring location centered at
exact opposition. Similarly, Hubble Space Telescope observations
during the exceptional 2005 opposition \citep{french2007b} revealed
that the brightness increase continues all the way to zero phase
angle: in 2005, the Earth was transiting the Sun as seen from Saturn,
implying a minimum $\alpha$ set by the finite solar radius of
$0.029^\circ$, with the brightness increasing by about 1/3 for
$\alpha<0.5^\circ$, in addition to a similar increase between
$0.5^\circ<\alpha<6^\circ$.

Two main explanations have been offered for the opposition brightening of
Saturn's rings: 1) the intrinsic brightening of the grainy ring
particle surfaces, and 2) the reduced amount of mutual interparticle
shadowing between ring particles, as the phase angle $\alpha
\rightarrow 0^\circ$. The main contributor to intrinsic brightening is
likely to be the coherent backscattering mechanism (CB), based on
constructive interference between the incoming and outgoing light rays
(Akkermans et al.~1988, Shkuratov 1988, Hapke 1990, Muinonen et al.~1991,
 Mishchenko 1992), although shadow-hiding (SH) at the particle's
surface (Hapke 2002) may also contribute.  Coherent backscattering, as
well as surface shadow hiding, are complicated functions of the
particle surface structure and optical properties of the grains; these
mechanisms are currently topics of extensive theoretical and
laboratory studies ({Nelson et al.~2000; Nelson et al.~2002; Hapke et
  al.~2006, Shepard and Helfenstein 2007, Shkuratov et al.~2007, Hapke
  et al.~2009}). In contrast, the interparticle shadowing
\citep[e.g.][]{hapke1986,irvine1966} contribution is not sensitive to
physical particle properties, but is primarily determined by the
optical depth and volume filling factor of the ring.  In what follows,
we will consistently call this latter effect ``interparticle
shadowing,'' rather than ``shadow hiding'' or ``mutual shadowing,'' in
order to avoid any possible misinterpretation in terms of shadows
associated with roughness of the surfaces of ring particles.

%DICK: add recent references

Classically, the strong and narrow opposition brightening of Saturn's
rings was interpreted in terms of interparticle shadowing in a low
volume density ring.  Near to opposition, the shadow a particle casts
on other particles becomes more and more hidden by the particle
itself.  The smaller the volume density, the longer the average shadow
cylinders are before hiding another particle: a more precise alignment
of illumination and viewing is thus needed for this effect to become
important.  In particular, \citet{lumme1983} calculated the
interparticle shadowing contributions for homogeneous B ring models,
and showed that the then-existing phase curves could be accounted for
solely by this effect, provided that the ring has a low volume density
of the order of $D \approx 0.02$. This corresponds in the case of
identical particles to a multilayer with a thickness of several tens
of particle diameters.  Such a low volume density seemed to contradict
dynamical models (e.g.~Araki and Tremaine 1986, Wisdom and Tremaine
1988, Salo 1992a) that, based on the laboratory measurements of the
elasticity of ice (Bridges et al.~1984), predicted that the rings
should flatten to a closely packed near-monolayer state with a
thickness of few particle diameters at most, indicating $D > 0.1$.
For such a large volume density, a homogeneous ring would have a much
wider opposition effect than the observed brightening. Therefore,
Mishchenko and Dlugach (1992) and Mishchenko (1993) suggested that the
brightening is instead due to CB (see also Muinonen et
al.~1991).{\footnote{The notion of an intrinsic opposition peak
    originated much earlier; see e.g. Cook et al.~1973,
    H\"ameen-Anttila and Vaaraniemi 1975, Irvine et al.~1988.}
  Mishchenko (1993) also argued that CB is strongly supported by the
  Lyot (1929) and Johnson et al.~(1980) measurements of negative
  linear polarization, whose magnitude drops rapidly within $\alpha <
  0.5^\circ$. Indeed, during 1990's the CB became accepted as the
  standard explanation for Saturn's rings opposition effect, and the
  Cassini VIMS observations have also been interpreted within this
  framework \citep{nelson2008}.

%{\bf (Nelson 2008 - COSPAR abstract -Laboratory
%    Investigations Relevant to Cassini VIMS Reports of Coherent
%    Constructive Interference in Saturn's Rings - add ref) }.

 However, the interpretation of the opposition effect solely in terms
 of intrinsic brightening has a severe problem: improved dynamical
 models of flattened rings {\em do in fact predict} a fairly narrow
 interparticle shadowing opposition peak, if a particle size
 distribution is taken into account (Salo \& Karjalainen 2003;
 hereafter SK2003). This is because, for a fixed $D$, the effective
 mean width of shadow cylinders drops faster than their mean length
 when the size distribution is broadened. Photometric simulations in
 SK2003 indicate that the effect is well-matched by Hapke (1986) size
 distribution models for semi-infinite particle layers.
 Interestingly, if the currently favored wide particle size
 distributions with width $R_{\rm max}/R_{\rm min} \sim 100$ (Marouf
 et al.~1983, French and Nicholson 2000) are assumed, interparticle
 shadowing can account for most of the opposition brightening for
 $\alpha >0.5^\circ$, and even have a significant contribution for
 $\alpha < 0.5^\circ$ (SK2003, French et al.~2007b). Nevertheless, the
 strong surge near $\alpha=0^\circ$ and the wavelength dependence of
 phase curves (French et al.~2007b) unambiguously show the intrinsic
 contribution to be present. Therefore, both intrinsic and
 interparticle shadowing mechanisms are likely to affect the
 opposition brightening, although it has been surprisingly difficult
 to disentangle their contributions to the phase curves.

{\bf \hskip 5cm INSERT FIG  1 HERE}

In this paper we propose that the {\em intrinsic brightening and
  interparticle shadowing can be reliably separated by a comparison of
  the opposition phase curves at different ring opening angles}.
Namely, whereas the intrinsic contribution should be the same
regardless of elevation, the interparticle shadowing contribution is expected
to be very sensitive to the viewing elevation.  This prediction
follows from dynamical simulations, which indicate a vertically
nonuniform particle distribution.  As explored in detail in SK2003,
the width of the modeled interparticle shadowing peak gets narrower
for more shallow illumination, since the reflection will be more and
more dominated by the low volume density upper layers; see Fig.~1 for
an illustration. This $B$ dependence of the effective volume density
should be further augmented by the particle size distribution, since
small particles are expected to have a larger scale height than the
larger particles. Moreover, an extended particle size distribution will
lead to a narrower opposition effect, in accordance with
theoretical calculations \cite{hapke1986}, although a broad particle size distribution
alone, without vertical structure, does not imply a $B$-dependent
opposition effect. To test these expectations, we will utilize the
extensive set of UBVRI observations of Saturn's rings, obtained with the
Hubble Space Telescope's WFPC2 \citep{french2007b}.  

In order to separate the intrinsic and interparticle contributions, we
will employ a set of dynamical simulation models performed with
different optical depths and widths of the size distribution. The
opposition phase curves are calculated for these models, covering the
range of viewing elevations accessible from the Earth.  We then match
the observed elevation angle dependence with the simulated one, using
the common phase curve range ($\alpha=0.5^\circ-6.0^\circ$) available
for all elevations and filters, and obtain a set of best-fitting size
distributions for the different ring components. The known
contribution of interparticle shadowing in these simulation models,
for any phase or elevation angle, can then be extracted from the
observed data points, to yield opposition phase curves representing
just the intrinsic contribution.  The success of the extraction
procedure can be tested by the requirement that the remaining
intrinsic contribution must depend only on wavelength and ring
location, and not on elevation angle.

%* possible complicatations: 

%- multiple scattering (separable via comparison of different filter)

%- particle properties + ring structure vary with r, not only with tau

%- rings not necessarily homogeneous:  wakes, OS

%- SK 2003: Lumme et al. (1983) observations could be fitted by
%extended size distribution models (accounted both oppo+tilt). However,
%if oppo was mainly CB/SH, would SK2003 explanation still be valid for the tilt
%effect?

An additional test for the importance of interparticle shadowing is
provided by the B ring tilt effect (the reflectivity $I/F$ increases
with elevation by 30\% for the ground based geometries), which is
traditionally interpreted as resulting from increased multiple
scattering at larger elevation angles (Esposito and Lumme 1977, Lumme
et al.~1983). In SK2003, we proposed that the tilt effect can also
arise due to an elevation-angle dependent opposition effect: 
at large
elevations (say, $B=26^\circ$ as in Fig.~1) the observed brightness is
enhanced by the wide interparticle shadowing opposition peak, which at
smaller elevations ($B=4^\circ$ in Fig.~1) 
becomes so narrow that it
is confined inside the typical observation phase angle of few
degrees. The full HST data set, with sufficient $B$ and $\alpha$
coverage, and excellent photometric accuracy, offers an ideal tool for
testing this hypothesis. As a side result, we also obtain an accurate
estimate for the amount of multiple scattering, and thus set
constraints on the intrinsic ring particle phase function.  Note that
Cassini imaging data available to date, though having superior spatial resolution and a
broader coverage of phase angles, do not enable such a systematic
study of the opposition effect at different elevations.

The plan of the paper is as follows. In Section 2, we compare HST data
at different elevations. We show that the observed phase curves are
steeper at smaller elevation angles, and moreover that this steepening
is independent of wavelength, consistent with what is expected for an
interparticle shadowing effect.  In Section 3, we devise a method for
extracting the elevation-dependent part of the opposition
brightening, utilizing dynamical and photometric simulations.  Section
4 then shows the results of fitting models to the observations, and
discusses the implications for the relative magnitudes of the intra-particle and
interparticle opposition effects.  The intrinsic opposition effect
phase curves are also presented in a tabular form, in terms of
parameters for two different fitting formulas (a linear-exponential fit,and a
simplified Hapke model including both CB and SH).  Section 5 discusses
the close interrelation between the interparticle mutual shadowing
opposition effect and the tilt effect, and uses the observed tilt
effect at different wavelengths to estimate the amount of multiple
scattering. Section 6 summarizes our conclusions.

\section{Elevation angle dependence of HST phase curves}
\label{sec:hst}

\subsection{Previous analysis of HST phase curves}

There are two previous studies of the opposition phase curves based on
a subset of the same HST observations used here.  In Poulet et al.~(2002), the
HST data obtained during Cycle 7 for $10^\circ$ elevation angle were
fitted with various models for the intrinsic opposition effect,
including the Hapke (1986) shadow hiding model, the Drossart (1993) fractal
phase function, and the Shkuratov et al.~(1999) model combining coherent
backscattering and shadow hiding.  However, no allowance was made for
a possible interparticle shadowing contribution, and subsequent HST
observations made it clear that the minimum
$\alpha \approx 0.3^\circ$ in the data utilized by Poulet et al.~(2002)
was too large to accurately
constrain the models. 
%{\bf 
French et al.~(2007b) combined the 2005 observations at exact opposition
  for $\sim 23^\circ$ elevation with the data from Cycles 10-12
  at comparable elevation ($\sim 26^\circ$), which allowed for much more
accurate fits of the phase curves than Poulet et al.~(2002) were able to 
obtain. %}  
In particular,
linear-exponential fits, and fits with the Hapke (2002) shadow
hiding/coherent backscattering model, indicated that the HWHM of the
opposition peak varies in the range $0.1-0.2^\circ$.  The effect of
interparticle shadowing was studied separately, using Monte
Carlo simulations, which indicated that the observed opposition surge
is stronger and narrower than what can be attributed even to a quite
extended particle size distribution (with width $R_{\rm max}/R_{\rm min} \lesssim
100$).  Nevertheless, the high-quality near-opposition phase curves
(see Fig.~2 in French et al.~2007b) give an impression of possibly two
superposed components, the more extended of which might represent the
interparticle contribution.

\subsection{HST observations at different elevations}

In the current study we use the full HST data set for $|B|=4.5
-26^\circ$ (See Table I in French et al.~2007b), which has been
processed and calibrated as described in \cite{french2007a}. Throughout
our analysis we use the {\em geometrically corrected} $I/F$
(H\"ameen-Anttila and Pyykk\"o 1970, Dones et al.~1993, and Cuzzi et
al.~2002), obtained by reducing the observations at slightly different
$B$ and $B'$ (the elevation angles of the Earth and the Sun,
respectively) to an effective common elevation angle $\Beff$:
\begin{equation}
  \sin \Beff \equiv \mu_{\rm eff} = \frac{2\mu\mu'}{\mu+\mu'},
\end{equation}

\noindent where $\mu \equiv |\sin B|$ and $\mu' \equiv |\sin B'|$, by
\begin{equation}
  \left( I/F \right)_{\rm corr} = \left( I/F \right) \frac{\mu+\mu'}{2\mu'}.
\end{equation}

\noindent This correction factor is exact for the reflected singly scattered light\footnote{%\bf 
The formula for the singly
  scattered reflected light is given by Eq.~(\ref{eq:iss}) in Section
  3.1. When the correction factor is applied to
  single scattering, $ \frac{\mu+\mu'}{2\mu'} \times (I/F)_{\rm ss}
  (\mu, \mu') = \frac{AP}{8} \left(1-\exp(-2\tau/ \mu_{\rm
    eff})\right)$, which equals $(I/F)_{\rm ss} (\mu =\mu'=\mu_{\rm
    eff})$ for all values of $\tau$.}  from a classical many-particle
thick ring (volume density $D \rightarrow 0$), and should hold quite
well even when multiple scattering is included \citep{lumme1970,
  price1973} or when scattering from a realistic geometrically thin
particle disk is considered.  Inclusion of this geometric correction
is very important in our case, where observations from different
elevation angles are compared with each other. For large $\alpha$
there can be significant differences in $B$ and $B'$, and the
correction factor $\frac{\mu+\mu'}{2\mu'}$ may amount to as much as
$20\%$ for low elevation observations.  If uncorrected, this spurious
effect of variable observing geometries would easily overwhelm the
true elevation angle dependence of ring brightness. During a single
HST Cycle, $\Beff$ is more or less constant (within a few tenths of
degree) although $B$ and $B'$ may vary by a few degrees (see Table I
in French et al.~2007a; Fig.~1 in French et al.~2007b). This will
allow us to group together all data from each individual HST Cycle. In
what follows, we will omit the subscript and denote the geometrically
corrected observed brightnesses simply as $I/F$.

{\bf \hskip 5cm INSERT FIG \ref{fig:HST_oe_profiles_4_22}  HERE}

To illustrate that a clear elevation angle dependent contribution is
indeed present in the full HST data set, 
Fig.~\ref{fig:HST_oe_profiles_4_22} compares the radial $I/F$ profiles at
large elevation ($\Beff \sim 23^\circ$) with those at
$\Beff=4.5^\circ$, which is the lowest elevation angle for which
observations are available. The solid lines indicate observations at
phase angles close to $6^\circ, 2^\circ$, and $0.5^\circ$. Indeed, for
$\beff=4.5^\circ$ the relative brightening as $\alpha$ decreases is
clearly stronger. For example, the typical B ring $I/F$ is enhanced by
about 25\% when $\alpha$ decreases from $6^\circ$ to $2^\circ$ for
$\beff=4.5^\circ$, but only by about 15\% for $\beff = 23^\circ$
(Fig.~\ref{fig:HST_oe_profiles_4_22_ratio}). A similar increase is seen
between $\alpha \sim 2^\circ$ and $\sim 0.5^\circ$. The fact that the
relative enhancement increases for lower $B$ is qualitatively in
agreement with the interparticle shadowing example of Fig.~1; in
Section \ref{sec:models} we will make a more detailed comparison to
our Monte Carlo models, after first characterizing the elevation
angle dependence of the observations.  The figure also shows the $I/F$ profile
for $B=22.9^\circ$, obtained at exact opposition (French et
al.~2007b), illustrating that a major part of the opposition
brightening takes place inside $\alpha\sim 0.5^\circ$.

{\bf \hskip 5cm INSERT FIG \ref{fig:HST_oe_profiles_4_22_ratio}  HERE}

The magnitude of opposition brightening and its dependence on
elevation (at least for the broader component outside
$\alpha=0.5^\circ$) are fairly similar for the C ring in comparison to
the optically much denser B and A rings (see Fig.~3). At first glance,
this might appear to contradict the importance of interparticle
shadowing, which is expected to be very sensitive to
$\tau$. Nevertheless, there seems to be a positive correlation between
brightness increase and the local optical depth, as required if at
least part of the brightening is due to reduced interparticle
shadowing.  A similar correlation between the slope of the
phase curve at $\beff \sim 26^\circ$ and the C ring optical depth was
discussed in French et al.~(2007b).

%COMPARE to French et al. 2007 - here and/or in Discussion

In French et al.~(2007b), the high elevation angle data set (Cycles
10-13) was analyzed in detail for three ring regions: $a= 78 \ 000 -83
\ 000 \ {\rm km}$ (the C ring; this region excludes most prominent
ringlets and plateaus), $a= 100 \ 000 -107 \ 000 \ {\rm km}$ (the B
ring), and $a= 127 \ 000 -129 \ 000 \ {\rm km}$ (the A ring; this is
the region where the azimuthal brightness asymmetry is strongest). 
%{\bf 
Since the data covered phase angles near to zero, %}, 
it was
possible to fit various detailed backscattering models to the phase
curves. In particular, besides the physically-motivated Hapke (2002)
models, it was shown that the data are quite well described by an
empirical linear-exponential model 
\begin{equation}
 \frac{I(\alpha)}{F} = a' \exp{(-\alpha/d')} + b' + k' \alpha,
\label{eq:lin-expo}
\end{equation}
\noindent commonly used for fitting of satellite and asteroid
near-opposition phase curves (Kaasalainen et
al.~2003). The parameters $a'$ and $d'$ describe the amplitude and
width of the narrow opposition peak, while $b'$ and $k'$ give the
background intensity and linear slope of the phase curve. The
half-width at half-maximum for the exponential component is HWHM $= d'
\ln 2$. In French et al.~(2007b) a detailed analysis of the model
parameters was presented as a function of ring location and
wavelength.

{\bf \hskip 5cm  INSERT TABLES 1,2,3  HERE}

Unfortunately, at the smaller elevation angles, the HST data are too
sparse to allow such a four-parameter fit (or a Hapke fit with seven
parameters).  The data we use, in addition to those shown in Tables
2--4 in French et al.~(2007b), are listed in Tables
\ref{table:cring}--\ref{table:aring}, representing measurements of
average geometrically corrected $I/F$ from Cycles 6--9 for the three
ring regions (C, B, A) defined above.  In some cases, there are
measurements for only 3--4 distinct phase angles.  Figure
\ref{fig:HST_loglin} collects the phase curves for six sets of
elevation angles, with the mean effective values of $\beff =4.5^\circ,
10.2^\circ, 15.4^\circ, 20.1^\circ, 23.6^\circ$, and $ 26.1^\circ$
(Cycles 6, 7a, 7b, 8, 9+13, and 10-12, respectively). The curves are
normalized to $I/F$ at $\alpha=6^\circ$, and the two most-widely
separated filters, F336W and F814W, are shown. For the lowest
elevation angle, only two B ring phase angle data points are available
for the F336W filter, due to the contamination of images by B ring
spokes.  Also shown in the plot are two-parameter log-linear fits of
the form
\begin{equation}
\frac{I(\alpha)}{F} = a \ln \alpha + b.
\label{eq:log}
\end{equation}
\noindent These fits were also used in the normalization of the data
to a common phase angle $\alpha=6^\circ$ in
Fig.~\ref{fig:HST_loglin}. In order to assure a uniform coverage of
phase angles at different elevations, the fits were made restricting
to values $\alpha >0.25^\circ$.  Such simple fits seem to match the
data quite well.  Nevertheless, for the most extensive data sets
($\Beff=26.1^\circ, 23.6^\circ$) it is evident that a steeper
logarithmic slope would be appropriate for $\alpha <0.5^\circ$,
justifying the use of more complicated fitting formulae when
sufficient data are available.  Note that the log-linear intensity
fitting formula is in practice almost indistinguishable from the
log-linear magnitude fits utilized by \cite{lumme1983} and Bobrov
(1970).

{\bf \hskip 5cm INSERT FIG \ref{fig:HST_loglin}  HERE}

{\bf \hskip 5cm INSERT FIG \ref{fig:HST_loglin_abc}  HERE}

Concentrating on the wider regime $\alpha>0.5^\circ$, Fig.~\ref{fig:HST_loglin}
suggests that the enhanced opposition brightening at smaller elevations
is well captured by the log-linear fits. A systematic increase of the
logarithmic slope with decreasing $\Beff$ is seen in all filters for all ring regions.  This is most clearly seen in
 Fig.~\ref{fig:HST_loglin_abc}, which compares the observations at different
elevation angles for the F555W filter. The values of the parameters
$a$ and $b$, for different filters and ring regions, are listed in
Table \ref{table:logfit_ab}.

{\bf \hskip 5cm INSERT TABLE 4  HERE}

In order to further characterize the elevation angle dependence of
opposition effect, and to separate the intrinsic and interparticle
contribution, we turn to modeling in the next section.

%%%%%%%%%%%%%%%%%%%%%%%%%%%%%%%%%%%%%%%%%%%%%%%%%%%%%%%%%%%%%%%%%%%%%%%%%%%%%%
%
%\section{Extraction of mutual shadowing contribution via simulation modelings}
\section{Disentangling the mutual interparticle shadowing and intra-particle contributions}
%
%%%%%%%%%%%%%%%%%%%%%%%%%%%%%%%%%%%%%%%%%%%%%%%%%%%%%%%%%%%%%%%%%%%%%%%%%%%%%%
\label{sec:models}

\subsection{Mathematical formulation}

We assume that the intensity at phase angle $\alpha$,
effective elevation angle $\beff$, normal optical depth $\tau$, and wavelength $\lambda$, can be written in the form
\begin{equation}
(I/F)(\alpha,\beff,\tau,\lambda) = [f_{i}(\alpha,\lambda) f_{e}(\alpha,\beff,\tau) + Q_{\rm ms}(\beff, \tau, \lambda)] (I/F)_{ss}(\alpha,\beff, \tau, \lambda),
\label{eq:if}
\end{equation}

\noindent where $f_i$ denotes the {\em intrinsic} (e.g. due to
coherent backscattering and/or shadow-hiding at the particles'
surfaces) and $f_e$ the {\em external} (due to reduced interparticle mutual
shadowing) contribution to the opposition brightening of the singly
scattered radiation, $Q_{\rm ms}$ is the fractional contribution of
multiple scattering, and $(I/F)_{ss}$ is the theoretical single scattering
intensity of reflected light,
\begin{eqnarray}
(I/F)_{ss} = \frac{A(\lambda) P(\alpha, \lambda) \mu'}{4(\mu+\mu')} ~ \left(1-\exp\left [ - \tau (\frac{1}{\mu} + \frac{1}{\mu'}) \right]\right),
\label{eq:iss}
\end{eqnarray}
\noindent where $A$ is the (possibly wavelength-dependent) Bond albedo
of the particles and $P$ is the particle phase function.\footnote{%\bf
  The separation of particle's intrinsic opposition brightening and
  its phase function is somewhat arbitrary; here, the phase function
  $P(\alpha, \lambda)$ stands for the overall angular distribution of
  the scattered radiation due to surface topography and
  illumination.} For a classical, zero volume density ring there is no
interparticle shadowing, and thus $f_e=1$. Likewise, $f_i=1$
corresponds to the absence of an intrinsic opposition peak; the theoretical maxima
for each of these factors is 2.  Note that here $Q_{\rm ms}$ includes
just the interparticle multiple scatterings, not the possible multiple
reflection events at the particle surface, thought to be responsible
for the coherent backscattering effect. Our goal is to separate the
intrinsic and external contributions $f_i$ and $f_e$. % {\bf 
Note that
  our Eqs.~(\ref{eq:if}) and (\ref{eq:iss}) assume that the ring can
  be described by a single uniform optical depth, whereas the actual
  rings are known to possess local density variations due to
  self-gravity wakes (Colwell et al.~2006, 2007, Hedman et
  al.~2007). Besides such local variations, the resolution element of
  HST observations is so large that it includes a superposition of
  different optical depths, due to large scale radial structure of
  rings. Because of this, all the parameters in the equations,
  including $\tau$, $f_i$, and $f_e$, must be considered as effective
  mean values.%}

In what follows we assume that $f_i$ depends on wavelength but is
independent of $\beff$ or $\tau$.  On the other hand, $f_e$ is
independent of wavelength, but is likely to depend on both $\beff$ and
$\tau$.  The multiply-scattered contribution vanishes for $\tau
\rightarrow 0$ and for $\beff \rightarrow 0$; in general its
contribution is expected to be small for all earth-based geometries
(Cuzzi et al.~2002). We shall therefore ignore $Q_{\rm ms}$ in this
section, an approximation that is justified in Section 5.

Assuming $Q_{\rm ms}=0$, the fractional brightness increase (denoted by
$OE$) in some interval $\amax \rightarrow \amin$ can be written as
\begin{eqnarray}
OE(\beff, \tau, \lambda) &=& \frac{I(\amin,\beff,\tau, \lambda)}{I(\amax,\beff,\tau, \lambda)}\nonumber \\ \nonumber \\
&\approx&\frac{f_i(\amin,\lambda) f_e(\amin,\beff,\tau) P(\amin,\lambda)}{f_i(\amax,\lambda) f_e(\amax,\beff,\tau) P(\amax,\lambda)}\nonumber \\ \nonumber \\
&=&OE_i(\lambda) \ OE_e(\beff,\tau)\frac{ P(\amin,\lambda)}{ P(\amax,\lambda)}, \\ \nonumber
\end{eqnarray}
\noindent where we have denoted $OE_i \equiv f_i(\amin)/f_i(\amax)$
and $OE_e \equiv f_e(\amin)/f_e(\amax)$.  To eliminate the intrinsic
brightening and the contribution from the particle phase function, we
normalize $OE$ by its value at some fixed elevation $\beff=B_{\rm
  norm}$.  This ratio contains only the external contribution,
\begin{eqnarray}
\frac{OE(\beff,\tau, \lambda)}{OE(B_{\rm norm},\tau, \lambda)}&=&\frac{OE_e(\beff,\tau, \lambda)}{OE_e(B_{\rm norm},\tau, \lambda)},
\label{eq:OE_e}
\end{eqnarray}
\noindent since the interparticle shadowing contribution $f_e$ is the only
factor that depends on $\beff$.

{\bf \hskip 5cm INSERT FIG  \ref{fig:HST_obs_oe} HERE}

In Fig.~\ref{fig:HST_obs_oe} we show the observed brightness
enhancement $OE_{\rm obs}$ as a function of elevation angle, for the
previously defined C, B, and A ring regions. The range $\amax=6^\circ$
and $\amin=0.5^\circ$ is chosen, as the brightening in this range is
likely to be due to the interparticle mutual shadowing effect, rather
than the more narrow intrinsic opposition peak.  For this range of phase angles, the
log-linear fits of the previous section fits can be used with good
accuracy, so that
\begin{equation}
  OE_{\rm obs} \equiv I(0.5^\circ)/I(6.0^\circ) = \frac{a \ln 0.5 +b}{a \ln 6.0 +b},
\end{equation}
\noindent where $a$ and $b$ are the fit parameters in
Eq.~(\ref{eq:log}).  As seen in Fig.~\ref{fig:HST_obs_oe} (upper row;
see also Table \ref{table:logfit_oe6}), the typical values of $OE_{\rm
  obs}$ for the C, B, and A ring regions are 1.25 -- 1.4 for
$\beff=26.1^\circ$, increasing to 1.5 -- 1.75 for
$\beff=4.5^\circ$. To show that an interparticle shadowing
contribution is indeed present, the lower row in
Fig.~\ref{fig:HST_obs_oe} shows $OE_{\rm obs}$ normalized to that at
$\beff=26.1^\circ$. If we assume, as in Eq.~\ref{eq:OE_e}, that the
intrinsic contribution is independent of $B$, the effect seen in the
lower row must be due solely to the dependence of interparticle mutual
shadowing on the opening angle. For the A and B ring regions the ratio
$OE_{\rm obs}(4.5^\circ)/OE_{\rm obs}(26.1^\circ)$ is about 1.2, and
about 1.1 for the C ring region.  That we are seeing an interparticle
shadowing effect is further supported by the fact that the ratio
$OE_{\rm obs}/OE_{\rm obs}(26.1^\circ)$ is similar for all filters, as
it should be if it arises from interparticle shadowing. It also
supports the assumption that multiple scattering $Q_{\rm ms}$ is
insignificant. That is, a significant multiple scattering contribution
would make the shape of the curves depend on the filter, since the
particle albedo increases toward longer optical
wavelengths.\footnote
{%\bf 
With the inclusion of multiple scattering,
  Eq.~(7) is modified to $OE_{\rm obs}(\beff) =
  \frac{f(0.5^\circ,\beff) +Q_{\rm ms}(\beff)}{f(6^\circ,\beff)+Q_{\rm ms}(\beff)} ~ \frac{P(0.5^\circ)}{P(6^\circ)}$ where
  $f(\alpha,\beff) \equiv f_i(\alpha,\lambda) f_e(\alpha,\beff,\tau)$.
  Assuming that $Q_{\rm ms} << f$ for all $\beff$, we may approximate $OE_{\rm obs}(\beff)/
  OE(\beff) = 1+Q_{\rm ms}(\beff)\big[1/f(0.5^\circ,\beff)-1/f(6^\circ,\beff)\big]$,
  where $OE$ in the denominator includes just the singly scattered
  component as in Eq.~(7). Assuming that $Q_{\rm ms}(4.5^\circ) << Q_{\rm
    ms}(26.1^\circ)$, we may further approximate $OE_{\rm
    obs}(4.5^\circ)/OE_{\rm obs}(26.1^\circ) \approx
  OE(4.5^\circ)/OE(26.1^\circ) (1+Q_{\rm
    ms}(26^\circ)(1/f(6^\circ,26.1^\circ)-1/f(0.5^\circ,26.1^\circ))$. Since
  $f(0.5^\circ)> f(6^\circ)$, the prefactor of $Q_{\rm ms}$ is
  positive.  This implies that significant multiple scattering would
  make the curves in the lower row of Fig.~\ref{fig:HST_obs_oe}
  steeper in the red filter than in the blue filter, since $Q_{\rm
    ms}({\rm red}) > Q_{\rm ms} (\rm blue)$.}

{\bf \hskip 5cm INSERT TABLE 5  HERE}

Similarly, to eliminate the interparticle shadowing contribution to
$OE$ we may normalize by its value at wavelength $\lambda=\lambda_{\rm
  norm}$,
\begin{eqnarray}
\frac{OE(\beff,\tau, \lambda)}{OE(\beff,\tau, \lambda_{\rm norm})}&=&\frac{OE_i(\lambda)}{OE_i(\lambda_{\rm norm})} 
\frac{P(\amin,\lambda)/P(\amin,\lambda_{\rm norm})}{P(\amax,\lambda)/P(\amax,\lambda_{\rm norm})}.
\end{eqnarray}
\noindent
Here, the first multiplier describes the wavelength-dependent
difference in the intrinsic opposition brightening, while the second
factor describes the change in the {\em color}
$P(\lambda)/P(\lambda_{\rm norm})$ between $\amax$ and $\amin$; these
are written separately, since the color change (Cuzzi et al.~2002)
might appear over a wider angular range than the narrow intrinsic
brightening peak. Since the intrinsic particle behavior should not
depend on opening angle, the above ratio should be independent of
$\beff$, as is also verified by HST data.

%OMITTED FIGURE -> REMOVE TEXT

%in Fig.~\ref{fig:HST_obs_oe_336norm},
%where $OE_{obs}$ is normalized by that in the F336W filter. Clearly,
%variations with $\Beff$, if any, are much smaller than that with
%wavelength, and the scatter is most likely due to slight errors in the
%absolute photometry of the WFPC2 images over 8 years.  The difference
%between filters illustrates both the dependence of color on phase
%angle (Cuzzi et al. 2002), and the dependence of intrinsic brightening
%on the wavelength.

In order to determine the actual interparticle mutual
shadowing contribution to the opposition brightening as a function of
$\alpha$, and not merely its relative contribution via
$OE_{\rm obs}/OE_{\rm obs}(26.1^\circ)$, we use simulation modeling in the
next sub-section.

\subsection{Monte Carlo simulation method}
Our photometric calculations are carried out with the method developed
in SK2003 and in Salo et al.~(2004), based on following a large number of
photons through a ring composed of discrete finite-sized particles.
The particle fields we use to model the rings are obtained from
dynamical simulation models. The particles are assumed to be
much larger than the wavelength so that geometric ray tracing can be
used.  The particle field, with periodic planar boundaries, is
illuminated by a parallel beam of photons, and the path of each
individual photon is followed in detail from one intersection with a
particle surface to the next scattering, until the photon escapes the
particle field; the new direction after each scattering is obtained
via Monte Carlo sampling of the particle phase function \citep[see
  e.g.][]{plass1968,salo1988}. The brightness at a chosen observing
direction is obtained by adding together the contributions of all
individual scatterings that are visible from this direction (not
blocked by any of the finite sized particles). Compared to direct
Monte Carlo estimates based on tabulating just the directions of
escaped photons, this indirect method gives significantly reduced
variance of the results (see Fig.~5 in SK2003).

Since we are dealing with low elevation angle observations, the
periodic boundaries must be treated very accurately, as described in
detail in SK2003. To reduce the effect of the discreteness of the
simulated particle fields, results from at least five separate particle
snapshots are combined in the phase curves. Also,  the particle 
fields are randomly rotated between successive
photons to avoid the
possibility that, for example, a single large particle separated from the main
particle field could dominate the results. This rotation is
allowed when particle fields that are homogeneous in the planar
directions are used. In contrast, in the self-gravitating examples,
the correct direction of viewing/illumination with respect to gravity
wakes must be maintained, in which case
 a larger number of simulation snapshots
(40) is averaged in order to reduce noise. Finally, since our main interest is in the opposition
effect, which represents a deviation from the classical zero volume
case, it is important that our method can reproduce very accurately
the classical results in the asymptotic limit $D \rightarrow 0$ (see
Fig.~4 in SK2003).

In the current study, two different particle phase functions are used:
the Lambert law
\begin{equation}
P_{L}(\alpha)= \frac{8}{3\pi} [\sin \alpha + (\pi -\alpha) \cos \alpha]
\label{eq:lambert}
\end{equation}
\noindent and a power-law phase function 
\begin{equation}
P_{\rm power}(\alpha)= c_n (\pi -\alpha)^{n_s},
\label{eq:power}
\end{equation}
\noindent where $c_n$ is a normalization constant ($\int_{4\pi}
P(\alpha)d\Omega=1$). For $n_s=3.09$, the latter formula gives a good
match to the phase function of Callisto \citep{dones1993}.  In the
case of Lambert scattering, we utilize the fact that the above given
{\em spherical-particle} phase function (van de Hulst 1980) follows from a very simple
{\em surface-element} scattering law $S_L(\cos e,\cos i)= \cos e/\pi$,
where $e$ and $i$ measure the emergence and incidence angles with
respect to the surface element's normal vector (this formula means
that the brightness of the Lambert surface element, $I = \pi F \cos i~
S_L /\cos e = F \cos i$, is independent of viewing angle, being just
proportional to the incoming flux). Thus, in each scattering we sample
from the distribution $S_L$ to obtain the new photon direction with
respect to the normal vector of the local surface element (see SK2003
for details). The advantage of using surface element scattering is
that the location of scatterings on the particle surface is correctly
sampled, which is crucial for accurate calculation of mutual interparticle shadowing
effects.  On the other hand, when using the power-law phase function,
we sample from Eq.~(\ref{eq:power}) the new direction with respect to
the direction of the incoming photon. Since the particles have a
finite size, this necessarily involves an approximation as compared to
using a proper surface element law. In SK2003 the Lambert surface
element scattering law and the Lambert spherical-particle phase
function treatments were compared in detail (see their Fig.~10), and
it was shown that both treatments give very similar results, provided
that the scattering location at the particle surface is accurately
sampled, so that the emerging photon is continued from the point of
scattering (instead of continuing from the particle center, which
would be conceptually more in accordance with the use of spherical
particle scattering law --- this alternative would, however, significantly
reduce the opposition brightening).

In the current study we use Lambert surface element scattering
whenever we want accurate estimates of the interparticle shadowing, i.e
the function $f_e(\alpha,\Beff, \tau)$ (the spherical-particle
treatment would also be sufficiently accurate near opposition,
provided that the photon path is continued from the intersection
point). The power-law phase function is mainly used in Section 5,
where we calculate the contribution of multiple scattering, and want
to compare the Lambert and power law phase functions. Although
the amount of multiple scattering itself, $I_{ms}$, for a given Bond
albedo $A$, is not strongly dependent on the phase function, the
fractional contribution $I_{ms}/(I_{ss}+I_{ms})$ will depend on $A$,
as a different $A$ is needed for a given phase function model to match
the low $\alpha$ observations dominated by $I_{ss}$. For example, the
ring brightness observed in F555W filter can be matched with the
standard $n_s=3.09$ power-law phase function if $A \sim 0.4$ is
adopted.  Since the Lambert phase function is less backscattering than
this power-law phase function, a larger $A \sim 0.7$ is needed to
obtain a similar low $\alpha$ brightness. As a consequence, the role
of multiple scattering will be more important in models using the
Lambert phase function.

The principal difference between our approach and the Porco et
al. (2008) ray tracing method is that we include scatterings to an
arbitrary order.  Additionally, our method uses Monte Carlo sampling of the
particle phase function (either the surface element law, or the spherical
particle model), so that after each scattering event a single emerging
photon is followed. The computational burden is thus {\em at most}
equally divided between each scattering order (in practice, the few
first orders dominate as the photon paths are terminated when they
leave the particle layer).  On the other hand, in Porco et al. (2008)
each successive scattering is represented by a bundle of emerging
photons, chosen according to a discretized phase function. This
implies that each successive scattering order requires more and more
computations (until they become computationally prohibitive; in
practice Porco et al. 2008 usually treat orders only up to 4),
although their contribution to the final result gets rapidly smaller.
Our Monte Carlo approach will lead to identical results, but with a
significantly reduced statistical variance for a given computational
effort.  The improved efficiency of our method might be quite
significant in some applications, in particular when dealing with
cases where multiple scattering is more important (transmitted
radiation, high phase angles, high particle albedo) than in the
current topic of opposition brightening, which is dominated by
first-order scattering.

%Also, many of the model checks in Porco et al (2008) (comparison to D=0)
%do not look very convincing. This might inidcate that insufficient number of
%scattering orders was treated, or that the number of photons was limited

\subsection{Grid of dynamical and photometric simulation models}

We study the effect of ring structure on the expected interparticle
mutual shadowing by performing simulations with different
dynamical optical depths ($\tau_{dyn} =0.1 - 2.0$) and particle size
distributions, assumed to follow a power law distribution
\begin{equation}
dN/dR \propto R^{-q}, \ \ R_{\rm min} < R < R_{\rm max}, \ \ q=3,
\end{equation}

\noindent with the ratio $R_{\rm min}/R_{\rm max}=0.02-0.2$. 
In all models the maximum radius $R_{\rm max}=5$ m. 
For the
elasticity of particles the Bridges et al.~(1984) velocity-dependent
coefficient of restitution is assumed
\begin{equation}
\epsilon_n(v_n) = \min[(v_n/v_c)^{-0.234},1],
\label{eq:bridges}
\end{equation}
\noindent where $v_n$ is the normal component of the relative velocity
of the impacting bodies and the scale parameter $v_c$ equals $v_B
=0.0077$ cm s$^{-1}$ in Bridges et al.'s measurements.  The
simulations are performed for the Saturnocentric distance $a= 100,000
$ km, with $\Omega = 1.94 \cdot 10^{-4} \ $s$^{-1}$.  Also, to keep
the models simple, self-gravity is not included in these simulations
(it is studied separately in Section 5 below).  The dynamical
simulations are performed with the local code, using the periodic
boundary conditions introduced by Wisdom and Tremaine (1988) and Toomre
and Kalnajs (1991); for more details of the code see Salo (1995) and
Salo et al.~(2001).  These models are then illuminated/viewed from the
elevation angles $B=4^\circ-26^\circ$ and the phase angle is varied
between $0^\circ$ to $90^\circ$.  Compared to SK2003, where several
examples of opposition brightening were given, and compared to
theoretical treatments of Lumme and Bowell (1981) and Hapke (1986), we
now cover a larger range of optical depths and viewing elevations in a
systematic manner, chosen to correspond to the range of HST
observations. %{\bf 
Also a larger range of $\alpha$ is explored, with
  future applications to spacecraft observations in mind (see Section
  6).
 % }. 
  Note that the simulation models are defined in terms of the
dynamical (geometric) optical depth, i.e.~the total fractional
area of particles. %{\bf 
As discussed in SK2003, in general
  $\tau_{\rm dyn}$ differs from photometric optical depth $\tau_{\rm phot}$,
  defined in terms of the probability $p$ for a light ray passing
  through the layer in the perpendicular direction,
  $p=\exp(-\tau_{\rm phot})$. However, at the classical limit $D=0$ we have
  $\tau_{\rm dyn}=\tau_{\rm phot}$, so that when referring to the
  theoretical $I_{ss}$ in Eq.~(\ref{eq:iss}) we make no distinction
between dynamical and photometric optical depths, and leave out the
subscript from $\tau$.%}

{\bf \hskip 5cm INSERT FIG  \ref{fig:model_grid_raw} HERE}

Figure \ref{fig:model_grid_raw} displays examples of calculated
interparticle shadowing curves, in terms of $f_e=
I_{ss}/I_{ss}(D=0)$. It is immediately evident that the mutual
interparticle shadowing opposition effect, measured as a deviation
from the classical single scattering result, %{\bf 
may extend to
  several tens of degrees.\footnote{In fact, the reflected $I_{ss}$ is
  enhanced for any phase angle: see SK2003 Fig.~7 and the discussion
  related to it, and Fig. \ref{fig:discussion_oppo_extended}
  below. This enhancement follows, since for a geometrically thin
  layer the illuminated upper layers are preferentially visible at
  every lit side geometry.} The maximum amplitude
$f_e(\alpha=0^\circ)$ is practically independent of the adopted size
distribution and approaches the theoretical maximum $f_e=2$
\citep{irvine1966} when the path optical depth $\tau_{\rm path}= \tau_{\rm
  phot}/\mu+\tau_{\rm phot}/\mu' =2 \tau_{\rm phot}/\mu_{\rm eff}$ is
large. Also, $f_e(0^\circ)$ is reduced for smaller $\tau_{\rm path}$,
regardless of the assumed particle size distribution: the maximum
amplitude is in good agreement with the theoretical estimate (SK2003)
\begin{eqnarray}
\label{eq:fe}
f_e(0^\circ) &=& \frac{2}{1+\exp(-\tau_{\rm path}/2)} \\[0.5cm]
&\approx&  1 +\frac{1}{4} \tau_{\rm path} - 0({\tau_{\rm path}}^3) ..., \nonumber
\end{eqnarray}
\noindent which follows from the theoretical treatment of
Lumme and Bowell (1981). In
Fig.~\ref{fig:model_grid_raw} this estimate is marked with a
horizontal line. Also shown in the upper right corner of the frames is
the photometric optical depth, which in dense homogeneous systems
exceeds $\tau_{dyn}$ by a factor $\sim (1+D)$ (SK2003; see also Peltoniemi and Lumme 1992).  Notice that
Eq.~(\ref{eq:fe}) implies a similar dependence on $\tau_{path}$,
regardless of which combination of $\tau_{\rm phot}$ and $\Beff$ produces it (see
Fig.~13 in SK2003 for detailed comparison for homogeneous systems;
their $E_{max}$ corresponds to $f_e(0^\circ)$).

Also shown in  Fig.~\ref{fig:model_grid_raw} are fits to the simulated shadowing
curves, using the Hapke (1986, 2002) formula for the single
scattering brightness enhancement due to SH in a semi-infinite
particle layer,
\begin{equation}
B_{SH}(\alpha)= 1 + \frac{B_{s0}}{1+\frac{\tan (\alpha/2)}{h_s}}.
\label{eq:hapke_sh}
\end{equation}

\noindent Here, $B_{s0}$ is the fractional amplitude, and $h_s$
describes the width of the effect. The fits, indicated by solid curves
in Fig.~\ref{fig:model_grid_raw} use Monte Carlo results for $\alpha <
10^\circ$. Very good agreement is seen (the mean RMS deviation is 0.01),
except for the simulations with the widest size distribution, in which
case the fit range extends furthest away from the peak of the function
(note that the Hapke SH formula implies $B_{SH}(\alpha) \rightarrow 1$
for large $\alpha$, whereas Monte Carlo simulations indicate $f_e >
1$). Similarly, the Monte Carlo curves are well-fitted with the
4-parameter linear-exponential formula, yielding a mean RMS $\sim
0.005$.

{\bf \hskip 5cm INSERT FIG  \ref{fig:model_hwhm} HERE}

As seen in Fig.~\ref{fig:model_grid_raw}, the angular width of the
simulated interparticle shadowing opposition peak depends strongly on
the adopted size distribution, and most importantly, also on the
elevation angle. On the other hand, it is nearly independent of
$\taud$.
 Fig.~\ref{fig:model_hwhm} shows the HWHM of the
interparticle shadowing enhancement factor, both as obtained directly
from $f_e(\alpha)$ curves (in the left), by setting
\begin{equation}
f_e(\alpha={\rm HWHM})-1 = \frac{1}{2}\left[f_e(\alpha=0^\circ)-1\right],
\end{equation}

\noindent and from the Hapke SH model (${\rm HWHM} \approx 2h_s$), as well as from the
linear-exponential model fits to the simulation curves. For the Hapke
model fits the HWHM's are close to the actual simulation values,
whereas for the linear-exponential fits the HWHM's are roughly 2-3
times smaller: the discrepancy is due to the fact that the HWHM for
the linear-exponential fit is calculated from the exponential part,
although a large part of the fit is due to the linear slope.

The fact that the widest size distribution has the narrowest
opposition peak, is consistent with the Hapke (1986) theoretical
model for semi-infinite particle layers. According to his formulae,
\begin{equation}
\frac{{\rm HWHM}}{D(\tau_s=1)} = \frac{3}{4} \frac{<\sigma> <R>}{<V>} =
\frac{3}{4} \frac{<R^2>^{1.5}}{<R^3>} =\frac{3}{4} Y,
\end{equation}

\noindent where $D(\tau_s=1)$ is the volume density at the layer where the slant
optical depth equals unity, $<\sigma>$ and $<V>$ are the average
scattering cross section and particle volume, and
$<R>=\sqrt{<\sigma>/\pi}$.  For a $q=3$ power law size distribution
the Hapke function $Y$ is

\begin{equation}
Y= \frac{\sqrt 2~(\ln W_s)^{1.5}~ W_s} {(W_s-1)~(W_s^2-1)^{0.5}},
\label{eq:hapke4}
\end{equation}

\noindent where $W_s=R_{\rm max}/R_{\rm min}$.  Applied to our size
distributions, and using the central plane volume filling factors
($D(z=0)=0.32 - 0.38$) for $D(\tau_s=1)$, this predicts that
HWHM=$3-12^\circ$, for $W_s=50-5$, respectively. In addition to
predicting correctly the relative change in HWHM, these width
estimates are quite close to the results of the calculations for
$B=26^\circ$, suggesting that for this elevation the reflection is
indeed from the dense equatorial layer. On the other hand, for lower
elevations the reflection is actually from the upper layers with
a substantially smaller effective D (see Fig.~1 for a schematic
illustration). According to Fig.~9 in SK2003, for the dynamical
simulations studied here, the filling factor at a given vertical
coordinate $z_o$ (on the side of illumination/viewing) is roughly
proportional to the path optical depth for a perpendicular
illumination reaching this layer,
\begin{equation}
D(z_0) \propto \tau_{\rm phot}(z>z_0).
\end{equation}
\noindent Since for an oblique view $\tau_s=\tau_{\rm phot}/\sin B$,
setting $\tau_s\approx1$ corresponds to reflection from the level
$z_0$ where $\tau_{\rm phot}(z>z_0) \approx \sin B$, at which elevation
the volume density $D(z_0) \approx \sin B \times D(z=0)$. This
explains the practically linear dependence,
\begin{equation}
{\rm HWHM} \propto B,
\end{equation}
coming in
addition to the size distribution dependence implied by the Hapke (1986)
formula.  Note that for a vertically uniform ring, the expected width
would be independent of $\beff$, regardless of size distribution.

%IDL> print,yhapke(5.)
%     0.552576
%IDL> print,yhapke(50)
%     0.167521

{\bf \hskip 5cm INSERT FIG  \ref{fig:model_f_60_05} HERE}

Figure \ref{fig:model_f_60_05} shows the enhancement factors
$f_e(0.5^\circ)$ and $f_e(6^\circ)$, which mark the range of phase
angles for which the simulated and observed opposition enhancements
will be compared in the next section. Two optical depths ($\taud=0.1$
and $1.5$) are compared: note that although for the small $\taud$ the
$f_e$'s are much smaller, the ratio $OE_e=f_e(0.5^\circ)/f_e(6^\circ)$
(right hand panel) can still be fairly large for low elevations, at
least qualitatively consistent with Fig.~7 showing significant
elevation angle dependence of $OE_{\rm obs}$ even for the C ring.  For the
larger optical depths, the enhancement factors, as well as their
ratio, are typically larger: however, the ratio
$OE_e=f_e(0.5^\circ)/f_e(6^\circ)$ depends in a quite complicated way
on the width of the assumed distribution and the elevation
angle. Essentially, this is due to the fact that for the wide
distributions, the width of the opposition peak sweeps through the
range $\alpha=0.5^\circ - 6.0^\circ$ when $B$ decreases.

{\bf \hskip 5cm INSERT Fig.~\ref{fig:model_oe_range}  HERE}

Finally, Fig.~\ref{fig:model_oe_range} shows examples of
opposition brightening for different optical depths and elevations,
both with and without the inclusion of multiple scattering. Instead of
normalizing to the theoretical single scattering values as in the previous
figures, the figure shows $I/I(6^\circ)$. In this figure, the $n_s=3.09$
power-law phase function with $A=0.5$ was used. The small difference
between the total intensity curves and those for single scattering
again underlines that, at least for these phase functions,
multiple scattering has only a minor role on the near-opposition phase
curves. In section 5, we show that this conclusion is also supported by
the tilt effect observations.

%%% 

%%%%%%%%%%%%%%%%%%%%%%%%%%%%%%%%%%%%%%%%%%%%%%%%%%%%%%%

\section{Extracting the elevation angle dependent component via model comparisons}
\label{sec:simucomp}

\subsection{Separating intrinsic and intraparticle contributions}

From the analysis presented in Section 3.2, we have not yet determined
the absolute contributions of intrinsic ($f_i$) and interparticle
($f_e$) contributions to the near-opposition brightness
increase. Rather, we have identified the amount of $OE_e(\beff)$
relative to $\beff= B_{\rm norm}$, or the relative amount of
$OE_i(\lambda)$ relative to $\lambda=\lambda_{\rm norm}$, respectively
(here, $OE_e \equiv f_e(\alpha_{\rm min})/f_e(\alpha_{\rm max})$ and $OE_i \equiv
f_i(\alpha_{\rm min})/f_i(\alpha_{\rm max})$).  In order to estimate the
interparticle shadowing enhancement factor $f_e(\alpha, \beff)$
we utilize simulation modeling: we compare the observed
$OE_{obs}(\beff)/OE_{obs}(B_{\rm norm})$ (using $\amin=0.5^\circ$,
$\amax=6^\circ$, $B_{\rm norm}=20^\circ$) to the
$OE_e(\beff)/OE_e(B_{\rm norm})$ ratios indicated by the above
described simulations performed for different optical depths and
widths of the size distribution.

The best match simulation then implies a particular $f_e$, and the
intrinsic contribution can be estimated from Eq.~(\ref{eq:if}) 
(ignoring the $Q_{\rm ms}$ term):

\begin{equation}
f_{i}(\alpha,\lambda) =
\frac{I_{obs}(\alpha,\beff,\tau, \lambda)}{I_{ss}(\alpha, \beff, \tau, \lambda)} \frac{1}{f_{e}(\alpha,\beff,\tau)},
\end{equation}

\noindent where $I_{ss}$ is the theoretical singly scattered intensity
for $D=0$.  Note that there is still some freedom here, since the
$I_{ss}$ contains the product $A P(\alpha)$, which, as being
independent of $\beff$ cannot be separated from $f_i$.  In practice
we will divide the observations with the simulated interparticle
shadowing contribution and determine the ratio

\begin{equation}
g_i(\alpha) \equiv \frac{f_i(\alpha) P(\alpha)}{f_i(6^\circ) P(6^\circ)}=\frac{I_{obs}(\alpha)}{I_{obs}(6^\circ)} : \frac {f_e(\alpha)}{f_e(6^\circ)}.
\end{equation}

\noindent The normalized intrinsic effect $g_i(\alpha)$ in the left
side is also a quite good approximation for $f_i(\alpha)$
itself. That is, since the intrinsic opposition effect has ${\rm
  HWHM}<1^\circ$, we have $f_i(6^\circ) \approx 1$. Also, the
variation in the phase function between $\alpha=0^\circ$ to $6^\circ$
is most likely very small ($P(0^\circ)/P(6^\circ)$ =1.005 and 1.11 for
the Lambert and the $n=3.09$ power law phase functions, respectively).
As a useful check of the extraction procedure, we can use the fact that
observations at all elevation angles should yield the same
$g_i(\alpha)$.

{\bf \hskip 5cm INSERT FIG  \ref{fig:simu_compare_oe_cba} HERE}

Figure \ref{fig:simu_compare_oe_cba} shows this procedure applied to
the C, B, A ring regions. In the left panels, the 
$OE_{obs}$ in the U (F336W) and I (F814W) filters is compared to simulations performed
with various widths of the size distribution, while in the right panels, both
observations and simulations have been normalized to the $OE$ at
$B_{norm}=20^\circ$.  In the left hand panels the observed $OE$'s are
clearly larger than the simulated ones, which contain just the
interparticle shadowing contribution: the excess is due to the
intrinsic opposition effect.  Also, the $OE$ in U is clearly larger
than in the I filter. However, at right, after normalization to
$B_{norm}=20^\circ$, the observed and simulated elevation angle trends
are much closer to each other.  Also note how well the normalization
removes the wavelength dependence of the observations in the right
panels, with the U and I filters behaving in a very similar manner.
% (the same
%data points were shown previously in Fig.~7, but separately for
%different filters). 

For the C ring (upper row in Fig.~\ref{fig:simu_compare_oe_cba}) a
detailed comparison is made to simulations with $\taud=0.1$, in which
case the magnitude of the elevation angle dependent $OE$ increases
monotonically with the width of the size distribution.  Clearly, the
best match to the elevation angle dependent part is obtained with the
widest studied distribution $W_s=50$; the curves also suggest that a
still larger $W_s$ would further improve the fit.  On the other hand, the B
ring (middle row) is compared to simulations with $\taud=2.0$, and now
the match is best for a much narrower size distribution $W_s=5-10$.
The main difference in the simulated
$OE_e(\Beff)/OE_e(B_{norm}=20^\circ)$ curves for small and large
$\taud$'s is the turning down at low $\beff$ values in the case of
large $\taud$ and large $W_s$. As mentioned earlier, the reason is
that in this case the width of the interparticle shadowing peak is so
small that it falls inside the studied $\alpha$ range.  For the A ring
region (lower row) the comparison is made to simulations with
$\taud=1$, and just as for the B ring region, a quite narrow size
distribution $W_s \sim 5$ is preferred.  Interestingly, for the A ring
the $\taud=1$ (or even $\taud=2$) case provides a slightly better
match than that with the nominal $\taud=0.5$, although the difference
is not large. This is not surprising, taking into account that the mid-A ring 
is the location where self-gravity wakes are strongest: the
actual amount of light reflection must result from  a superposition of dense wakes and rarefied
gaps, with the wakes having much higher optical depth than the nominal
$\tau \sim 0.5$ \citep{colwell2006, Hedman07a}. The fact that we are comparing the A ring to
non-gravitational simulations in the first place might seem
suspect. However, French et al.~(2007b) showed that there is very
little difference in the phase curve between wake and non-wake
simulations.

{\bf \hskip 5cm INSERT FIG  12 HERE}

%%%%%%%%%%% SIMPLIFIED - COMP WITH PREVIOUS VERSIONS

In order to demonstrate that the above extraction procedure works as
intended, Fig.~\ref{fig:oppofit_residuals} compares the original HST
F336W phase curves (left panels) with those after the removal of the
elevation angle dependent part (middle).  All the curves are shown
normalized to $\alpha=6^\circ$. Clearly, the residual curves in the
middle panel, $g_i(\alpha)=f_i(\alpha)
P(\alpha)/[f_i(6^\circ)P(6^\circ)]$, are all very close to each other,
indicating that the removal of the elevation dependent part has been
successful (similarly for the other filters; see
Fig.~\ref{fig:oppofit_residuals_5colors}). The removed interparticle
shadowing contributions themselves, $f_e(\alpha)/f_e(6^\circ)$, for
the various $\beff$'s are also shown (right panels).

%for the various values of
%$\beff$ 
%Figure \ref{fig:oppofit_residuals_5colors} is
%similar to the middle frames of Fig.~\ref{fig:oppofit_residuals},
%except that the residuals in all five different filters are compared.

{\bf \hskip 5cm INSERT FIG  13 HERE}

\subsection{Fitting the intrinsic component}

%{\bf 
Once the elevation angle dependent part of the opposition effect has
been removed, it is interesting to make model fits to the residual
curves, which presumably represent the true intrinsic opposition
effect. In this subsection, the fit parameters are given for both
linear-exponential and Hapke (2002) models. We also compare our
results to French et al.~(2007b), to see how much the deduced
intrinsic parameters differ from those obtained from fits to original
high elevation angle data points, where the effect of interparticle
shadowing is least pronounced.
%}

Figure~\ref{fig:oppofit_residuals_5colors} shows the deduced residual
intrinsic components (large symbols), together with model
fits (thick blue curves), for all five different filters. These data,
collecting the observations from all elevations, are also compared
with the original high elevation angle data ($\beff \sim 23^\circ$;
indicated by small symbols and thin orange curves\footnote{\label{footnote_26_23} The
  $\beff=26^\circ$ data are also included, after they have been
  corrected to correspond to $\beff=23^\circ$, by first dividing by
  $f_e(26^\circ) I_{ss}(26^\circ)$ and then multiplying by
  $f_e(23^\circ) I_{ss}(23^\circ)$.}): since from these latter points
the interparticle shadowing contribution has not been removed, the
ratio between the data sets measures the amount of interparticle
shadowing correction for $\beff=23^\circ$.

The model fits shown in Fig.~\ref{fig:oppofit_residuals_5colors} are made with the Hapke (2002)
formulation, which includes both intraparticle shadow hiding and coherent
backscattering contributions,
 \begin{equation}
  g_i(\alpha) = A_i B_{SH}(\alpha) B_{CB}(\alpha).
\end {equation}

\noindent Here $B_{SH}(\alpha)$ describes the intrinsic shadow hiding
part, assumed to be similar in form to Eq.~(\ref{eq:hapke_sh}) used above for fitting the
simulated interparticle shadowing, and $B_{CB}(\alpha)$ is the
coherent backscattering contribution (with fractional amplitude
$B_{c0}$ and ${\rm HWHM} \approx 0.72 h_c$),

\begin{equation}
B_{CB}(\alpha)= 1 + B_{c0} ~ \frac{1+\displaystyle \frac{1-e^{-\tan({\alpha}/{2})/h_c}}{\tan(\alpha/2)/h_c}}
{2 \Big[1+\tan(\alpha/2)/h_c\Big]^2}
\label{eq:hapke_cb}.
\end{equation}

\noindent Compared to Hapke (2002), we have omitted from shadow hiding
the part containing the Henyey-Greenstein phase function of regolith
grains, as well as the multiple scattering at the particle surface
regolith, since the parameters related to these contributions cannot
be reliably determined from the near-opposition data alone. Here, this
part is absorbed into a single parameter $A_i$ related to the
unspecified optical properties of regolith grains.  Note, however,
that we apply this fit to data normalized to $\alpha=6^\circ$: in this
case $A_i$ is not an independent parameter but is determined by the
normalization and the other parameters.  A similar model, except for
fitting the original high elevation $I/F$ measurements, and including
the full Hapke (2002) formulas for the grain phase function and
multiple scattering, was used in French et al.~(2007b). As in French
et al.~(2007b), here we also take into account the finite size and
limb darkening of the solar disc in the calculation of model
brightness for the near-to opposition phase angles.

The parameters of the intrinsic effect fits shown in
Fig.~\ref{fig:oppofit_residuals_5colors} are collected in Table
\ref{table:intrinsic_hapke}: the typical RMS residuals of the fits
normalized to $\alpha=6^\circ$ are of the order of 0.015-0.02 (similar
to those obtained when using original uncorrected high elevation angle
data). For comparison, fits to the intrinsic $g_i(\alpha)$ using the
linear-exponential formula (Table \ref{table:intrinsic_linexp}) yield
residuals comparable in magnitude to those using the simplified Hapke
model.

{\bf \hskip 5cm Insert Table \ref{table:intrinsic_hapke}}   %6

{\bf \hskip 5cm Insert Table \ref{table:intrinsic_linexp}}  % 7

Figure \ref{fig:oppofit_residuals_5colors} also shows separately the
intraparticle SH contribution: in the fitted models, this is
conveniently quantified by the enhancement factor at the zero phase,
$SH(0^\circ)= A_i (1+B_{s0})$.  Interestingly, for the A and B rings,
the amount of the deduced intra-particle SH is almost negligible
($SH(0^\circ) \sim 1.0$), except for F336W, where $SH(0^\circ) \sim
1.1$.  For the A and B ring regions, the intrinsic brightening can
thus be accounted for almost entirely by the coherent backscattering, with a
typical value $CB(0^\circ)=(1+B_{c0}) \approx 1.4-1.5$. On the other
hand, for the C ring the SH brightening is clearly stronger in all
filters, with $SH(0^\circ) \sim 1.3$; in contrast, the CB contribution
is quite similar to the B and A ring regions. Note that this kind of
separation is not possible when using the original data without
removal of the interparticle contribution.

{\bf \hskip 5cm INSERT FIG  14 HERE}

{\bf \hskip 5cm INSERT FIG  15 HERE}

%%%%%%%%%%%%%%%%%%%%%%%
%THE FOLLOWING IS PERHAPS UNNECESSARY?
%%% - I THINK IT IS FINE TO INCLUDE THIS - Dick

The linear-exponential and Hapke model fit parameters are further
displayed in Figs.~\ref{fig:oppofit_residuals_linexp_fits} and
\ref{fig:oppofit_residuals_hapke_fits}, respectively.  Altogether,
fits to three different data sets are compared:

1) The original high elevation angle data set used in French et
al.~(2007b), combining the Cycle 13 exact opposition point for
$\beff=22.9^\circ$ with the $\beff=26^\circ$ data (upper row).

2) The original $\beff=23^\circ$ (Cycles 13 and 9 combined ) and
$\beff=26^\circ$ (Cycles 10-12) data sets, the latter normalized to
$\beff=23^\circ$ as described in footnote \ref{footnote_26_23} (middle row).

3) The combined data set from all $\beff$'s, containing
just the intrinsic component (lower row).

\noindent Data sets 2) and 3) are those discussed in connection to
Fig.~\ref{fig:oppofit_residuals_5colors}.  A comparison of data
sets 1) and 2) (the two first rows in
Fig.~\ref{fig:oppofit_residuals_linexp_fits}) shows that the A and B
ring fits are nearly identical: both sets indicate that HWHM is about
$0.1^\circ$ for the BVRI filters, rising to $0.15-0.20^\circ$ toward
U. However, for the C ring region, although the estimated HWHM is more
or less the same in all filters for both data sets, there is a nearly
two-fold difference in the fitted values.  A similar difference
between the data sets 1) and 2) is also seen in the Hapke model
parameters (compare the two uppermost rows of
Fig.~\ref{fig:oppofit_residuals_hapke_fits}).  The difference arises
from the inaccurate inclusion of the Cycle 13 exact opposition
point among the $26^\circ$ data without accounting for the different
elevation (data set 1). The strong elevation angle
dependence when combining such nearby $\beff$'s is due to the low
optical depth of the C ring, making its $I/F$ decrease sharply
with elevation angle. That is, for $\tau=0.1$ the geometrical factor
in the singly scattered $I/F$ (the exponential term in
Eq. (\ref{eq:iss})) is about 1.1 times larger for $\beff=23^\circ$
than for $\beff=26^\circ$.  For the higher optical depth B and A rings the
dependence is much weaker (for $\tau=0.5, 1.0, 2.0$ the difference in
$(I/F)_{ss}$ is 1.03,1.005, 1.0001, respectively) and the nearby
$\beff$'s can be safely combined.\footnote{In principle, the fact that
  we are fitting $I/I(6^\circ)$ instead of $I/F$ itself could affect
  our fits. However, we confirmed that a similar change in the C ring
  fit parameters is seen in the French et al. (2007b) original fits,
  if the $I/F$ for the Cycle 13 point is divided by
  $I_{ss}(23^\circ)/I_{ss}(26^\circ) \approx 1.1$ before combining
  with the $26^\circ$ data.}

Comparing the fits for the intrinsic opposition effect data, and for the
original $\beff=23^\circ+26^\circ$ data set properly normalized to
$\beff=23^\circ$ (third and second rows, respectively, in 
Figs.~\ref{fig:oppofit_residuals_linexp_fits} and
\ref{fig:oppofit_residuals_hapke_fits}), we can see that the main
difference is in the linear slope (for the linear-exponential fits),
and in the $SH(0)$ amplitude (for the simplified Hapke model fits). For the
linear-exponential model the slope is significantly reduced, in
particular for the B and A rings, and the same is true for the $SH(0)$
in the Hapke model. This quantifies the difference seen in 
Fig.~\ref{fig:oppofit_residuals_5colors} between the two sets of model
curves.  On the other hand, the HWHM's are almost unaffected by
the removal of interparticle shadowing contribution. 
%The fits to the
%intrinsic opposition effect are listed in Tables
%\ref{table:intrinsic_linexp} (linear-exponential model) and
%\ref{table:intrinsic_hapke} (Hapke model).

{\bf \hskip 5cm Insert Table \ref{table:intrinsic_external_23} here}  %8

Table \ref{table:intrinsic_external_23} lists the intrinsic (SH +CB)
and external (interparticle shadowing) contributions to $OE(0.5^\circ)
\equiv I(0.5^\circ)/I(6^\circ)$ and $OE(0^\circ) \equiv
I(0^\circ)/I(6^\circ)$, for different filters and ring regions. The
internal contribution $OE_i$ is calculated from the Hapke fit
parameters (solid curves in Fig.~\ref{fig:oppofit_residuals_5colors}),
and the interparticle contribution $OE_e$ has been calculated for
$\beff=23^\circ$, using the best fit simulation models shown in
Fig. 12.  The modeled total opposition enhancement is then $OE = OE_i
\times OE_e$; for $OE(\alpha=0^\circ)$ this can be compared to the
observed (Cycle 13) $I(0^\circ)/I(6^\circ)$ listed in the last column
(showing agreement to within 2\%).  Table
\ref{table:intrinsic_external_4} is similar, except that $OE_e$ has
been calculated for $\beff=4.5^\circ$: combined with $OE_i$ the
predicted total enhancement $I(0^\circ)/I(6^\circ)$ is about 2.7, 2.5,
and 2.6 for the C, B, and A rings, respectively. Unfortunately there
are no HST comparison data, since the minimum phase angle during the
low elevation angle opposition (Cycle 6) was $\approx 0.3^\circ$ (the
smallest observed $\alpha =0.46^\circ$).

%%%%%%%%%%%%%%%%%%%%%%%%%%%%%%%%%%%%%%%%%%%%%%%%%%%%%%%%

\section{Tilt effect: interparticle shadowing or multiple scattering?}
\label{sec:tilt}

%FROM ASY-REMOVED

\subsection{HST observations of the tilt effect at different filters and phase angles}

Traditionally, the term {\em tilt effect} refers to the brightening of
the B ring with increasing elevation, amounting to as much as $30\%$
for the ground-based range of $\Beff$'s \citep[e.g.][]{lumme1983}, in
contrast to the nearly constant brightness one would expect for an
optically thick classical multilayer ring dominated by single
scattering. On the other hand, for the A ring the observed $I/F$ was
found to be nearly constant or slightly decreasing with increasing
ring tilt (Lumme and Irvine 1976b), more consistent with the single
scattering prediction, Eq.~\ref{eq:iss} ($I_{ss}/F$ is a decreasing
function of $\beff$ in the case of $\tau \lesssim 0.5$, and
practically constant for $\tau \ge 0.5$ ).
Note that these early measurements actually refer just to
the brightest innermost portion of the A ring: the HST images (Cuzzi
et al.~2002) revealed that for the mid-A ring the brightness in fact
decreases quite markedly with elevation, much more than predicted by
Eq.~\ref{eq:iss}.

 In Salo et al.~(2004)
this A-ring negative tilt effect was attributed to the increased
visibility of the gaps between self-gravity wakes at larger
elevations. Indeed, this explanation in terms of gaps/wakes is now
fully supported by the Cassini occultation measurements (Colwell et
al.~2006, 2007, Hedman et al.~2007).

The B ring tilt effect has been viewed as a consequence of multiple
scattering, which becomes more important with increasing ring
elevation: this is also supported by the fact that the tilt
effect is pronounced for the optically thick B ring \citep{lumme1976b,
  esposito1977, lumme1983}. However, from analysis of HST observations
showing the lack of significant color variations with respect to ring
elevation, \cite{cuzzi2002} and \cite{poulet2002} concluded that
multiple scattering must be quite weak in the backscattering geometry
of Earth-based observations.  SK2003 proposed that the tilt effect is
a consequence of the variation in the effective filling factor with
opening angle, taking place for vertically non-uniform
rings. According to this view, based on N-body simulations and Monte
Carlo scattering calculations, the observed reflection at low
elevations is dominated by the rarefied upper ring layers, which
should have a very narrow opposition peak.  Thus the tilt-effect
observations, made typically at a phase angle of a few degrees, fall
outside the opposition peak. However, as the elevation angle
increases, the reflection is more and more dominated by the dense
equatorial ring layer. This should exhibit a much wider opposition
peak, which increases the observed brightness. 
The magnitude of this effect should also increase with  increasing $\tau$.
  However, in SK2003 no
suitable data were available for testing this hypothesis.  Note that
this explanation is intimately tied to the mechanism generating the
opposition effect of the rings: in the previous section we have shown
that this contains both intrinsic and interparticle contributions. It
is therefore important to test the SK2003 hypothesis for the tilt
effect, using the HST data.

% Additionally, it does not require that the opposition
%effect is due solely to mutual shadowing, but only that interparticle
%shadowing is significant. 

{\bf \hskip 5cm INSERT FIG  16 HERE}

The observed tilt effect is illustrated in Fig.~\ref{fig:tilt}, which displays
radial profiles of the ring ansa brightness at several elevations,
normalized to that at $\Beff=4.5^\circ$. Observations with the filter
F555W are shown separately for two phase angles: $\alpha=6^\circ$
(upper panel) and $\alpha=0.5^\circ$ (lower panel). The
$\alpha=6^\circ$ plots correspond to Fig.~8b in \cite{cuzzi2002},
except for the normalization. They show the strong positive tilt effect
($I/F$ increasing with increasing $\beff$) for the brightest part of
the B ring, and a weaker but still positive effect in the innermost A
and B rings. In contrast, the mid-A ring ($124,000-133,000$ km) has a
negative tilt effect, due to the aforementioned wakes.

{\bf \hskip 5cm INSERT FIG  17 HERE}

At low phase angle ($\alpha=0.5^\circ$), the behavior changes quite
markedly. For the A ring, the negative tilt effect is even more
pronounced at $\alpha=0.5^\circ$ than at $\alpha=6^\circ$.  In the
inner B ring ($93,000-99,000$ km), the tilt effect is now also
negative, and a positive effect is prominent only in the region
$105,000 - 110,000$ km.  
%In contrast to other ring regions, the
%outermost A ring beyond the Encke gap behaves in a similar manner for
%$\alpha=0.5^\circ$ and $6^\circ$. 
 To
exclude the possibility that the differences could be due to anomalous
behavior of $\Beff=4.5^\circ$  images (some of which were
affected by spokes, though not the ones included to
 Fig.~\ref{fig:tilt}),
 Fig.~\ref{fig:tilt_336_814} compares two other filters,
this time normalized to $\Beff=10^\circ$. Overall, a very similar
behavior is seen as in the previous figure, although the B ring
$\alpha=6^\circ$ tilt effect appears a bit stronger for F814W than for
F336W.

\subsection{Modeling the B ring tilt effect}

In principle, both interparticle shadowing and multiple scattering can
cause a positive tilt effect. Our goal in this section is to provide a
quantitative estimate of how much these factors contribute to the observed B
ring tilt effect. For interparticle shadowing, our estimate follows
directly from the opposition effect models of the previous section.  An
estimate for the fractional amount of multiple scattering, $Q_{\rm ms}$,
can be obtained by comparing the magnitude of the tilt effect at long
and short wavelengths. The particle albedo increases
significantly with wavelength in the visual regime, which also
increases the relative amount of multiple scattering: thus the
contribution of multiple scattering, if significant at all, should
result in a strong wavelength dependence in the tilt effect. Our
approach here, for multiple scattering, is thus very similar
to the Cuzzi et al.~(2002) color analysis.

{\bf \hskip 5cm INSERT FIG  18 HERE}

We quantify the tilt effect by the ratio of scaled intensities,
$\hat{I}=I/I_{ss}$, measured at a given elevation, normalized to that
of $\hat{I}$ at $\beff=4.5^\circ$.  Using the ratio of scaled
intensities, instead of intensities, simply removes the geometric
contribution to the tilt effect arising from the factor
$1-\exp(-2\tau/\sin \beff)$ in $I_{ss}$ (i.e. at small $\tau$,
$I_{ss}(B)$ is a decreasing function of $B$ while at larger $\tau$ it
is practically constant). The difference is not large: compare the two
uppermost frames in Fig.~\ref{fig:tilt_ms}, showing the A and B ring
tilt effect, by plotting either the ratio of intensities (upper
frame), or that of scaled intensities (middle frame) as a function of
\em Voyager} PPS (Photopolarimeter Subsystem) optical depth,
obtained from the NASA Planetary Data System Rings Node \citep{showalter1996}.
Both profiles highlight the strong $\tau$ dependence of the tilt effect.

Denoting the ratio $\hat{I}(\Beff)/\hat{I}{(\beff=4.5^\circ)}$  by $R_B$, we have
\begin{eqnarray}
R_B(\alpha,\beff, \tau, \lambda) &=&
\frac{(I/I_{ss})(\alpha,\beff,\tau, \lambda)}
     {(I/I_{ss})(\alpha,\beff=4.5^\circ,\tau, \lambda)} \cr \nonumber
     \\[0.2cm] \nonumber &\approx& \frac{f_i(\alpha,\lambda)
       f_e(\alpha,\beff,\tau) + Q_{\rm ms}(\beff, \tau, \lambda)}
        {f_i(\alpha,\lambda) f_e(\alpha,\beff=4.5^\circ,\tau)},
\end{eqnarray}

\noindent where we have utilized the fact that $Q_{\rm ms}$ in Eq.~(\ref{eq:if}) is
insignificant at $\beff =4.5^\circ$ \citep{cuzzi2002}. Furthermore, when evaluated at
$\alpha=6^\circ$, we can safely assume that $f_i \approx 1$ (since its
HWHM in Section 4 found to be $<< 6^\circ$). Therefore,

\begin{eqnarray}
R_B(\alpha=6^\circ, \beff, \tau, \lambda) &\approx& \frac{f_e(\alpha=6^\circ,\beff,\tau) + Q_{\rm ms}(\beff, \tau, \lambda)}{f_e(\alpha=6^\circ,\beff=4.5^\circ,\tau)}. 
\end{eqnarray}

 \noindent Since $Q_{\rm ms} << 1$,  we may further approximate
\begin{eqnarray}
R_B(\alpha=6^\circ, \beff, \tau, \lambda) &\approx& \frac{f_e(\alpha=6^\circ,\beff,\tau)}{f_e(\alpha=6^\circ,\beff=4.5^\circ,\tau)} + Q_{\rm ms}(\beff, \tau, \lambda),
\end{eqnarray}

\noindent since according to Fig.~\ref{fig:model_f_60_05}, the factor
$f_e(\alpha=6,\beff=4,5^\circ,\tau)$ should be close to unity. This
approximation illustrates that the tilt effect can indeed be partly
due to interparticle shadowing (the $f_e(\beff)/f_e(4.5^\circ)$ term), and
partly due to multiple scattering ($Q_{\rm ms}$): both imply enhanced
ring brightness at larger $\beff$, and also at larger $\tau$.
Concerning the interparticle shadowing contribution, the various size
distribution models of Fig.~\ref{fig:model_f_60_05} in Section 3 imply
an enhancement by 1.25--1.35 for $\Beff=26^\circ$, $\tau_{\rm dyn}=1.5$, all in qualitative
agreement with the B ring observations.  The interparticle shadowing
effect is thus very robust, in the sense that it does not require very
specific ring models in order to be able to account for the observed
strong tilt effect.
 
To isolate the $Q_{\rm ms}$ contribution, we form the difference of $R_B$ at
two different filters, with wavelengths $\lambda_1$ and $\lambda_2$,
\begin{equation}
\Delta_{\lambda}R_B= R_B(\alpha=6^\circ,\beff,\tau, \lambda_2)-R_B(\alpha=6^\circ,\beff,\tau, \lambda_1) \approx Q_{\rm ms}(\lambda_2)-Q_{\rm ms}(\lambda_1).
\end{equation}
\noindent
Following Cuzzi et al.~(2002; their Appendix) we assume that
the multiply-scattered flux $I_{ms}(\lambda) \propto [A(\lambda)]^n$, where
$n \approx 2-3$ indicates the typical order of scattering responsible 
for multiply scattered light.
The ratio of fractional multiple scattering contributions at different
wavelengths is then
$Q_{\rm ms}(\lambda_2)/Q_{\rm ms}(\lambda_1) \sim
[A(\lambda_2)/A(\lambda_1)]^{n-1}$.
Eliminating $Q_{\rm ms}(\lambda_1)$ leads to
an estimate
\begin{equation}
Q_{\rm ms}(\lambda_2) \sim \frac{\Delta_{\lambda}R_B}{1-[A(\lambda_1)/A(\lambda_2)]^{n-1}}.
\label{eq:qms-estimate}
\end{equation}

\noindent Since the backscattered flux is dominated by single scattering, the
ratio of albedos can be roughly estimated as
$A(\lambda_1)/A(\lambda_2) =
I_{ss}(\lambda_1)/I_{ss}(\lambda_2)\approx I(\lambda_1)/I(\lambda_2)$
evaluated at $\alpha=6^\circ, \beff=4.5^\circ$ (having minimal
contribution of multiple scattering and opposition brightening).  For the studied B ring region the observed $I/F$'s at
$\beff=4.5^\circ$, $\alpha=6^\circ$ are $0.16$ and $0.44$ at F336W and
F814W, respectively.  Equation (\ref{eq:qms-estimate}), together with
$n \sim 2$ then indicates $Q_{\rm ms}(F814W) \sim 1.5
\Delta_{\lambda}R_B$.  According to lowermost frame of
Fig.~\ref{fig:tilt_ms}, the maximum of $\Delta_{\lambda}R_B$ is $\sim
0.1$, suggesting a maximal $Q_{\rm ms} \sim 0.15$ for the F814W filter
(and $\sim 0.05$ for F336W). This justifies the omission of $Q_{\rm
  ms}$ in previous sections (the fractional error of the derived $f_i$
is of the order of $Q_{\rm ms}$).  Comparing to the interparticle
shadowing contribution (about 30\% enhancement in $R_B$) we can now
estimate that the relative contribution of multiple scattering to the
tilt effect should be about 1/3 of the total effect, at most, and practically negligible at shorter wavelengths. The
multiple scattering contribution corresponds roughly to the difference
between the F814W and F336W trends in Fig.~\ref{fig:tilt_ms} (upper or
middle frame, whereas the difference of F336W points from the
theoretical single scattering curve represents the interparticle
shadowing factor).

%The multiple scattering and
%interparticle shadowing factors correspond roughly to the difference
%between the F814W and F336W trends in  Fig.~\ref{fig:tilt_ms}, and the
%difference of F336W points from the theoretical single scattering curve,
%respectively.

The consistency of the estimated $Q_{\rm ms}$ can also be checked by a
direct comparison with simulation models. For example, the observed
$I/F$ in the B ring region (for $\Beff=26^\circ, \alpha=6^\circ$) can
be reproduced with an $n_s=3.09$ power law phase function by assuming
$A_{336}=0.21$ and $A_{814}=0.57$, when the dynamical model with
$\taud=2.0,~ W_s=10$ is assumed.  The same model implies
$R_B(814)-R_B(336) \sim 0.04$ (with the simulated maximum $Q_{\rm ms} \sim
0.06$), which is smaller but still in fair agreement with the
$\Delta_{\lambda}R_B \sim 0.06$ implied by the observations for the B
ring region.  On the other hand, for a Lambert phase function, albedo
values 0.36 and 0.87 would be required for these two filters. This in
turn would imply $R_B(814)-R_B(336) \sim 0.2$ (and $Q_{\rm ms} \sim
0.35$), a factor of three larger than the observed difference between
the two filters.  In Fig.~\ref{fig:tilt_ms} (bottom row) we also plot
the modeled $\Delta R_B$ as a function of $\tau$, for the two phase
functions, using the above albedo values.  We conclude that the
particles are significantly more backscattering than Lambert spheres,
and are more similar to those implied by the $n_s=3.09$ power law.
This is consistent with previous studies based on direct fitting of
large range Voyager phase curves (Dones et al. 1993).

{The reduction of the B ring tilt effect from $\alpha=6^\circ$ to
$0.5^\circ$, which was shown in  Fig.~\ref{fig:tilt}, is just what is
expected if the elevation-dependent interparticle shadowing is primarily
responsible for the observed tilt effect.  To see this, we may form

\begin{eqnarray}
R_B(\alpha=0.5^\circ,\beff,\tau, \lambda) &\approx& \frac{f_i(\alpha=0.5^\circ,\lambda) f_e(\alpha=0.5^\circ,\beff,\tau) + Q_{\rm ms}(\beff, \tau, \lambda)}{f_i(\alpha=0.5^\circ,\lambda) f_e(\alpha=0.5^\circ,\beff=4.5^\circ,\tau)}\nonumber \\ \nonumber \\ 
 &=& \frac{f_e(\alpha=0.5^\circ,\beff,\tau)}{f_e(\alpha=0.5^\circ,\beff=4.5^\circ,\tau)} +\\ \nonumber \\ \nonumber &&\frac{Q_{\rm ms}(\beff, \tau, \lambda)}{f_i(\alpha=0.5^\circ,\lambda) f_e(\alpha=0.5^\circ,\beff=4.5^\circ,\tau)}, 
\end{eqnarray}

\noindent Close to opposition, we can approximate $f_i f_e \sim 2$ in the divisor of the $Q_{\rm ms}$ term, leading to
\begin{eqnarray}
R_B(\alpha=0.5^\circ,\beff,\tau, \lambda) &\approx& \frac{f_e(\alpha=0.5^\circ,\beff,\tau)}{f_e(\alpha=0.5^\circ,\beff=4.5^\circ,\tau)} + 0.5 \, Q_{\rm ms}(\beff, \tau, \lambda).
\label{eq:21}
\end{eqnarray}

\noindent 
 According to Fig.~\ref{fig:model_f_60_05}, the
interparticle shadowing factor (the first term in Eq.~(\ref{eq:21}))
is reduced for $\alpha=0.5^\circ$ when compared to $\alpha=6^\circ$: in particular, for the narrow
$W_s=5-10$ size distributions, favored by the opposition phase curve fits
of Section 4, the modeled $f_e(26^\circ)/f_e(4^\circ) \sim 1.15$,
significantly smaller than the $\sim 1.3$ for $\alpha=6^\circ$. This
drop in the magnitude of the interparticle shadowing tilt effect, from
$\sim 30\%$ to $\sim 15\%$, is in remarkable agreement with
observations shown in Fig.~\ref{fig:tilt}.  Also, according to
Eq.~(\ref{eq:21}) the contribution from multiple scattering should be
reduced to roughly one-half for the smaller phase angle. This is
consistent with observations showing a weaker wavelength dependence of
tilt effect when $\alpha$ is reduced (see
Fig.~\ref{fig:tilt_336_814}).

In summary, the interparticle shadowing mechanism, by which the
interparticle opposition peak widens at larger elevations, can account
quite well for the observed positive tilt effect of the B
ring. Additionally, its phase angle dependence -- the differences seen
between $\alpha=0.5^\circ$ and $6.0^\circ$ -- are accounted for; this
effect had not been considered in earlier studies, which concentrated
on $\alpha=6.0^\circ$.  The increased amount of multiple scattering
with elevation seems to be a secondary effect, accounting primarily
for the slightly stronger tilt effect at longer wavelengths.

\subsection{Self-gravity wakes and the negative A ring tilt effect}

So far, our model comparisons have been made with non-gravitating
simulations. On the other hand, Saturn's A and B rings are known to
possess self-gravity wake structures (Salo 1992b, see also Toomre and
Kalnajs 1991, Colombo et al.~1976), responsible for the optical
brightness asymmetry (Camichel 1958, Lumme and Irvine 1976a, Thompson
et al.~1981, Franklin et al.~1987, Dones et al.~1993) and the optical
depth variations detected with various Cassini instruments (Colwell et
al.~2006, 2007, Hedman et al.~2007, Ferrari et al.~2009). Here we
address the connection of wakes to the negative tilt effect observed
in the mid A ring.  We model this region with the same two standard
self-gravity models that were used in Salo et al.~(2004) and French et
al.~(2007a) for studies of the A ring azimuthal brightness
asymmetry. In the first model, identical particles are assumed (IDE),
while the second model (SIZE) employs a q=3 power law with
$W_s=10$. In both cases $\tau_{\rm dyn}=0.5$, internal particle density
$\rho=450$ kg m$^{-3}$ is assumed together with the Bridges et al.~(1984)
elasticity law. As shown in French et al.~(2007a), the asymmetry
amplitude implied by these two models brackets the observed asymmetry
in HST observations: the IDE-model yields about 15\% too large
asymmetry amplitude, while that of the SIZE-model is about 40\% too
small. In other respects the IDE model is also clearly better: it
matches nicely the elevation angle dependence of the asymmetry
amplitude, and moreover yields the correct minimum longitude, whereas
for the SIZE model the minimum longitude is off by about $5^\circ$;
this mismatch in minimum longitude for the supposedly more realistic
size distribution models was recently confirmed by Porco et
al.~(2008). The transmission properties of the models are also in
accordance with low elevation ($B=3.45^\circ$) VIMS occultation
studies (Hedman et al.~2007): the IDE and SIZE models imply maximum
transmission probabilities $T=0.09$ and $0.02$, respectively, while
the observations indicate $T \approx 0.08$.  Although one can fine
tune the strength of asymmetry in the simulation models (see Fig.~9 in
French et al.~(2007a), displaying the effect of changing the
elasticity law or the underlying dynamical optical depth), these two
models are probably sufficient to cover the qualitative effects of the
wake structure on the tilt effect. For comparison, we will also show
results from non-gravitating size distribution simulations with
$W_s=10$, both for $\taud=0.5$ and for $\taud=2.0$.

{\bf \hskip 5cm INSERT FIG  19 HERE}

As we discussed in the previous sub-section, the $\alpha=0.5^\circ$
observations lie largely inside the interparticle shadowing opposition
peak, regardless of the observing elevation, so there should be no
significant increase of brightness with $ \Beff$ due to the improved
visibility of the dense central layer. This dependence on phase angle
is displayed in a more quantitative way in Fig.~\ref{fig:tilt_simu},
showing the observed $(I/F)$ vs. $\Beff$ in the same B and A ring
regions for which the opposition phase curves were studied in the
previous sections. The dense B ring behavior is plotted in the upper
left corner, matched reasonably well by the simulation model (similar
to the SIZE model, but with no self-gravity and with $\taud=2.0$), at
both $\alpha=6^\circ$ and $\alpha=0.5^\circ$.

The importance of including self-gravity when modeling the A ring is
clearly seen in the upper right corner of
Fig.~\ref{fig:tilt_simu}. Here, the agreement of the non-gravitating
$\taud=0.5$ model with the mid-A ring tilt curve (upper right corner)
is far from satisfactory. The model curves are almost flat, whereas
the observed $I/F$ are monotonically decreasing with $B$.  The
observed brightness {\em difference} between the $\alpha=6^\circ$
(open circles) and $\alpha=0.5^\circ$ (filled circles) ring
brightnesses is, however, well-described by the difference between the
corresponding model curves (dashed and solid lines,
respectively). This suggests that the opposition effect-related
brightening with $\Beff$ acts in the A ring just as in the B ring, and
that the systematic decline is due to an additional effect missing
from the homogeneous non-gravitating model. Indeed, the strong
asymmetry attributed to gravity wakes is expected to be accompanied by
a negative tilt effect (Salo et al.~2004). The two bottom panels
display the results for the two previously introduced self-gravitating
models, which again bracket the observed behavior (note that no
attempt was made to fit the data points). Curiously, for the tilt
effect the SIZE model seems to be closer to observations than the IDE
model (the opposite was true for asymmetry and transmission
amplitudes; this probably implies that some ingredient is still
missing from current simulation models for the A ring gravity wakes).

Gravity wakes have also been inferred for the B ring, but occultation
studies (Colwell et al.~2006, 2007) suggest that the gaps in the B ring are
relatively more narrow in comparison to A ring wakes. Thus their
influence on the surface area and the reflection properties is not so
pronounced. This is in accordance with the weaker reflection asymmetry
in HST and radar observations \citep{nicholson2005, french2007a}. Only
in the less dense inner B ring is the asymmetry amplitude noticeable
(French et al.~2007a). Interestingly, in this same region the tilt
effect seems to be much smaller than in the other parts of the B ring
(see Fig.~\ref{fig:tilt}). Still, the interparticle shadowing
mechanism seems to be important in the inner B ring, evidenced by the
reduction of the tilt effect for $\alpha=0.5^\circ$ (in fact, it turns
into a negative tilt effect).  It thus seems that the interparticle
shadowing mechanism (promoting positive tilt effect) is important for
both the A and B rings, though for the mid A ring, and to lesser
degree also in the inner B ring, the effect of gravity wakes/gaps
(providing a negative contribution to tilt effect) needs to be taken into
account as well.

\section{Discussion and conclusions}

 The analysis of Hubble Space Telescope near-opposition phase curves
 obtained for $\beff=4.5^\circ - 26.1^\circ$ shows unambiguously that
 the opposition brightening of Saturn's rings depends on the ring
 elevation. This is most strikingly evidenced by Fig. \ref{fig:HST_loglin_abc}, showing the
 systematic steepening of the slope of $I/F$ vs $\ln{\alpha}$ when
 $\beff$ gets smaller.  This previously unreported dependence
 demonstrates the unique value of the 1996-2005 HST data set that
 spans a full Saturn season (Cuzzi et al.~2002, French et
 al.~2007b). Comparison of the different filters indicates that,
 although the magnitude of the total opposition effect increases toward
 shorter wavelengths, the elevation dependent part is practically the
 same in all filters.  This elevation dependence, and its independence
 of wavelength, provide strong observational confirmation for the
 presence of an interparticle shadowing opposition effect, in accordance
 with dynamical/photometric simulations (SK2003).

 In contrast to the present study, it is not possible from
 single-elevation reflection data alone to disentangle the intrinsic
 (coherent backscattering and/or shadow hiding at particle surfaces)
 and interparticle shadowing contributions, since the expected
 functional forms are rather similar (e.g.~Hapke 2002). This
 difficulty is particularly true for the intrinsic SH contribution,
 which is described by the same function that accurately fits the
 modeled near-opposition {\it interparticle} shadowing, although they
 might have different amplitudes and HWHM's. In principle, the
 wavelength dependence of CB might be used to disentangle the various
 contributions, but unfortunately there is no current theoretical
 agreement about what kind of wavelength dependence to expect.  Also,
 the observational picture is not clear, probably because of
 differences in available $\alpha$ ranges, and the fitting functions
 used. Analysis of high-elevation HST data (French et al.~2007b)
 implies that the opposition effect has a nearly constant HWHM $\sim
 0.1^\circ$ for BVRI filters, increasing slightly at U. This is
 somewhat in contrast to the analysis of Cassini ISS data (Deau et
 al.~2009; similarly referring to high elevation $\beff \sim
 22^\circ$) that indicates a roughly 2-fold larger HWHM, reaching a maximum in the
 green filter.  On the other hand, Cassini VIMS phase curves suggest
 that at near-infrared the HWHM increases rapidly with wavelength,
 from $0.2^\circ$ to $>1^\circ$ between 1.5 $\mu$m  and 3.5 $\mu$m
 (Hapke et al.~2006).
 
Polarization measurements would be helpful, since models predict
(Hapke 1990, Mishchenko 1993, Rosenbush et al.~1997) that the CB
intensity peak should be accompanied by a similar narrow peak in the
degree of polarization (both circular and linear).  Existing
ground based measurements of linear polarization (Lyot 1927, Johnson
1980, Dollfus 1996), although not ruling out such a peak, do not have
sufficient accuracy or wavelength coverage -- or mutual agreement -- for
quantitative comparison to intensity light curves. Unfortunately, the
polarization capabilities of Cassini are inadequate for such studies.

 Independent support for the interparticle shadowing opposition effect
 is provided by the Cassini CIRS measurements, showing a pronounced
 opposition effect in the ring's thermal phase curves (Altobelli et
 al.~2009). CB is ruled out, since there can be no interference
 between the incoming visual photons heating the particle and the
 infrared photons reradiating the heat. Compared to the strongly
 peaked visual phase curves the thermal opposition effect extends over
 several tens of degrees. However, quantitative comparison to the
 optical phase curve must await detailed thermal modeling that extends
 beyond
 the current models such as those of Ferrari and Leyrat (2006) and
 Morishima et al.~ (2009).

As demonstrated in Section 4, the elevation-dependent part of the
opposition effect in the HST data can be removed via model comparisons.  In practice,
we used a set of photometric/dynamical simulations performed for
various optical depths and widths of size distributions, and used a
fixed phase angle range $\alpha=0.5^\circ - 6^\circ$, where the
difference in the observed brightening $I(6^\circ)/I(0.5^\circ)$ for
different $\beff$'s was compared with that predicted by
simulations. The fact that interparticle multiple scattering is not
significant simplified these comparisons, as it was sufficient to
compare with the enhancement factor $f_e$, giving the ratio of the
simulated single scattering contribution with respect to the theoretical
$D=0$ formula.  The deduced interparticle contribution was divided out
from the observations, leaving what presumably represents the intrinsic
contribution $f_i$.

Comparison to simulations, with $\tau_{\rm dyn}=0.1-2.0$, indicated that
the interparticle enhancement $f_e$ is quite well fitted with the
functional form of the Hapke (1986) shadowing model for semi-infinite
layers. In practice, the finite optical depth affects the maximum
amplitude of $f_e$, which is reduced when the path optical depth
decreases (this decrease in turn is in good accordance with Lumme and
Bowell (1981) theoretical calculations, see SK2003).  The dependence
of HWHM on the width of the size distribution is at least
qualitatively consistent with Hapke's Y-function.  Regardless of the
good agreement with theoretical treatments, the self-consistent
dynamical simulations together with photometric modeling are still
indispensable, in order to account correctly for the
elevation-dependent interparticle shadowing, which is sensitive to the
vertical structure of the ring, via the effective volume density
$D_{\rm eff}$ at the layer mainly responsible for scattering.
Simulations indicate roughly linear dependence, HWHM $\propto \beff$,
which acts in addition to the size distribution dependence
implied by the Hapke's formula.  Additionally, simulations indicate
that HWHM is practically independent of $\taud$ (see Fig.~\ref{fig:model_hwhm}).

From their analysis of Cassini ISS phase curves, typically extending
to $\alpha \sim 25^\circ$, Deau et al.~(2009) found that the slope of
the phase curve outside the opposition peak shows a clear correlation
with optical depth (they used a linear-by-parts fit, and this outer
slope corresponds to linear component beyond $\alpha>0.3^\circ$). They
conclude that this steepening is contrary to what would be expected
from interparticle shadowing: they reason that higher $\taud$
generally implies larger volume density and thus presumably also more
extended interparticle shadowing opposition effect, e.g. phase curves
should have less steep slope outside the central peak due to the
intrinsic opposition effect.  Deau et al.~(2009) then conclude that
the $\tau$ dependence of the slope must follow from different particle
surface properties at low and high $\tau$ environments, rather than be
a result of interparticle shadowing.  However, according to our
detailed calculations, the variation of the interparticle shadowing
effect with optical depth seems consistent with the observations. For
example, Fig.~\ref{fig:model_oe_range} indicates that the slope beyond the central maximum
is generally steeper for larger $\taud$'s. The reason for this
behavior is that, although the maximum central plane filling factor
$D(z=0)$ does indeed increase with $\taud$, the optical properties are
determined by $D_{\rm eff}$ at the layer dominating the
scattering. The fact that we found the HWHM of $f_e$ to be nearly
independent of $\taud$ suggests that the variations in $D_{\rm eff}$
are much smaller than those in $D(z=0)$.  On the other hand, the maximum
amplitude of $f_e$ does increase with $\taud$, until saturation is
reached at large $\tau_{\rm path}$. This, together with nearly
constant HWHM, accounts for the increased slope.  Therefore, our
conclusion is that the $\tau$ dependence of the outer slope gives
additional support for the interparticle shadowing effect.  Of course,
there may be additional indirect correlations between particle surface
properties and local optical depth as suggested by Deau et al.~(2009).

Based on the elevation-dependent part of the observed opposition
effect, we find that the C ring region we study is best described by
an extended size distribution with $W_s \gtrsim 100$, whereas for the
A and B rings a significantly narrower $W_s \lesssim 5-10$ is
deduced. The estimated C ring lower bound is consistent with
  \cite{french2000}, who found $W_s \sim 1000$ from the analysis of
  forward scattered light in ground based stellar occultation
  data. However, for the A and B rings our upper bound is definitely
  smaller than $W_s \sim 70$ found by \cite{french2000}: most likely
  this discrepancy follows from the fact that the uniform ring models
  we have studied are too simple to describe all aspects of ring
  reflection and transmission.

In our models, the Bridges et al.~(1984) coefficient of restitution
and a power law size distribution with $q=3$ was assumed.  The
dominant factor affecting the fit is the volume density $D_{\rm eff}$,
which depends not only on the size distribution, but also on the
elasticity of particles. Less dissipative particles lead to
collisional energy balance corresponding to geometrically thicker
rings, whereas increased dissipation flattens the rings, until a
minimum thickness corresponding to few times the maximum particle
radii is achieved (see e.g. Schmidt et al.~2009). Thus, for
  significantly less dissipative particles\footnote{Such systems are
    potentially quite interesting, as they can be susceptible to
    viscous instability (see Salo and Schmidt 2010).}, say with the
  scale factor $v_c=10v_B$ in Eq (\ref{eq:bridges}) instead of
  $v_c=v_B$, the best fit for the C ring would be obtained for a
  size distribution with $W_s \gtrsim 10$ (combined with the resulting
  vertically thicker ring, $W_s=10$ would lead to roughly the same
  $D_{\rm eff}$ as the original Bridges et al.~elasticity formula in
  combination with $W_s = 50$). Clearly, such a narrow distribution
  would be very hard to reconcile with \cite{french2000} estimates.}
On the other hand, even if the particles were more dissipative than
implied by the Bridges formula (say, having a constant
$\epsilon_n=0.1$), the best fit for B and A rings would still imply a
fairly narrow $W_s\sim 10$. In conclusion, we cannot claim that any
unique formula for the elastic properties of particles could be
deduced from the matching of the elevation-dependent opposition effect,
taking into account the uncertainties in the size
distribution. However, it seems that the match in terms of a Bridges et
al.~(1984) type, frosty particle elasticity model, is a fairly robust
one, and suggests a significantly wider size distribution in the C
ring in comparison to B and A rings. Most importantly for the current
goal, the deduced intraparticle opposition effect contribution is not
overly sensitive to which particular simulation model is used in the
extraction of $f_e$, as long as it can correctly account for the
elevation angle dependence.

% we find that C has wide and B and A have narrow distributions
% assuming vb=1  -> W(C) > 100  ; W(A,B) < 10

% depends on elasticity: more dissipative en=0.1 -> flatter with larger
% D and thus wider peak. Can be compensated by having somewhat wider
% distribution

% similarly, more elastic VB=10  -> narrower distribution for C ring would be ok

%---> can not separately W or epsilon.
%     On the other hand, this means that the extraction process
%     is fairly robust: the intrinsic componen

After removal of the interparticle opposition effect, fits to the
intrinsic opposition effect were made, using both linear-exponential
and Hapke-model fits (the SH part of the latter models was simplified,
as only the near-opposition part of phase curve is fitted).  Both types of
fits imply that the intrinsic effect is mainly due to CB
(in linear-exponential model this can be identified with the
exponential component). In fact, for the $B$ and $A$ rings the intrinsic SH
contribution is almost negligible: the fitted SH component is very close
to what is implied by the $n_s=3.09$ power law phase function used in
Dones et al. 1993, when extrapolated to near-opposition phase angles.
For the C ring, the deduced intraparticle SH is less than half of
the implied coherent backscattering contribution.  Compared to the
earlier fits based on the original high elevation HST data, without
separation of the interparticle effect (French et al.~2007b), there is
rather little difference in the deduced CB parameters. In particular,
the amplitude $B_{c0} \approx 0.4$ for all ring components, as in
fits to the original data, whereas the HWHM $\approx 0.1^\circ$ is
slightly increased. Nevertheless, in any careful analysis of the intrinsic
opposition effect aiming to deduce, for example, the regional variations in
the properties of regolith-grains covering particle surfaces, the
elevation dependent part should first be excluded.  For example, the
difference in the deduced SH components between the C ring and the
higher density B and A rings, not distinguishable without exclusion of
the interparticle shadowing, might reflect the different collisional
environments in the ring components.

%* After removal of external component
%  combined CB-SH fit suggests that the effect is mainly CB

%  intraparticle SH only imoportant for C ring.

%  This is further illustrated in Figs...FIGS
%which show C and B both in log and linear

%In particular for B ring, the fitted SH component is failry close to
%Callisto...

%*what does this mean in terms of C and B particles?

{\bf \hskip 5cm INSERT FIG  20 HERE}

{\bf \hskip 5cm INSERT FIG  21 HERE}

The various modeled contributions are best illustrated in 
Figs.~\ref{fig:discussion_cb_sh_cring} and \ref{fig:discussion_cb_sh_bring},
for the C and B rings, respectively. (The A ring case would be almost
indistinguishable from the B ring.) The CB and SH contributions are
shown separately by the black and gray shaded regions; these are
identical in the upper and lower frames, corresponding to
$\beff=23.5^\circ$ and $4.5^\circ$. The modeled interparticle
shadowing contribution is shown by the dashed region, which is much more
pronounced for the
lower elevation. Also shown are the HST data points corresponding to
the indicated $\beff$. (Note that the the models are based on fitting
simultaneously the whole range of $\beff$'s.)  To emphasize the
extremely narrow CB peak, the phase curves are also shown on a linear
scale in the right-hand frames.

{\bf \hskip 5cm INSERT FIG \ref{fig:discussion_oppo_extended} HERE}

For comparison with spacecraft observations, we predict the behavior
of the intraparticle shadowing effect for phase curves beyond the
regime accessible by ground based studies ($\alpha < 6.37^\circ$) in
Fig.~\ref{fig:discussion_oppo_extended}. This shows the single
scattering contribution for our adopted B ring model, now covering a
full range $0^\circ \le \alpha \le 180^\circ$. Three different
elevations are compared, as well as the elevation-independent
theoretical $D=0$ curve, following in form the spherical-particle
Lambert phase function. According to Fig.~\ref{fig:discussion_cb_sh_bring}, 
the intrinsic peak would affect just
a very narrow portion near zero phase angle. On the other hand, as
mentioned previously, the interparticle shadowing contribution extends
to all phase angles. In fact, the relative brightening over the classical
($D=0$) multilayer starts to increase again beyond $\alpha \approx
50^\circ$. Clearly this is no longer an ``opposition effect," but
relates to a general reduction of shadowing in geometrically thin
layers of particles (see footnote 5). Nevertheless, in practice this enhanced $I_{ss}$
has little significance, since it is offset by an even larger
reduction in the multiple scattering contribution (Fig. \ref{fig:discussion_oppo_extended} right panel; see also Fig.~7 in
SK2003). Since multiple scattering dominates at high phase angles,
the total brightness $I/I(D=0)$ is reduced. It was shown in SK2003
that this combination of geometrically thin layers appearing brighter
in backscattering and dimmer at forward-scattering makes it possible
to match simultaneously both the low and high phase angle brightnesses
of the inner A ring. Indeed, for this ring region, Dones et al.~(1993)
found that the phase curve from Voyager images cannot be accounted for
by classical radiative transfer models, and suggested that this might
be due to geometrically thin rings.  Clearly, a similar effect needs to
be taken into account in the interpretation of Cassini ISS
observations. In particular, by combining observations from several
different elevations, one can eliminate the uncertainties related to
particle phase function and albedo, and deduce constraints for the
ring vertical profile and size distribution.\footnote{Porco et
  al.~(2008) claim to have deduced accurate ring thicknesses, based on
  small deviations between the observed and modeled $D=0$ phase
  curves, repeating calculations such as described in SK2003. However,
  there are problems in their approach: for example, they assume that
  the ring particle phase function is determined precisely by a
  power-law phase function, and use two observations to deduce
  simultaneously three unknown parameters: the ring thickness, the
  particle albedo, and the index $n_s$ of a power law phase
  function. Moreover, the ring models they use are not dynamically
  self-consistently calculated size distribution models: they assume a
  Gaussian distribution of identical particles, which assumption is
  prone to affect the resulting model brightness (see e.g.~Fig.~21 in
  SK2003). Apart from these considerations, there also seem to be some
  problems in the convergence of their model results when the
  classical limit should be reproduced exactly (see their Figs.~10-13).}

The characteristics of the B and A ring tilt effects were explored in
Section 5. We showed a new observational result: the strong positive B
ring tilt effect seen at $\alpha=6^\circ$ (the phase angle most often
addressed in ground based studies) is significantly weakened when
observations at $\alpha=0.5$ are compared. We showed that the tilt
effect itself, as well as its smaller amplitude as $\alpha \rightarrow
0^\circ$, follow in a natural manner from the same models which match
the elevation-dependent opposition effect.  Briefly summarized, at low
$\beff$ the width of interparticle opposition peak is much less than
$6^\circ$, so that near-opposition brightening has no contribution to the ring
brightness. On the other hand, when $\beff$ increases the opposition
peak gets wider, leading to an increased contribution to the brightness.
For example, the best-matching B ring interparticle shadowing model
has HWHM $ \sim 1^\circ-10^\circ$, for $\beff=4^\circ - 26^\circ$ (see
Fig. 8).  Clearly, at phase angles smaller than the minimum width,
the interparticle contribution is present regardless of $\beff$: this
accounts for the reduced tilt effect for $\alpha=0.5^\circ$.  The A
ring negative tilt effect was attributed to gravitational wakes, which
have a larger effect on the reflection in the moderate-$\tau$ A ring
(and the inner B ring) in comparison to the densest part of the $B$
ring. Nevertheless, the shadowing contribution was also present,
evidenced by the difference between $\alpha=6^\circ$ and
$\alpha=0.5^\circ$.

{\bf \hskip 5cm INSERT FIG  23 HERE}

%TANNE
The dependence of the expected B ring tilt effect as function of phase
angle is further illustrated in
Fig.~\ref{fig:discussion_tilt_extended}: here, the second and third
frames show the  $\alpha=0.5^\circ$ and $6^\circ$ cases discussed just above.
 (Note  the arbitrary normalization to $\beff=1^\circ$).  The first
frame is for exact opposition, where the predicted interparticle
contribution to the tilt effect practically vanishes (the small
residual effect follows from the weak dependence of maximum $f_e$ on
path optical depth). In the figure, the modeled tilt effect is shown
separately for single scattering and for total brightness
(single+multiple scattering): the difference between these indicates
the contribution due to increased multiple scattering when $\beff$
increases.  Also shown in Fig.~\ref{fig:discussion_tilt_extended} is
the expected B ring tilt effect for phase angles $> 6^\circ$. Clearly,
if the brightening of $I_{ss}$ in the case of non-zero D were limited
to near opposition, there would be no shadowing contribution to the tilt
effect for $\alpha$ larger than the width of the opposition
peak.  However, due to the aforementioned general brightening
of geometrically thin layers, the tilt effect due to enhanced $I_{ss}$
is present for all $\alpha's$:  the strength of the effect is
proportional to the difference between the various $\beff$ curves in
Fig.~\ref{fig:discussion_oppo_extended}.  Additionally, for $\alpha
\gtrsim 120^\circ$ the multiple scattering contribution to the tilt effect
should become more and more important.

The effect of elevation-dependent interparticle shadowing, seen in the
HST observations, should be present also in the Cassini ISS, VIMS, and
CIRS data, particularly for the recent epoch with low solar elevation.
In fact, the effect might be noticeable also for  $\beff \sim
15-25^\circ$, for which range Cassini data have already been analyzed.  For
example, often phase curve data from different tilt angles are combined
together: it will be important to determine what influence this has on
the fitted parameters (e.g.~on the HWHM of the opposition peak).
Also, removing the wavelength-independent interparticle opposition
effect will affect the relative amplitudes of the intrinsic
opposition peak deduced at different filters: obtaining an unbiased
view of the wavelength trends of opposition peak is
important for the physical interpretation of the observations.
Our plan for the future is to expand our photometric modeling to cover
a larger range of observing geometries relevant for Cassini,
and moreover cover a larger set of dynamical models.
Clearly, this will help to provide improved constraints for 
both the photometric properties of ring particles, as well as for the
local structure of rings, influenced by the particles' physical properties
and their size distribution.

%1) Summarize what you think the logical next steps are for modeling, based
%on the data sets already published. There seem to be some inconsistencies
%betweeen ISS,VIMS, and HST regarding the wavelength dependence and size of
%the narrow true-opposition peak, for example. How can these best be
%resolved?
%2) If we want to move beyond the current IDE/Size models and try to
%constrain things like the internal density of the particles and their
%coefficient of resititution, what is the best way to do this? 

%CASSINI: take these models, calculated for the exact geometries
%and after this compare

\section{Acknowledgements}

\vspace{1cm}

%------\acknowledgments

We thank Luke Dones and Linda Spilker for detailed and constructive
reviews. Our results are based on observations with the NASA/ESA
Hubble Space Telescope, obtained at the Space Telescope Science
Institute (STScI), which is operated by the Association of
Universities for Research in Astronomy, Inc.~under NASA Contract
NAS5-26555. This work was supported in part by grants from STScI, by
the Academy of Finland and by Wellesley College.

\clearpage

%%%%%%%%%%%%%%%%%%%%%%%%%%%%%%%%%%%%%%%%%%%%%%%%%%%%%%%%%%%%%%%%%%%%%%%%%%%%%%%%%%%%%%%%%%%%%%%%%%%%%%%%%
%------------------------------------------------
%Fig.~1 SCHEMATIC - DPS2005
%------------------------------------------------
%/home/heikki/MCCODE/RESULTS/HST_OPPOTILT_FIGS/

\begin{figure}[b]
  \includegraphics[width=1\textwidth]{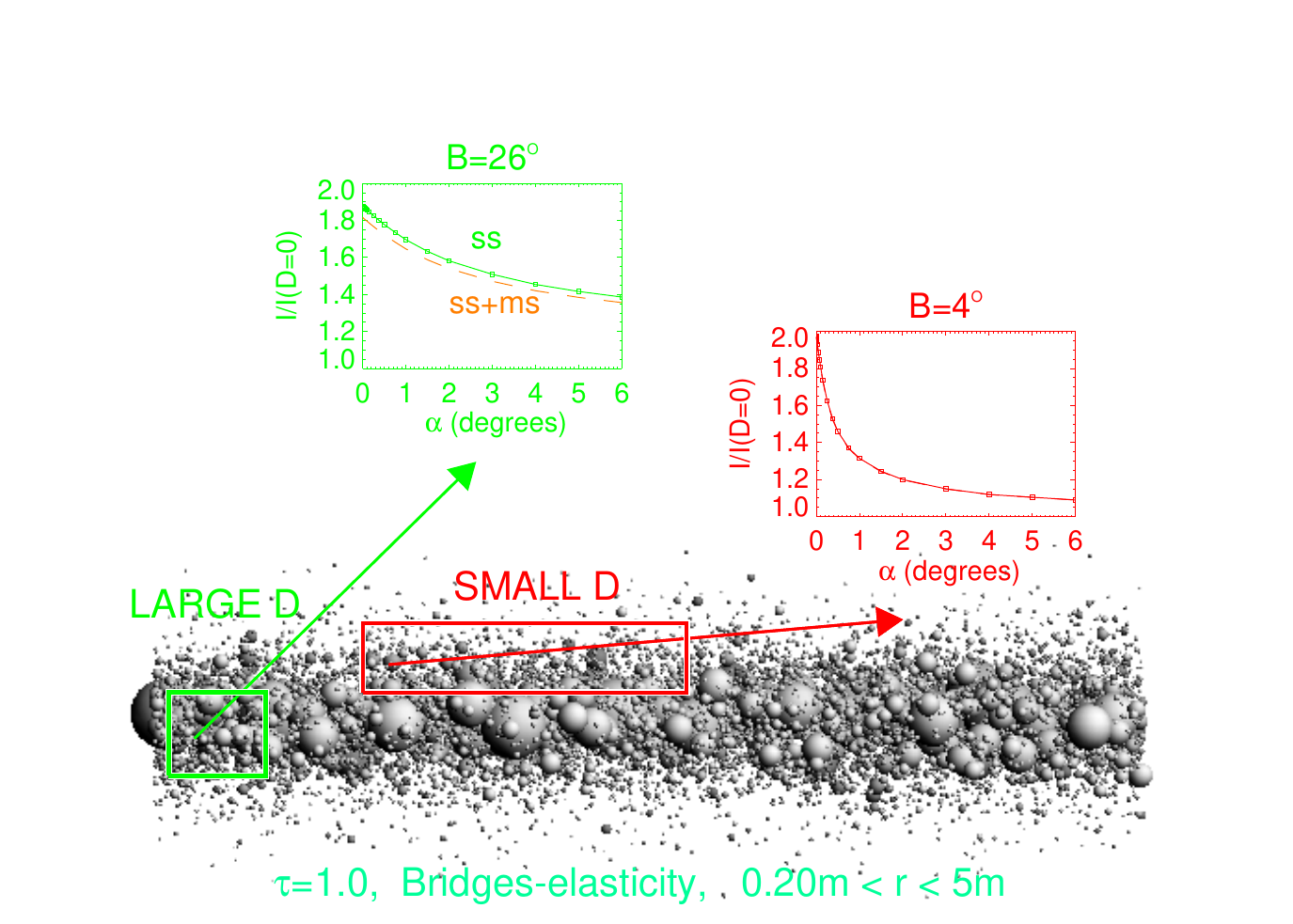}
  \caption{An illustration of the opposition brightening due to
    reduced interparticle shadowing. A side view of a dynamical
    simulation model is displayed, together with opposition phase
    curves calculated for two different observing elevations. An
    $n_S=3.09$ power law phase function with Bond albedo 0.5 is
    assumed.  The curves display $I/I(D=0)$, where $I(D=0)$ is the
    theoretical brightness for a classical zero volume density
    ring. The curves for the single scattering (ss) component and
    total brightness including multiple scattering (ss+ms) are shown
    separately, with solid and dashed lines (for $B=4^\circ$ the
    multiply-scattered component is negligible).  At large elevation
    angles ($B=26^\circ$) the light rays are able to penetrate to
    central layers, where the typical particle separations are
    comparable to particle size: such a high volume density leads to a
    broad opposition brightening curve. On the other hand, at small
    elevation angles ($B=4^\circ$) the reflection happens mainly in
    the rarefied upper layers, where particle separations are large
    compared to their size: such a small effective volume density
    leads to a much narrower opposition effect.
    \label{fig:model_cartoon}}
\end{figure}

%%%%%%%%%%%%%%%%%%%%%%%%%%%%%%%%%%%%%%%%%%%%%%%%%%%%%%%%%%%%%%%%%%%%%%%%%%%%%%%%%%%%%%%%%%%%%%%%%%%%%%%%%
%------------------------------------------------------------------------
% FIG. 2: OBSERVED I/F profiles at 4.5 and 23.5
%------------------------------------------------------------------------

\begin{figure}[b]
  \includegraphics[width=1\textwidth]{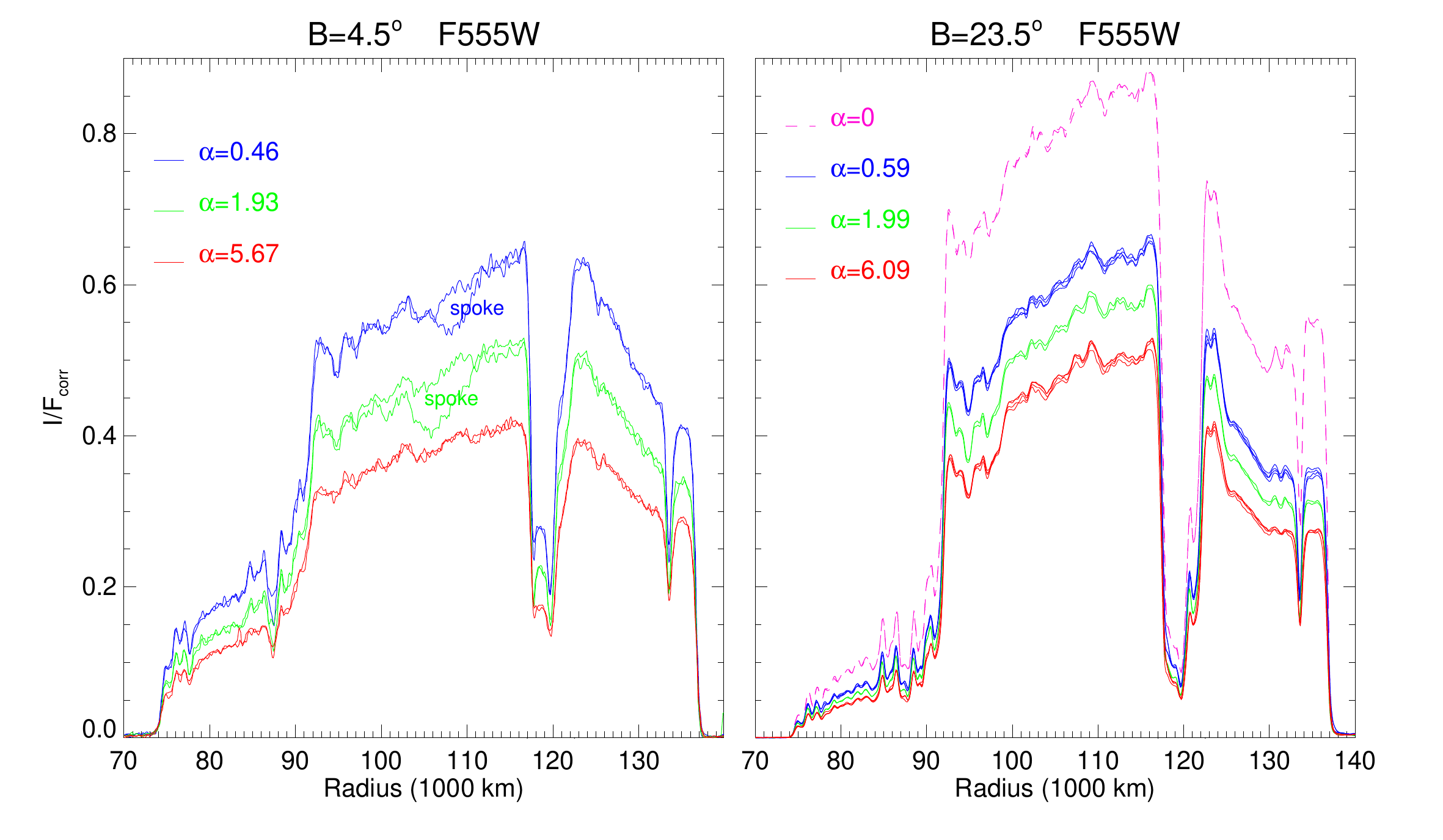}
  \caption{Examples of the ansa $I/F$ profiles for $\beff=4.5^\circ$
    (left) and for $\beff=23.5^\circ$ (right). Solid lines collect the
    observations in F555W filter close to common phase angles $\alpha
    \sim 6^\circ$, $\sim 2^\circ$, and $\sim 0.5^\circ$. The almost
    overlapping curves correspond to profiles extracted from adjacent
    east/west ansa images: note that for $\beff=4.5^\circ$ the images
    for the two lowest phase angles had spokes in the east ansa \citep{mcghee2005},
    affecting the mid B ring profiles (the affected portions of these
    images are omitted from all subsequent analyses).  At right, the
    profile at exact opposition is also shown (for $B=22.9^\circ$). Here, as in
    all the subsequent plots, the corrected $I/F$ is shown,
    compensating for the small differences in $B$ and $B'$ during each
    subset of observations.
    \label{fig:HST_oe_profiles_4_22}}
\end{figure}

%%%%%%%%%%%%%%%%%%%%%%%%%%%%%%%%%%%%%%%%%%%%%%%%%%%%%%%%%%%%%%%%%%%%%%%%%%%%%%%%%%%%%%%%%%%%%%%%%%%%%%%%%
%------------------------------------------------------------------------
% FIG. 3: B, A ring ratios
%------------------------------------------------------------------------

\begin{figure}[b]
  \includegraphics[width=1\textwidth]{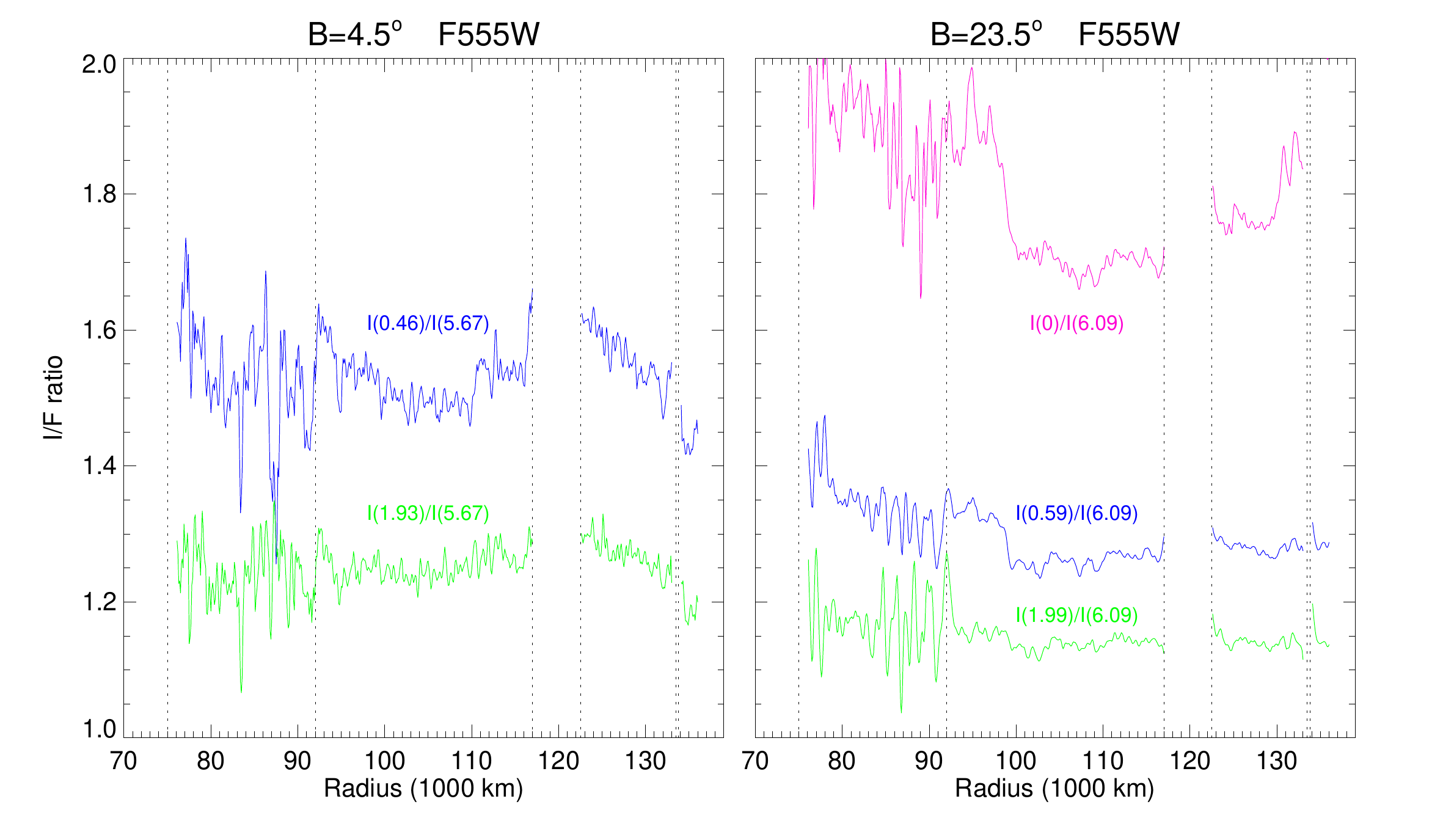}
  \caption{Same as  Fig.~\ref{fig:HST_oe_profiles_4_22}, except that
    the profiles have been divided by that at $\alpha \sim
    6^\circ$. Comparison of the two opening angles illustrates a clear
    elevation angle dependence in the magnitude of relative opposition
    brightening (at least for the common interval $0.5^\circ < \alpha
    < 6^\circ$).
    \label{fig:HST_oe_profiles_4_22_ratio}}
\end{figure}

%%%%%%%%%%%%%%%%%%%%%%%%%%%%%%%%%%%%%%%%%%%%%%%%%%%%%%%%%%%%%%%%%%%%%%%%%%%%%%%%%%%%%%%%%%%%%%%%%%%%%%%%%
%%               REMOVED IN REVISED VERSION
%------------------------------------------------------------------------
% FIG. 4: C ring ratios
%------------------------------------------------------------------------

%%\begin{figure}[b]
%%  \includegraphics[width=1\textwidth]{HST_oppo2008_oe_profiles_4_22_ratio_cring.pdf}
%%  \caption{Same as  Fig.~\ref{fig:HST_oe_profiles_4_22_ratio}, except 
%% for the C ring region. Also shown is the Voyager PPS optical 
%%depth profile, smoothed to the resolution of the HST images.
%%    \label{fig:HST_oe_profiles_4_22_ratio_cring}}
%%\end{figure}

%%%%%%%%%%%%%%%%%%%%%%%%%%%%%%%%%%%%%%%%%%%%%%%%%%%%%%%%%%%%%%%%%%%%%%%%%%%%%%%%%%%%%%%%%%%%%%%%%%%%%%%%%
%------------------------------------------------------------------------
% FIG. 5: Log-lin fits to observations
%------------------------------------------------------------------------

\begin{figure}[b]
  \includegraphics[width=1\textwidth]{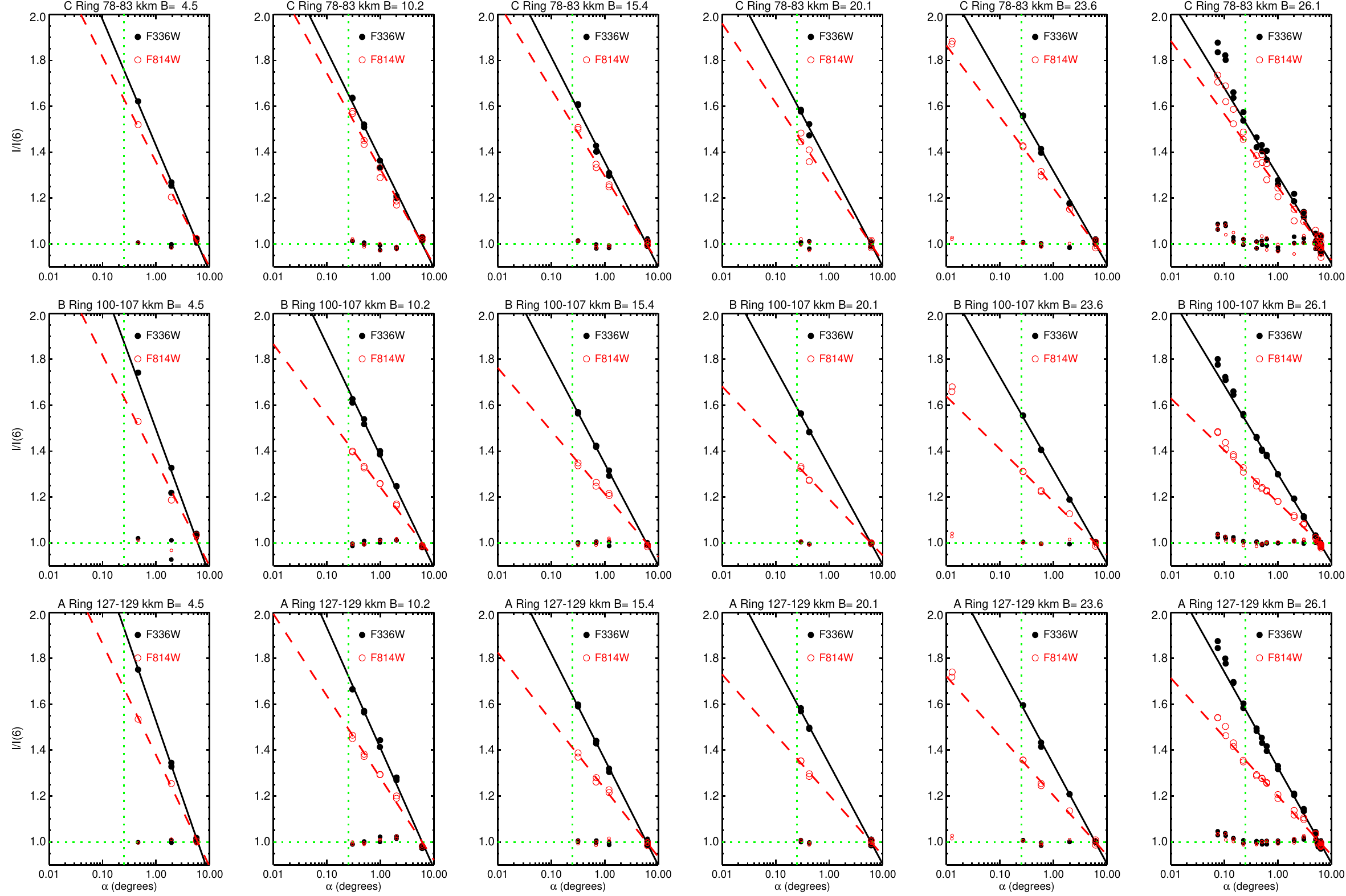}
  \caption{ Opposition phase curves of different ring regions at
    different elevation angles, with $I/F$ normalized to that at
    $\alpha=6^\circ$. The C ring (upper row), the B ring (middle) and
    the A ring (lower) regions are the same as those studied in French
    et al. (2007b).  From left to right the elevation angle
    increases from $\beff=4.5^\circ$ to $26.1^\circ$. Curves for two
    different filters are shown, F336W and F814W.  The lines indicate
    log-linear fits of the form $I/F = a \ln \alpha +b $, obtained
    using values for $\alpha >0.25^\circ$. Also shown are the residuals of
    the fits (deviation of small symbols from unity).  This fit range,
    excluding the near to opposition data points, was chosen in order
    to give a similar coverage of phase angles for all elevations:
    note that for $\Beff=23.6^\circ$ there are additional small $\alpha$
    measurements falling outside the fitted range.
    \label{fig:HST_loglin}}
\end{figure}

%%%%%%%%%%%%%%%%%%%%%%%%%%%%%%%%%%%%%%%%%%%%%%%%%%%%%%%%%%%%%%%%%%%%%%%%%%%%%%%%%%%%%%%%%%%%%%%%%%%%%%%%%
%------------------------------------------------------------------------
% FIG. 6: OBSERVED PHASE CURVES NORMALIZED to 6 degrees, different BEFF's
%------------------------------------------------------------------------

\begin{figure}[b]
\includegraphics[width=1\textwidth]{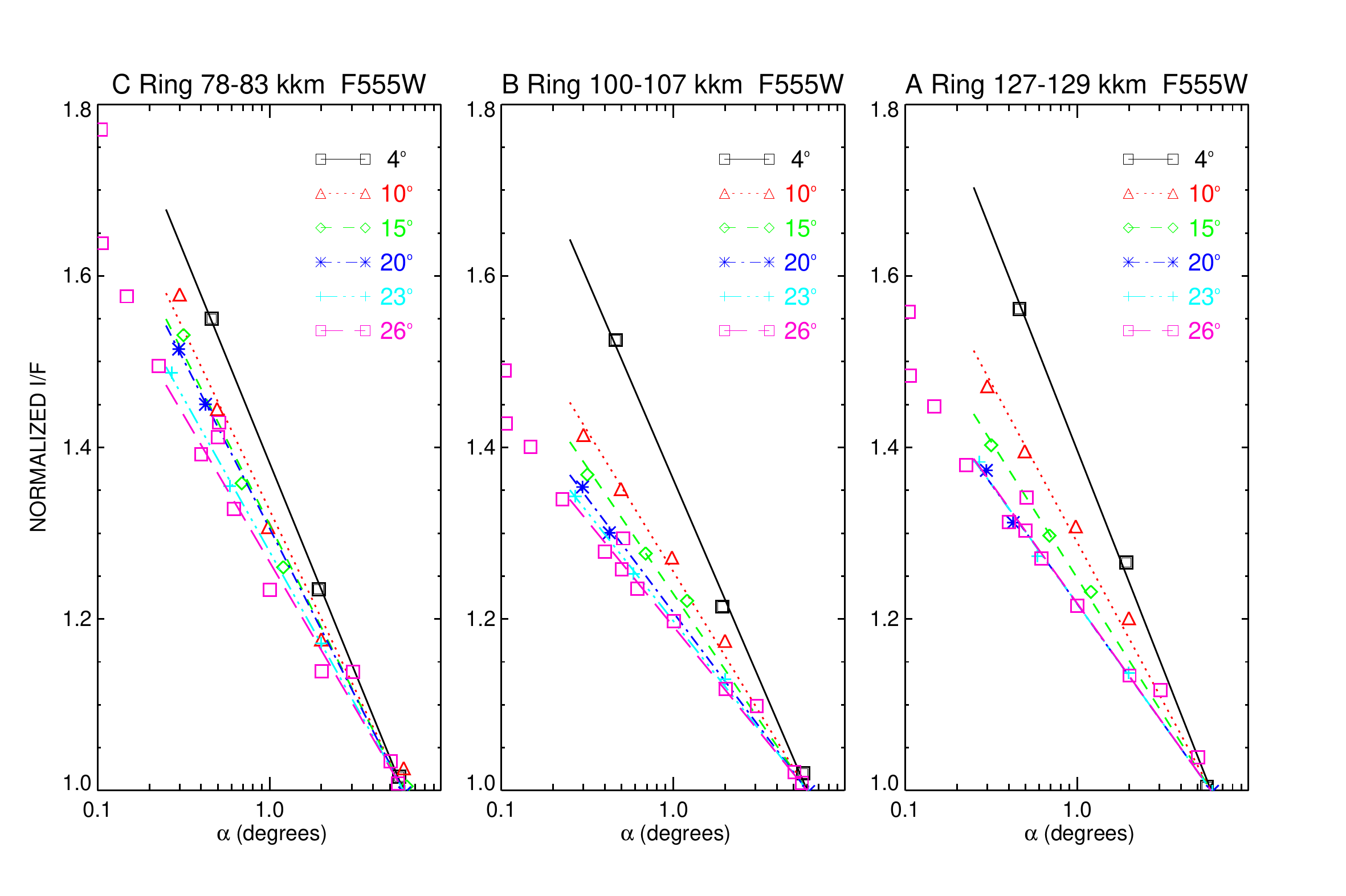}
  \caption{Phase curves for the C, B, and A ring regions in the F555W
    filter, with the data from different elevation angles collected in
    each frame. The lines indicate the log-linear fits obtained
    using data values with $\alpha >0.25^\circ$ (same as in
     Fig.~\ref{fig:HST_loglin} for F336W and F814W filters).
    \label{fig:HST_loglin_abc}}
\end{figure}

%%%%%%%%%%%%%%%%%%%%%%%%%%%%%%%%%%%%%%%%%%%%%%%%%%%%%%%%%%%%%%%%%%%%%%%%%%%%%%%%%%%%%%%%%%%%%%%%%%%%%%%%%
%------------------------------------------------------------------------
% FIG. 7: OBSERVED OE 0.5-6.0 normalized to 26 degrees
%------------------------------------------------------------------------

\begin{figure}[b]
\includegraphics[width=1\textwidth]{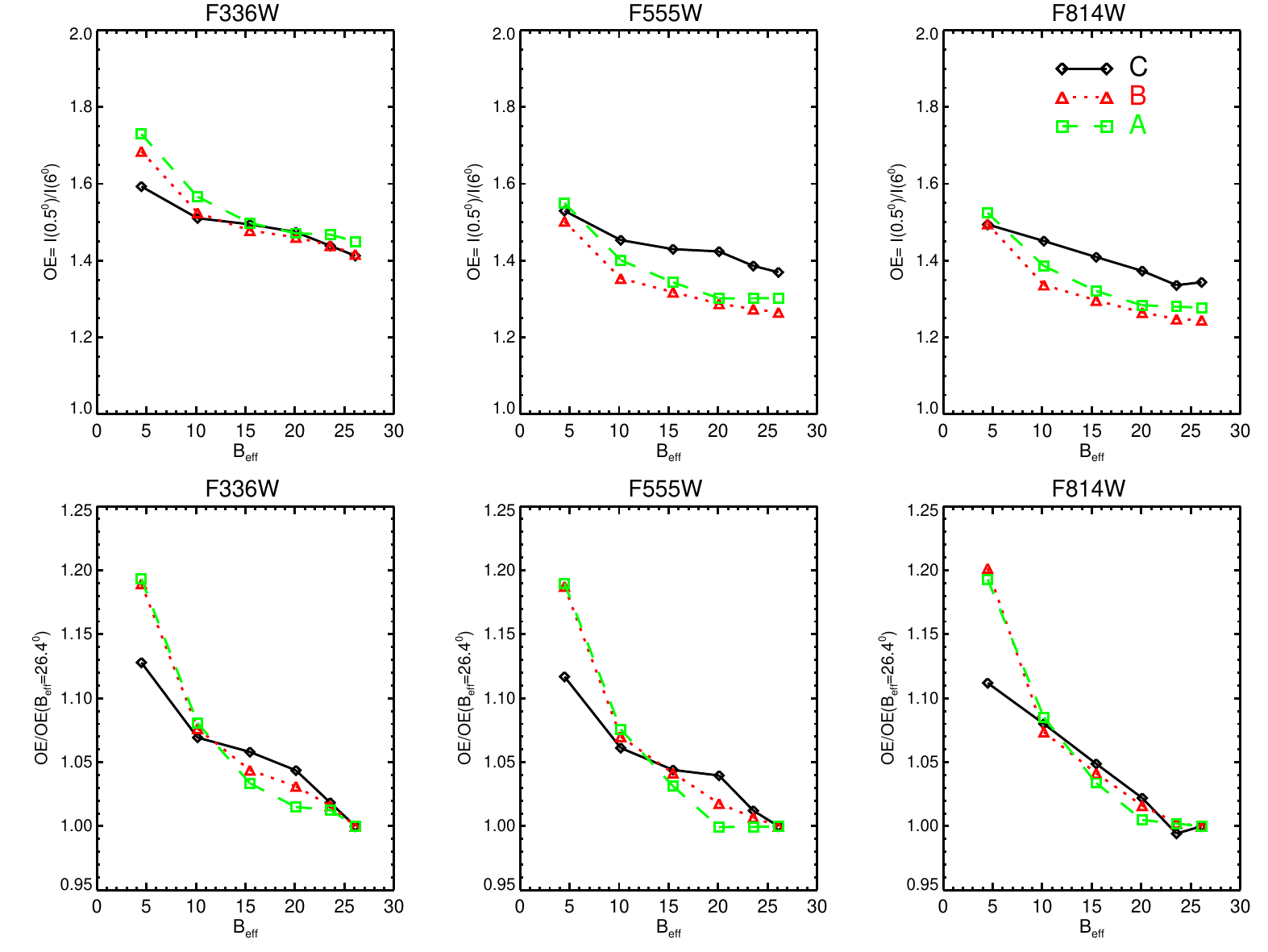}
      \caption{The wide component of the opposition brightening
        measured in terms of $OE=I(0.5^\circ)/I(6^\circ)$, obtained
        from log-linear fits. The upper row shows OE as a function of
        $\beff$ for the three studied C, B, and A ring regions, for
        three different filters. In the lower row, OE has been
        normalized to that at $\beff =26.1^\circ$.
      \label{fig:HST_obs_oe}}
\end{figure}

%%%%%%%%%%%%%%%%%%%%%%%%%%%%%%%%%%%%%%%%%%%%%%%%%%%%%%%%%%%%%%%%%%%%%%%%%%%%%%%%%%%%%%%%%%%%%%%%%%%%%%%%%
%%               REMOVED IN REVISED VERSION
%------------------------------------------------------------------------
% FIG. 8: OBSERVED OE 0.5-6.0 normalized to 336
%------------------------------------------------------------------------

%%\begin{figure}[b]
%%\epsscale{1}
%%  \includegraphics[width=1\textwidth]{HST_oppo2008_obs_save_oe_336norm_compact.pdf}  
%%\caption{Like  Fig.~\ref{fig:HST_obs_oe}, except that $OE=I(0.5^\circ)/I(6^\circ)$ has been
%%normalized to that at $\lambda=336$ nm (F336W). Note that any systematic dependence of $OE/OE(336)$ 
%%on $\beff$ is smaller than the scatter of the data.
%%\label{fig:HST_obs_oe_336norm}}
%%\end{figure}

%
%------------------------------------------------
% Simulations
%------------------------------------------------

%%%%%%%%%%%%%%%%%%%%%%%%%%%%%%%%%%%%%%%%%%%%%%%%%%%%%%%%%%%%%%%%%%%%%%%%%%%%%%%%%%%%%%%%%%%%%%%%%%%%%%%%%
%------------------------------------------------
%Fig.~9 grid of MC simulation
%------------------------------------------------
\begin{figure}[b]
\includegraphics[width=1\textwidth]{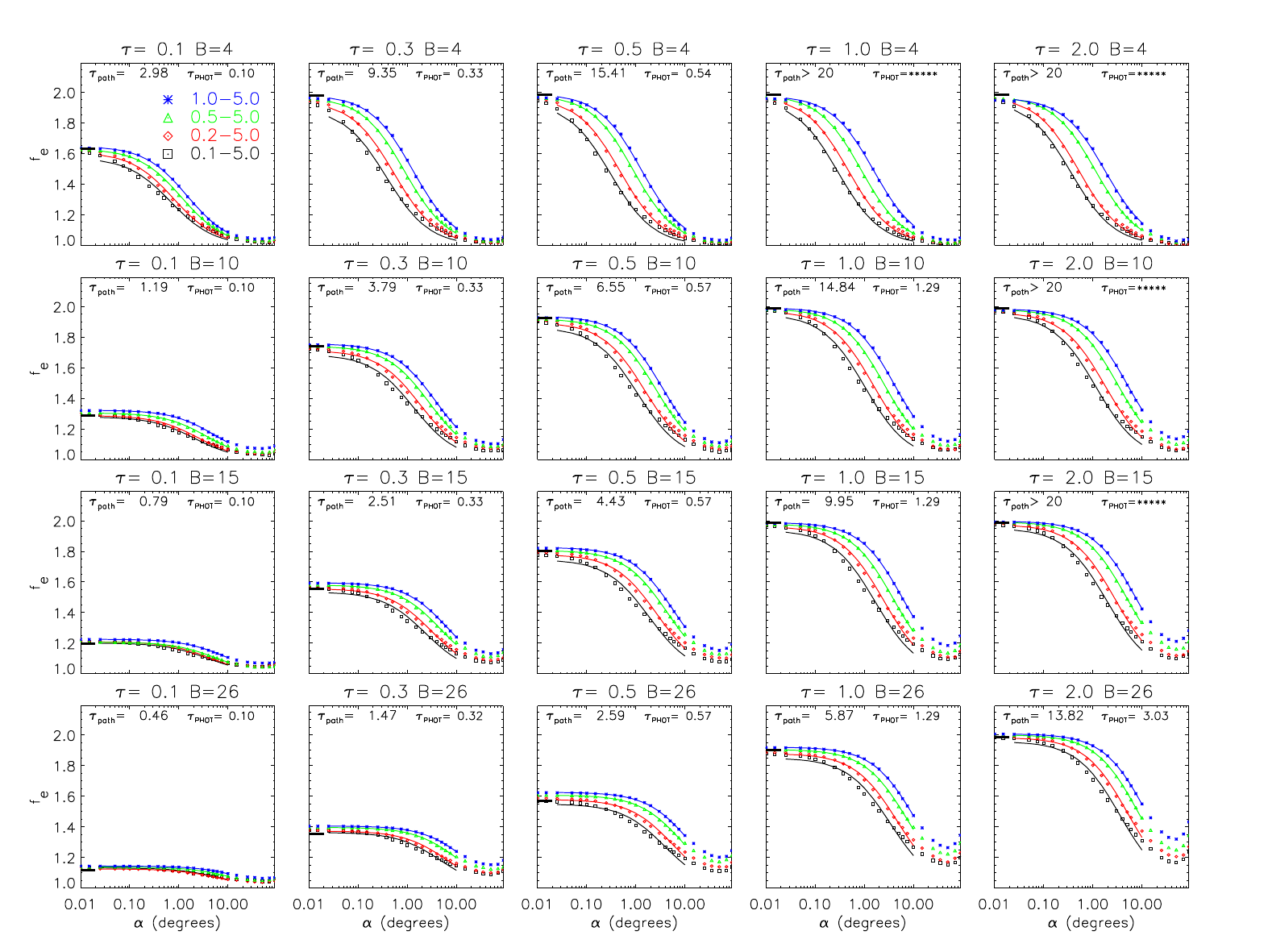}
  \caption{
\baselineskip 0.6cm
Grid of dynamical and photometric simulation models
    performed with different optical depths and widths of particle
    size distributions.  The interparticle shadowing enhancement of
    the single scattering, $f_e=I_{\rm ss}/I_{\rm ss}(D=0)$, is
    plotted as a function of phase angle $\alpha$; here $I_{\rm
      ss}(D=0)$ is the theoretical single scattering intensity for
    classical zero volume density ring, whereas $I_{\rm ss}$ is the
    simulated value including the shadowing effects between finite
    sized particles.  The simulation results are indicated by symbols,
    while the solid curves indicate Hapke (2002) SH fits to them.  The
    dynamical simulations use the Bridges et al.~(1984) elasticity
    law, and a power law size distribution $dN/dR = R^{-q}$, with
    $q=3$ and $R_{\rm max}=5$~m; the minimum size is $R_{\rm min}
    =0.1- 1.0$ m. Simulations performed with dynamical optical
    depths $\taud=0.1- 2.0$ are shown. Self-gravity is not included,
    and thus the systems remain homogeneous in all planar
    directions. Photometric Monte Carlo calculations are performed for
    elevations $B=\beff=4^\circ, 10^\circ, 15^\circ, 26^\circ$, using
    a Lambert surface element scattering law.  The numbers in the
    frames indicate the path optical depth $\tau_{\rm path}=-\ln p$, where
    $p$ is the probability of a photon to pass through the particle
    layer, and the calculated normal optical depth $\tau= \tau_{\rm path}
    \sin B$.
    \label{fig:model_grid_raw}}
\end{figure}

%%%%%%%%%%%%%%%%%%%%%%%%%%%%%%%%%%%%%%%%%%%%%%%%%%%%%%%%%%%%%%%%%%%%%%%%%%%%%%%%%%%%%%%%%%%%%%%%%%%%%%%%%
%------------------------------------------------
%Fig.~10 SHOW simulated HWHM + fits
%------------------------------------------------

\begin{figure}[b]
\includegraphics[width=1\textwidth]{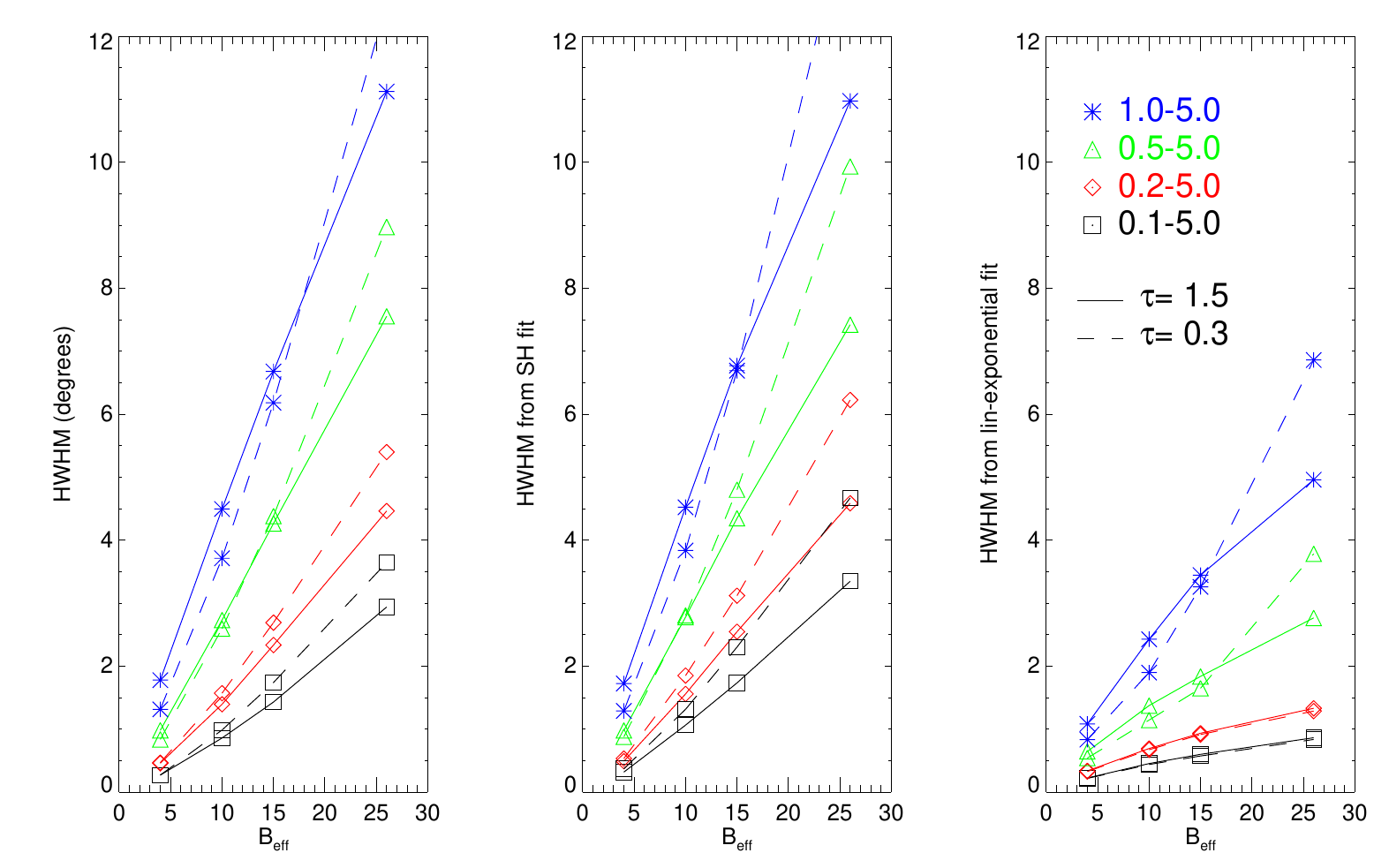}
  \caption{The half-width half-maximum (HWHM) for the interparticle shadowing
    effect in the single scattered component $f_e=I_{\rm ss}/I_{\rm
      ss}(D=0)$, obtained directly from the simulated $f_e(\alpha)$
    curves (left), from Hapke SH fits (middle), and from
    linear-exponential fits (right) to some of the simulations displayed in
     Fig.~\ref{fig:model_grid_raw}.
    \label{fig:model_hwhm}}
\end{figure}
%%%%%%%%%%%%%%%%%%%%%%%%%%%%%%%%%%%%%%%%%%%%%%%%%%%%%%%%%%%%%%%%%%%%%%%%%%%%%%%%%%%%%%%%%%%%%%%%%%%%%%%%%
%------------------------------------------------
%Fig.~11 SHOW fe 0.5, 6.0, simulated OE_e: how do different models behave
%------------------------------------------------

\begin{figure}[b] 
\includegraphics[width=1\textwidth]{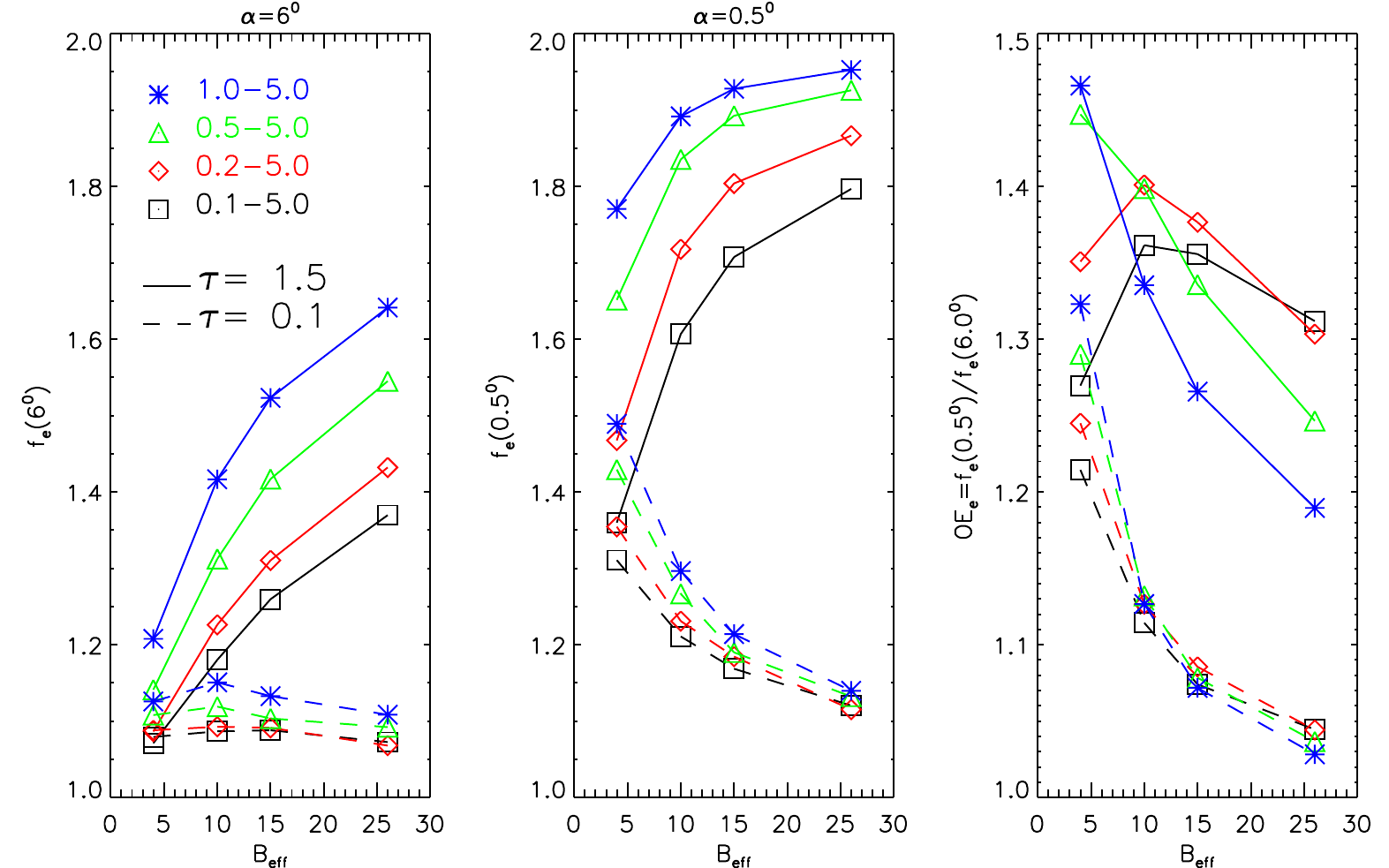}
    \caption{ The interparticle shadowing enhancement of the single
      scattering, $f_e=I_{\rm ss}/I_{\rm ss}(D=0)$ evaluated at
      $\alpha=6^\circ$ (left) and $\alpha=0.5^\circ$ (middle). At the
      right, the ratio $OE_e=f_e(0.5^\circ)/f_e(6^\circ)$ is shown.
  \label{fig:model_f_60_05}}
\end{figure}

%%%%%%%%%%%%%%%%%%%%%%%%%%%%%%%%%%%%%%%%%%%%%%%%%%%%%%%%%%%%%%%%%%%%%%%%%%%%%%%%%%%%%%%%%%%%%%%%%%%%%%%%%
%------------------------------------------------
% Fig.~12 with multiple scattering
%------------------------------------------------

\begin{figure}[b]
\includegraphics[width=1\textwidth]{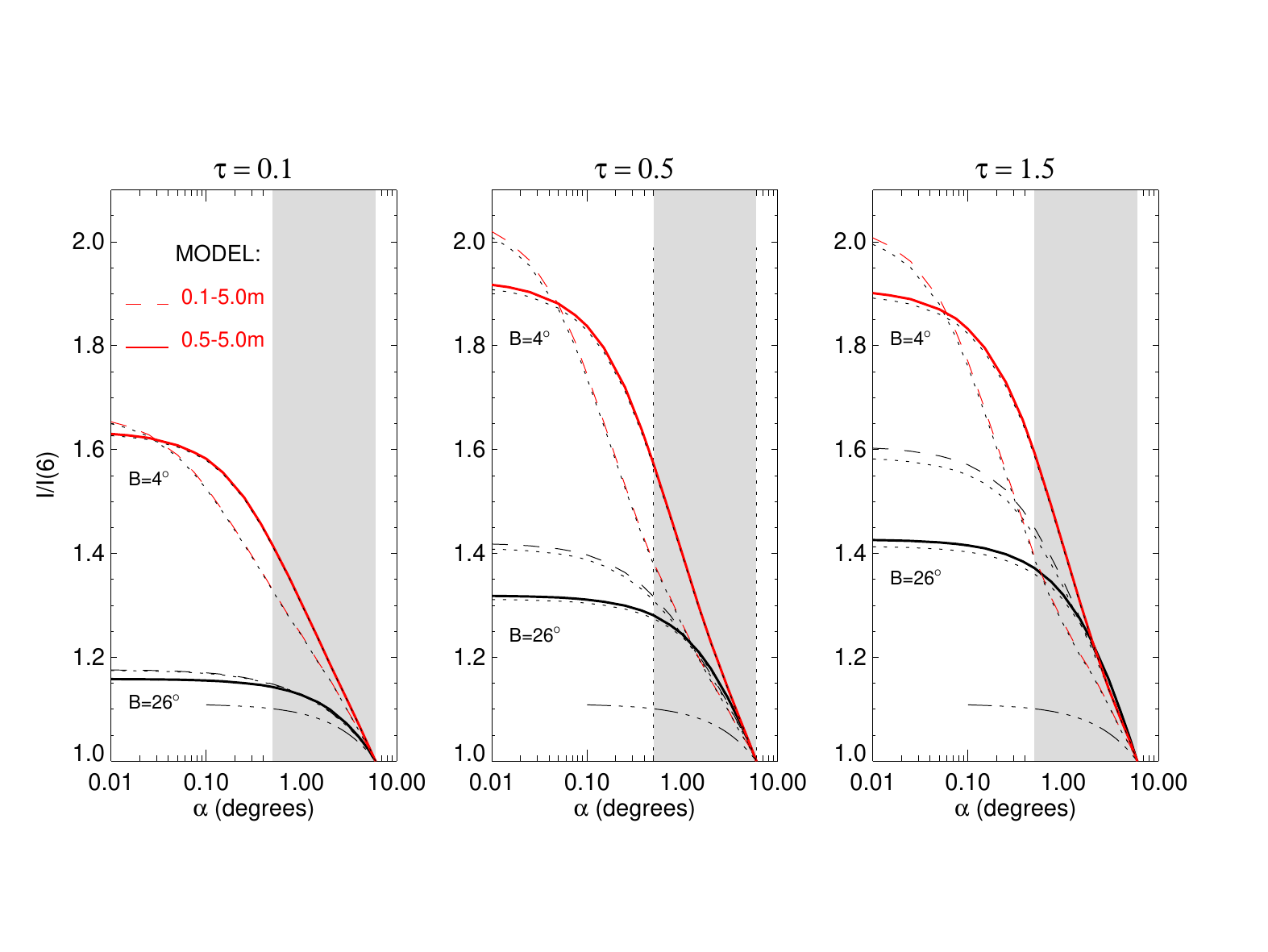}
  \caption{Opposition phase curves in selected simulation models,
    including the multiply-scattered contribution: solid and dashed
    lines indicate the single scattered intensity normalized to
    $\alpha=6^\circ$, while dotted lines indicate the same for the
    total singly + multiply-scattered radiation. The $n_s=3.09$ power-law phase
    function is used, with Bond albedo 0.5; the gray dash-dotted curve
    indicates the contribution from the power-law phase function
    alone, amounting to about 1.11 for the interval $\alpha =0^\circ$
    to $6^\circ$.
 The shaded region indicates the range $\alpha=0.5-6^\circ$
    used in the comparison of simulated and observed
    intensities in Section \ref{sec:simucomp}.
    \label{fig:model_oe_range}}
\end{figure}

%%%%%%%%%%%%%%%%%%%%%%%%%%%%%%%%%%%%%%%%%%%%%%%%%%%%%%%%%%%%%%%%%%%%%%%%%%%%%%%%%%%%%%%%%%%%%%%%%%%%%%%%%
%------------------------------------------------
%fig 13. 
%------------------------------------------------

\begin{figure}[b]
\includegraphics[width=0.8\textwidth]{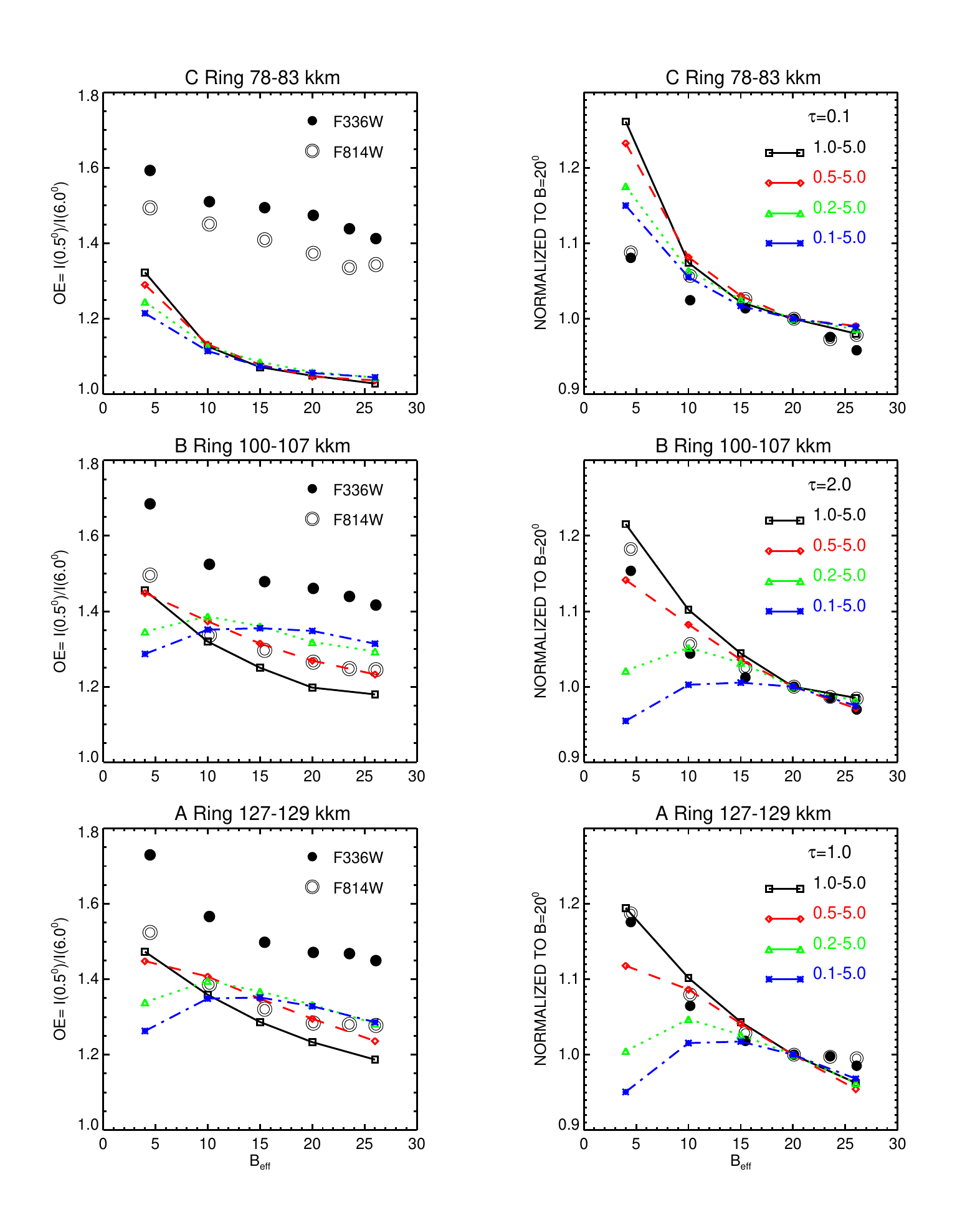}
  \caption{The left panels show the extended opposition effect as a function of $B_{\rm eff}$,
    measured by $OE=I(\alpha=0.5^\circ)/I(\alpha=6^\circ)$, for the
    three different ring regions: solid and open symbols stand for
    observations in the F336W and F814W filters, respectively. The right
    panels show the same normalized to that at $B_{\rm eff}=20^\circ$.
    Also shown is the interparticle shadowing effect in simulations
    performed with various widths of the size distribution: the C, B,
    and A ring data are compared to simulations with dynamical optical
    depth $\taud=0.1, 2.0, $ and $1.0$, respectively.
    \label{fig:simu_compare_oe_cba}}
\end{figure}

%%%%%%%%%%%%%%%%%%%%%%%%%%%%%%%%%%%%%%%%%%%%%%%%%%%%%%%%%%%%%%%%%%%%%%%%%%%%%%%%%%%%%%%%%%%%%%%%%%%%%%%%%
%------------------------------------------------
%fig 14. remove elevation dependence of observed OE, match to simulation models
%------------------------------------------------

\begin{figure}[b]
\includegraphics[width=1\textwidth]{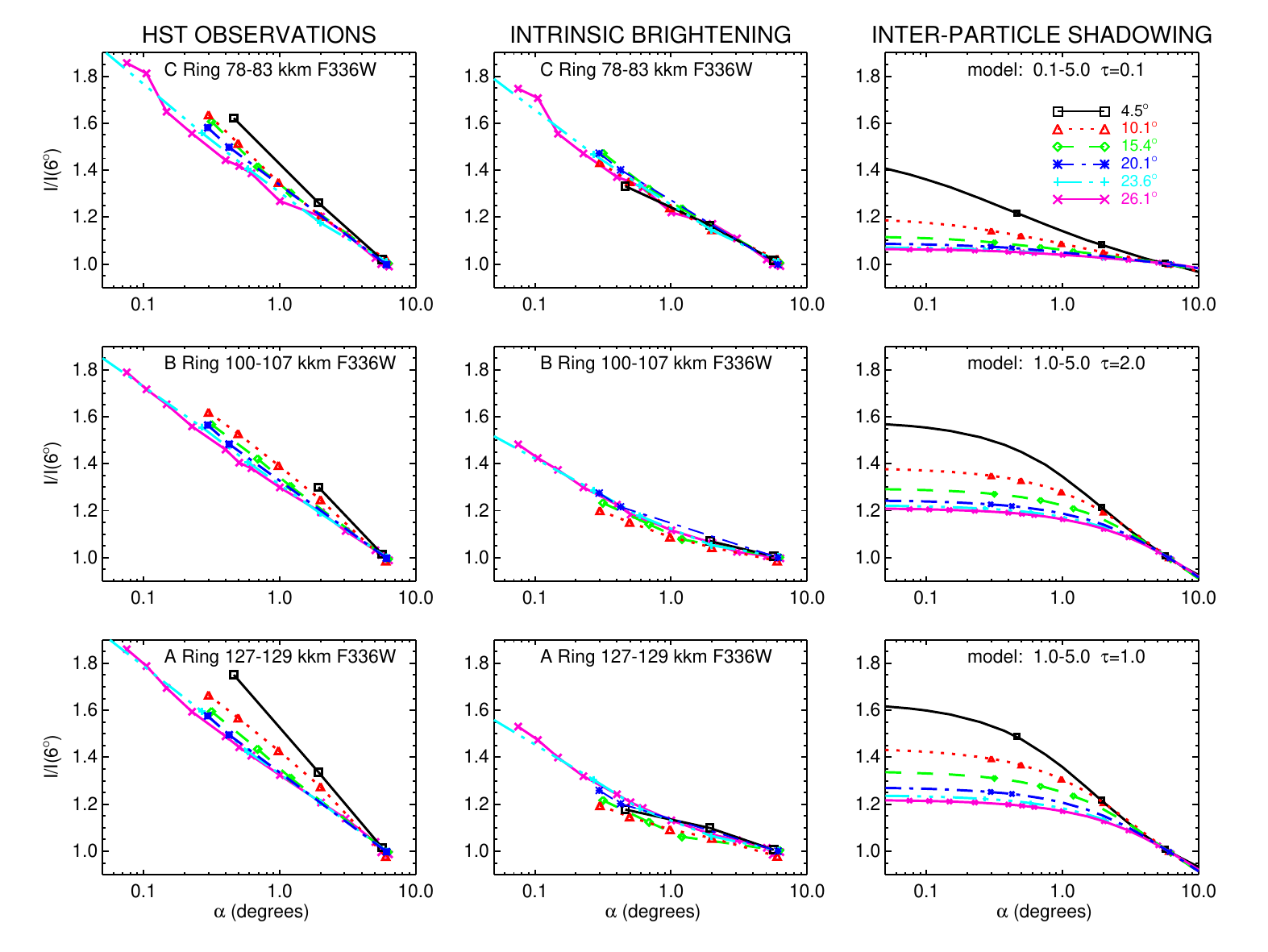}
  \caption{The left column shows the original HST phase curves for 6
    different sets of $B_{\rm eff}$, all normalized to
    $\alpha=6^\circ$. In the middle column, the intrinsic component
    $g_i(\alpha)=f_i(\alpha)P(\alpha) \, /  \,[f_i(6^\circ)P(6^\circ)]$ is shown,
    obtained by dividing out the interparticle shadowing contribution,
    determined by the best match to the elevation angle dependence of
    $OE$ in  Fig.~\ref{fig:simu_compare_oe_cba}: this interparticle
    shadowing contribution ($f_e(\alpha) \, / \, f_e(6^\circ)$) is shown
    at right.  The C, B, and A ring data are compared to simulations
    with dynamical optical depths $\taud=0.1, 2.0, $ and $1.0$,
    respectively, using a size distribution $0.1 - 5.0$ m for the C
    ring, and $1.0-5.0$ m for the B and A rings. Only the F336W filter
    is shown. The lines do not represent fits to the data, but simply
    connect the observations for each $\beff$.
    \label{fig:oppofit_residuals}}
\end{figure}

%%%%%%%%%%%%%%%%%%%%%%%%%%%%%%%%%%%%%%%%%%%%%%%%%%%%%%%%%%%%%%%%%%%%%%%%%%%%%%%%%%%%%%%%%%%%%%%%%%%%%%%%%
%Fig 15 residual 5 colors

%
\begin{figure}[b]
\includegraphics[width=1\textwidth]{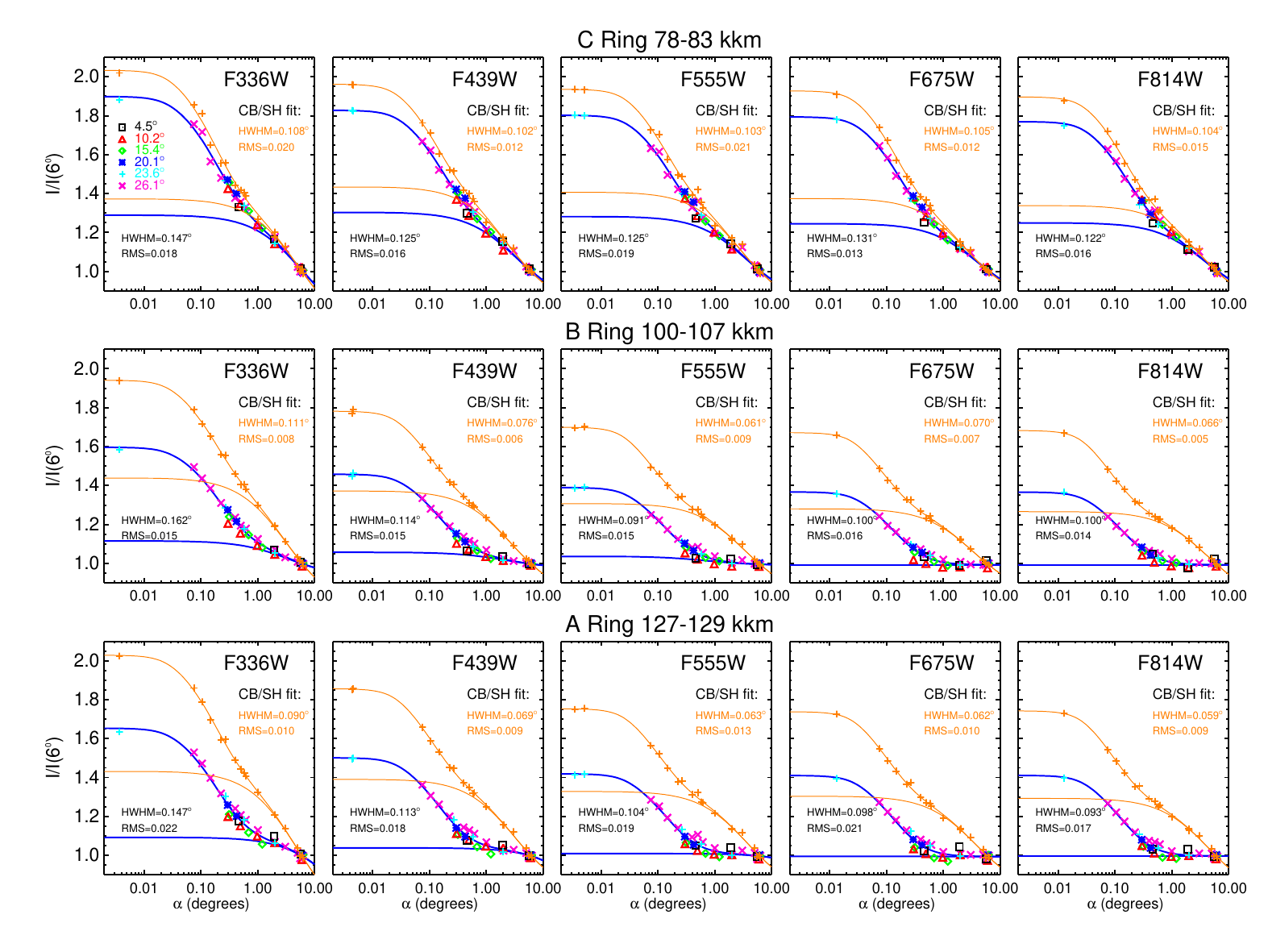}
  \caption{ The intrinsic component of opposition brightening in all five
    filters, after removing the interparticle shadowing contribution
    displayed in  Fig.~\ref{fig:oppofit_residuals} (for F336W the
    points are the same as the middle column of
     Fig.~\ref{fig:oppofit_residuals}).  Thick blue curves indicate the
    CB/SH (simplified Hapke 2002 model) fits to the data; the lower curve
    shows the SH contribution separately, while the CB contribution
    corresponds to the ratio between the two curves.  For
    comparison, the original HST phase curves at $B_{\rm eff}
    =23^\circ - 26^\circ$ are also shown (thin curves, small orange
    crosses); the ratio between orange and blue curves indicate
    the magnitude of the interparticle shadowing contribution. The HWHM
    and residual RMS for both fits are indicated.
    \label{fig:oppofit_residuals_5colors}}
\end{figure}

%%%%%%%%%%%%%%%%%%%%%%%%%%%%%%%%%%%%%%%%%%%%%%%%%%%%%%%%%%%%%%%%%%%%%%%%%%%%%%%%%%%%%%%%%%%%%%%%%%%%%%%%%
\clearpage
%Fig 16
%HST_oppo2009_linexp_hapke.pro
\begin{figure}[b]
\includegraphics[width=1\textwidth]{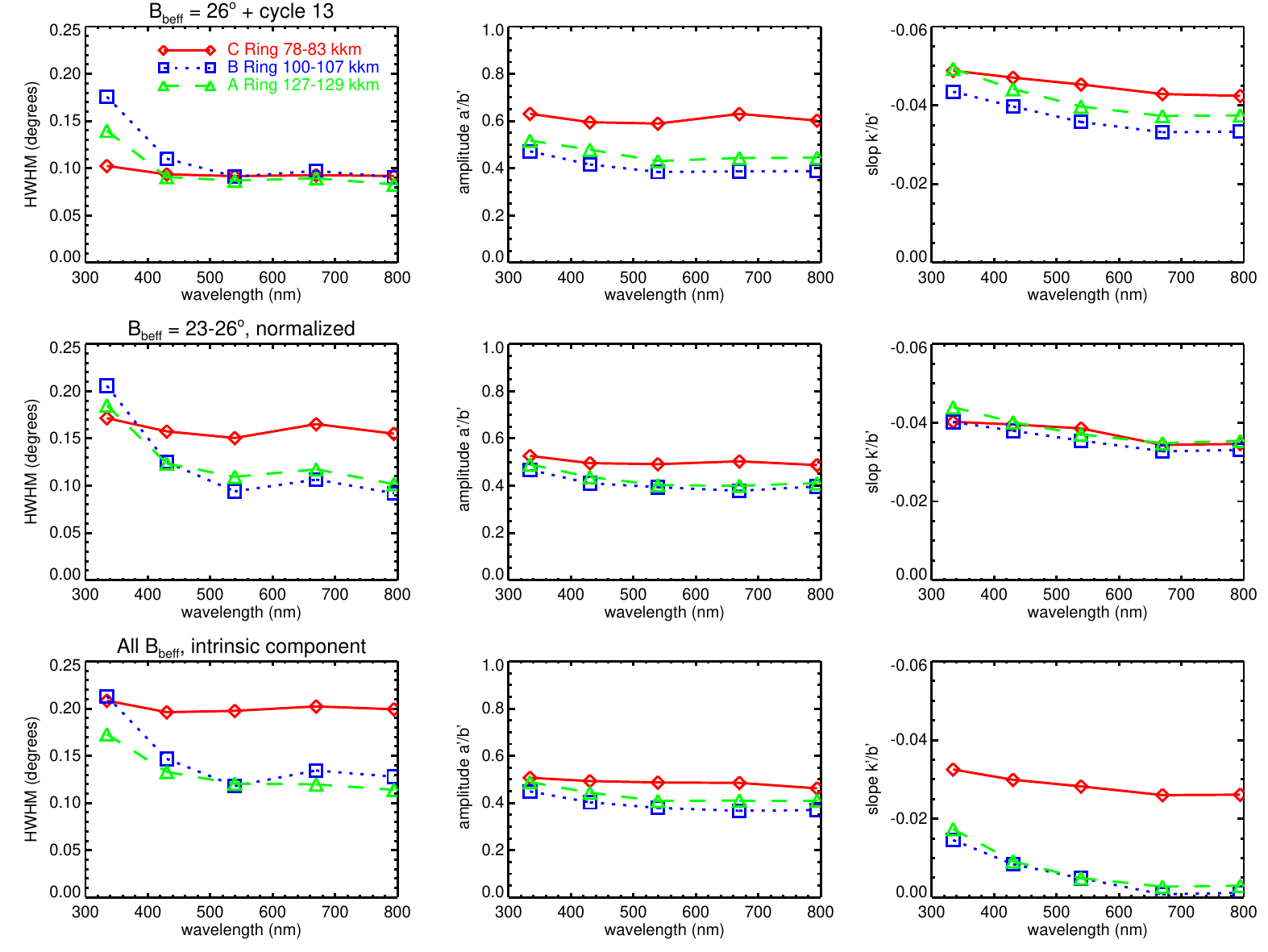}
 \caption{Wavelength dependence of linear-exponential model fits for
   the C, B, and A ring regions. The upper row corresponds to
   the data set used in French et al.~(2007b): the $B_{\rm eff}
   \approx 26^\circ$ data (Cycles 10-12) are combined with the Cycle 13
   opposition data point for $B_{\rm eff} \approx 23^\circ$, without
   any normalization of the $I/F$ levels. The frames display the HWHM,
   amplitude, and normalized slope from the fits. In the middle row,
   all original data for $B_{\rm eff} \sim 23^\circ$ and $\sim
   26^\circ$ are combined (Cycles 9-13), after proper normalization of the $I/F$
   levels to $\beff=23^\circ$; the interparticle shadowing
   component has not been eliminated (this corresponds to the thin curves and
   small orange symbols in Fig.~\ref{fig:oppofit_residuals_5colors}). The lower row combines the
   data from all elevation angles, after removal of the interparticle
   shadowing component (corresponding to the thick curves and large symbols
   in Fig.~\ref{fig:oppofit_residuals_5colors}).
    \label{fig:oppofit_residuals_linexp_fits}}
\end{figure}
\clearpage

%%%%%%%%%%%%%%%%%%%%%%%%%%%%%%%%%%%%%%%%%%%%%%%%%%%%%%%%%%%%%%%%%%%%%%%%%%%%%%%%%%%%%%%%%%%%%%%%%%%%%%%%%
%%               REMOVED IN REVISED VERSION
%Fig 17
%HST_oppo2009_linexp_hapke.pro
\begin{figure}[b]
\includegraphics[width=1\textwidth]{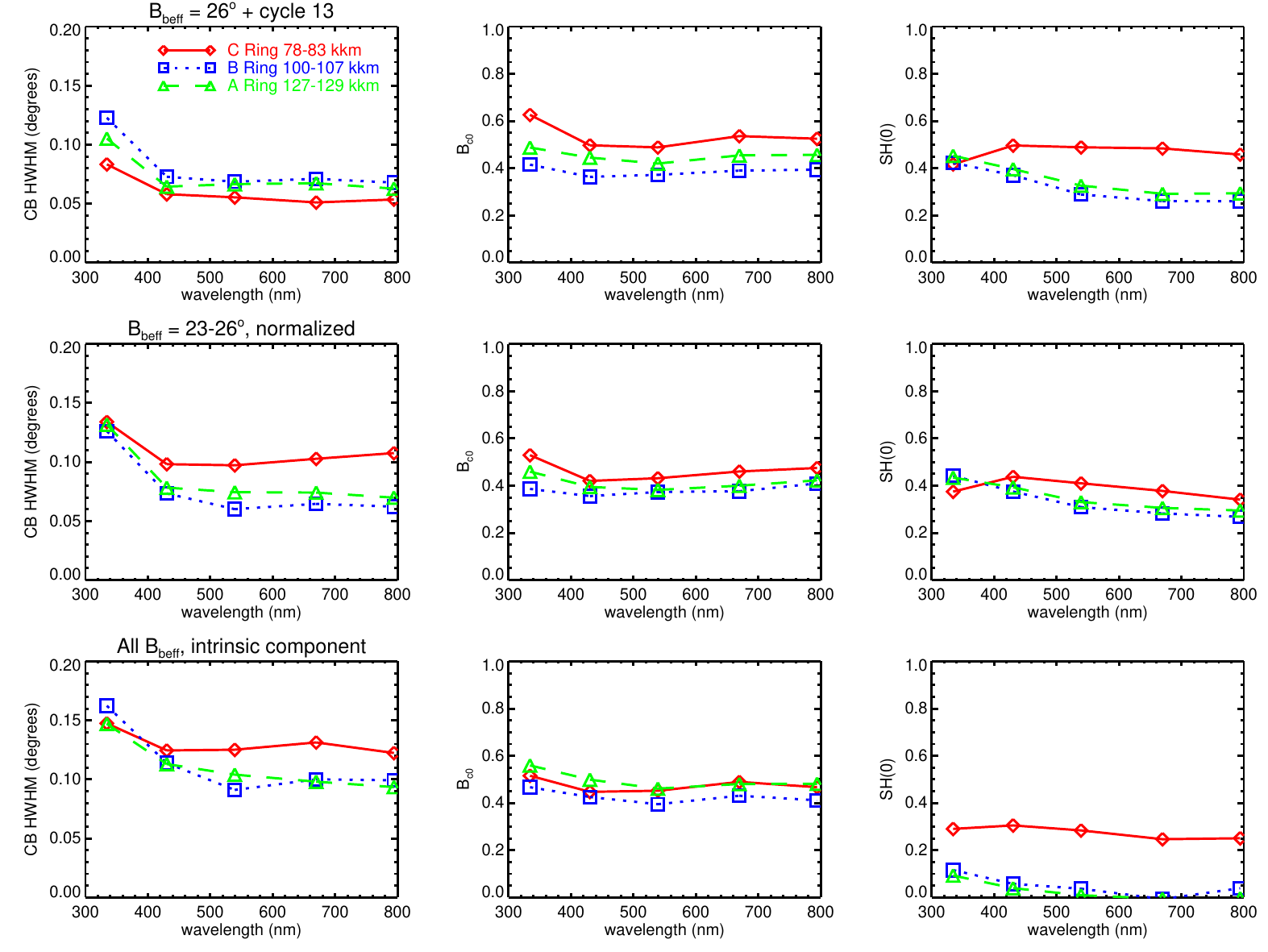}
 \caption{Same as  Fig.~\ref{fig:oppofit_residuals_linexp_fits}, except
   for CB/SH fits. The frames display the coherent backscattering HWHM$=0.72 h_{cb}$ (left column) and amplitude $B_{c0}$ (middle column), and
   the shadow hiding contribution at zero phase angle, $SH(0^\circ)=A_0(1+B_{s0})$ (right column).
    \label{fig:oppofit_residuals_hapke_fits}}
\end{figure}

%--------------------------------------
%--------------------------------------
%        TILT EFFECT FIGURES
%--------------------------------------
%--------------------------------------

%%%%%%%%%%%%%%%%%%%%%%%%%%%%%%%%%%%%%%%%%%%%%%%%%%%%%%%%%%%%%%%%%%%%%%%%%%%%%%%%%%%%%%%%%%%%%%%%%%%%%%%%%
%Fig 18
%F555W normalized to 4.5
%HST_oppo2009_tilt_save.pro

\begin{figure}[b]
\includegraphics[width=1\textwidth]{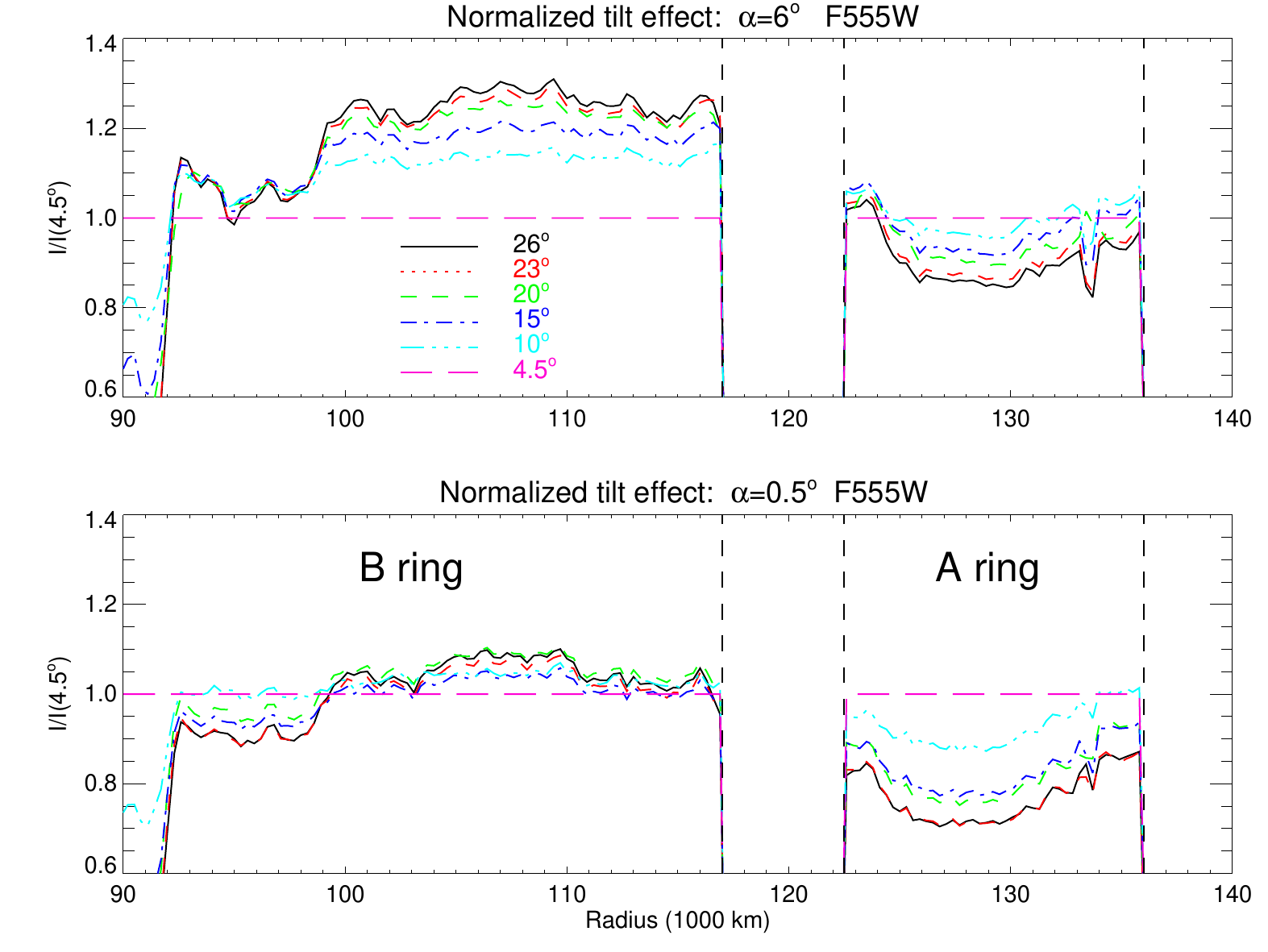}
  \caption{The tilt effect on the brightness of the A and B rings,
 for $\alpha \sim 6^\circ$ (upper frame) and $\alpha \sim
 0.5^\circ$ (lower frame). Radial F555W brightness profiles at the
 ansa have been grouped together by $B_{\rm eff}$ and averaged, and
 then normalized by the low ring elevation profile with $\Beff=4.5^\circ$. Note the positive tilt effect for the B ring
 and the negative (inverse) tilt effect for the A ring.
    \label{fig:tilt}}
\end{figure}
\clearpage

%%%%%%%%%%%%%%%%%%%%%%%%%%%%%%%%%%%%%%%%%%%%%%%%%%%%%%%%%%%%%%%%%%%%%%%%%%%%%%%%%%%%%%%%%%%%%%%%%%%%%%%%%
%Fig 19
%F336W F814W normalized to 10
%HST_oppo2008_tilt_336_814.pro
\begin{figure}[b]
\epsscale{0.7}
\includegraphics[width=1\textwidth]{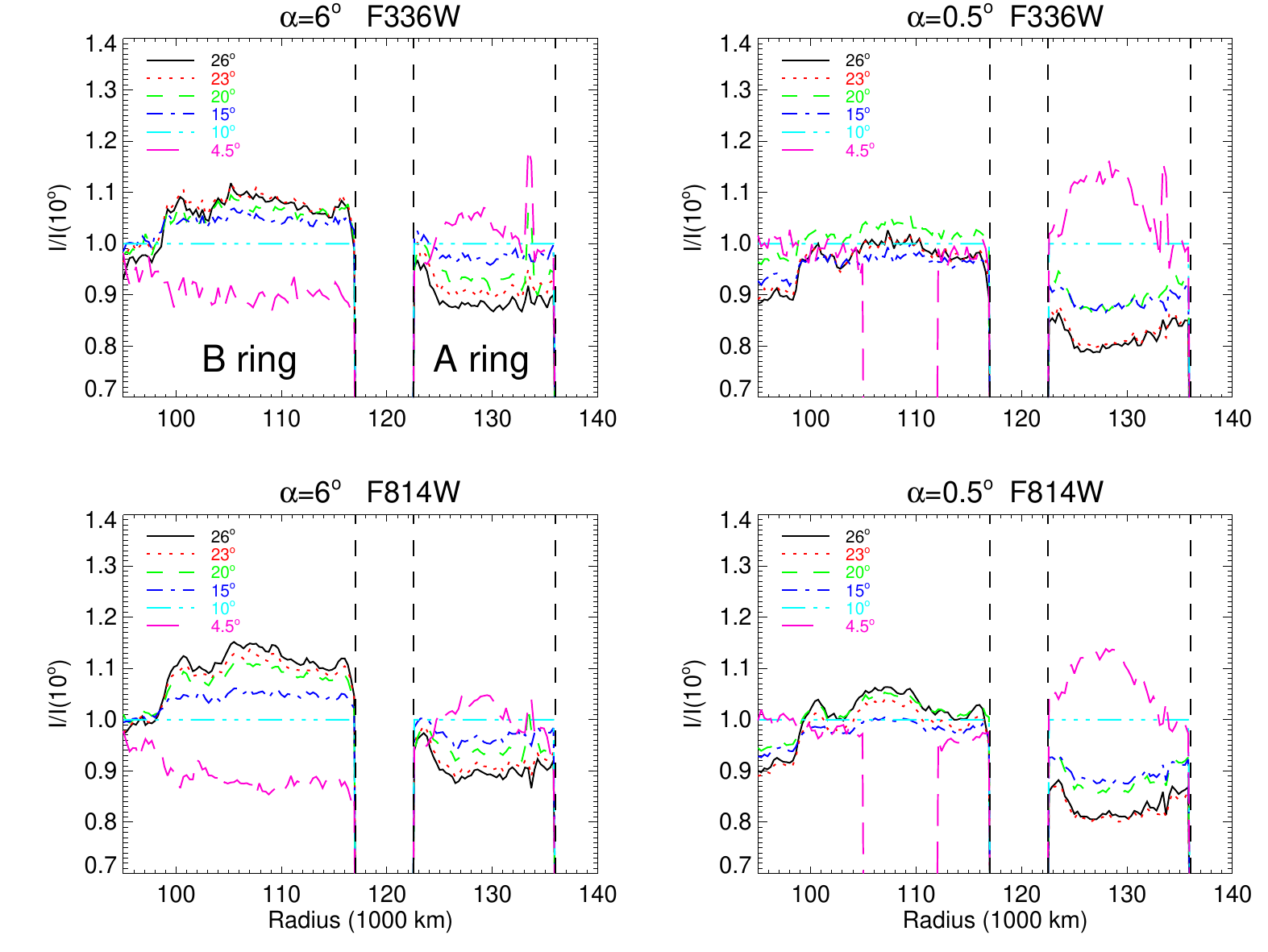}
\figcaption{Comparison of the B and A ring tilt effects, for
  non-opposition ($\alpha\sim 6^\circ$) and near opposition
  ($\alpha\sim 0.5^\circ$), for two filters. Since the low phase
  angle $\beff=4.5^\circ$ images were contaminated by spokes, we have
  made the normalization to $\beff=10^\circ$, and omit the affected
  region (105, 000 - 112, 000 km). Note how the B ring positive tilt
  effect is marginally larger for the larger $\lambda$ in the case of
  $\alpha \sim 6^\circ$, indicating that some fraction of the brightness
  increase is due to multiple scattering increasing with elevation, as
  proposed by Lumme et al.~(1983); however, the contribution is small
  compared to that of elevation-dependent opposition brightening.
\label{fig:tilt_336_814}}
\end{figure}

%%%%%%%%%%%%%%%%%%%%%%%%%%%%%%%%%%%%%%%%%%%%%%%%%%%%%%%%%%%%%%%%%%%%%%%%%%%%%%%%%%%%%%%%%%%%%%%%%%%%%%%%%
%Fig 20
%HST_oppo2009_tilt_ms_3panel_tarkistus
%savefile made by HST_oppo2009_tilt_save.pro

\begin{figure}[b]
\epsscale{0.7}
\includegraphics[width=1\textwidth]{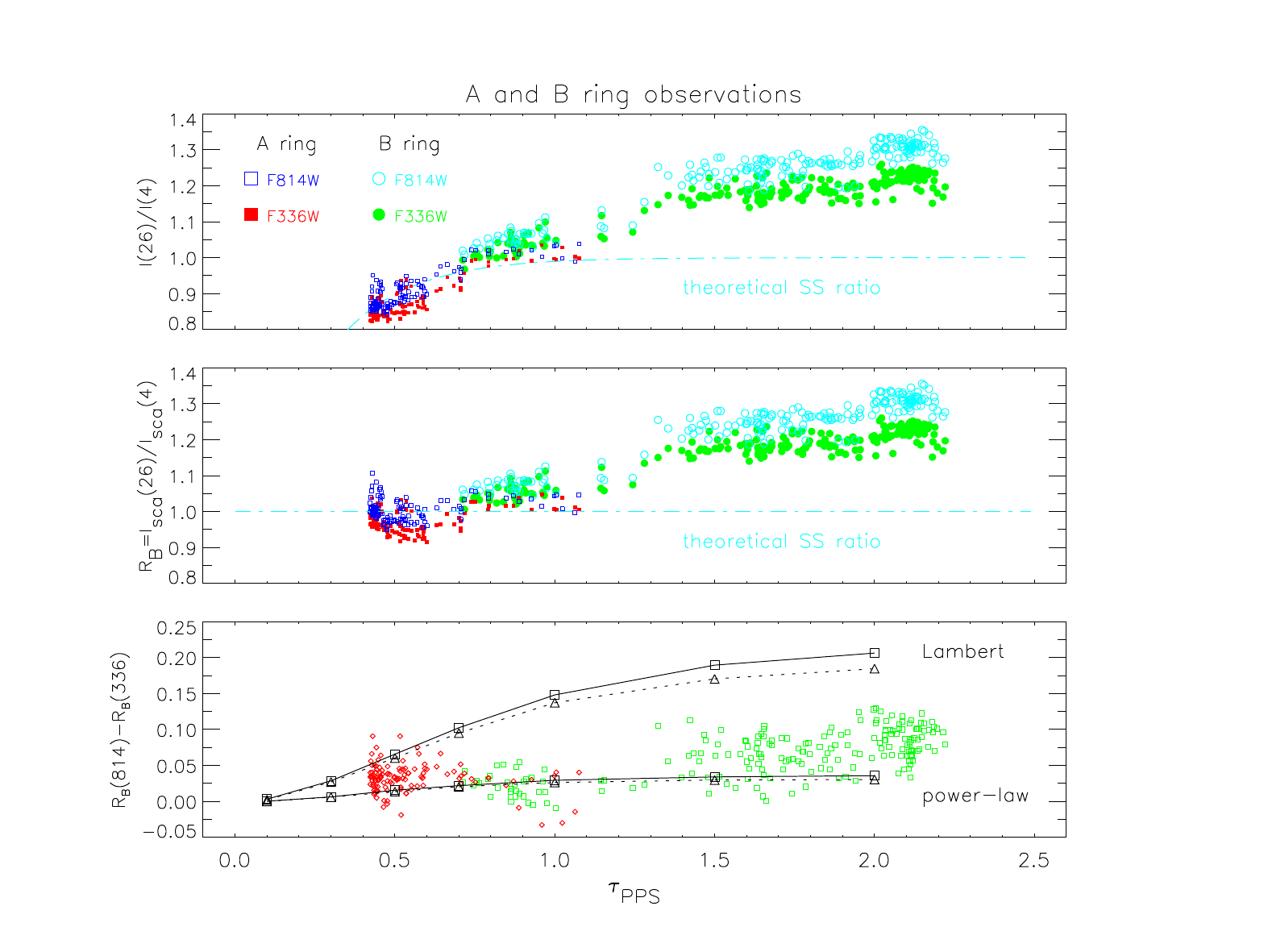}
\figcaption{Upper panel: Observed
  $I(\beff=26^\circ)/I(\beff=4.5^\circ)$ versus Voyager PPS optical
  depth, plotted in F814W and F333W filters for
  $\alpha=6^\circ$.  The dash-dotted curve indicates the theoretical ratio
  of singly scattered intensities: the values exceeding this curve
  indicate positive (B ring) or negative (mid A ring) tilt effect.
  The middle panel frame shows the the same, but using scaled intensities:
  $R_B={\hat I}(26^\circ)/\hat{I}(4.5^\circ)$, where
  $\hat{I}(r)=I(r)/I_{ss(D=0)}(\tau_{pps}(r))$; here, the theoretical
  singly scattered ratio is unity.  Lower row: the difference $R_B
  (F814W) - R_B(F336W)$, which represents the contribution of multiple
  scattering. Also shown are simulation models using both Lambert
  phase and $n_s=3.09$ power law phase functions: solid and dotted
  lines stand for simulations with size distributions of 0.1 -
  5.0 m and 0.5-5.0 m, respectively. The assumed particle
  albedos for F333W and F814W filters are 0.36 and 0.87, respectively,
  when the Lambert phase function is used, and 0.21 and 0.57 in connection
  with the power law phase function: with these assumptions the
  modeled $I/F$ values for 0.5-5.0 m size distribution match
  the observations at $\beff=26^\circ, \alpha=6^\circ$.
\label{fig:tilt_ms}}
\end{figure}

%%%%%%%%%%%%%%%%%%%%%%%%%%%%%%%%%%%%%%%%%%%%%%%%%%%%%%%%%%%%%%%%%%%%%%%%%%%%%%%%%%%%%%%%%%%%%%%%%%%%%%%%%
%%               REMOVED IN REVISED VERSION
%Fig 21
%HST_oppo2008_models_tarkistus

%%\begin{figure}[b]
%%\epsscale{0.7}
%%  \includegraphics[width=1\textwidth]{HST_oppo2008_models_tarkistus_f06_05_tilt_4panel.pdf}
%%  \caption{Upper row: the interparticle-shadowing enhancement factor $f_e$
%%    for different elevations, normalized to that at $\Beff=4^\circ$.  At left, the phase angle $\alpha=6^\circ$, 
%%    and at right
%%    $\alpha=0.5^\circ$.  The different size distribution models are
%%    compared, for two dynamical optical depths.  Lower row: the ratio
%%    $f_e(B=26^\circ)/f_e(B=4^\circ)$, shown as a function of optical
%%    depth.  The tick marks at the right edges of the frames indicate the maximum observed
%%    tilt effect in the mid B ring.
%%    \label{fig:tilt_simu_fe}}
%%\end{figure}

%%%%%%%%%%%%%%%%%%%%%%%%%%%%%%%%%%%%%%%%%%%%%%%%%%%%%%%%%%%%%%%%%%%%%%%%%%%%%%%%%%%%%%%%%%%%%%%%%%%%%%%%%
\begin{figure}[b]
\includegraphics[width=1\textwidth]{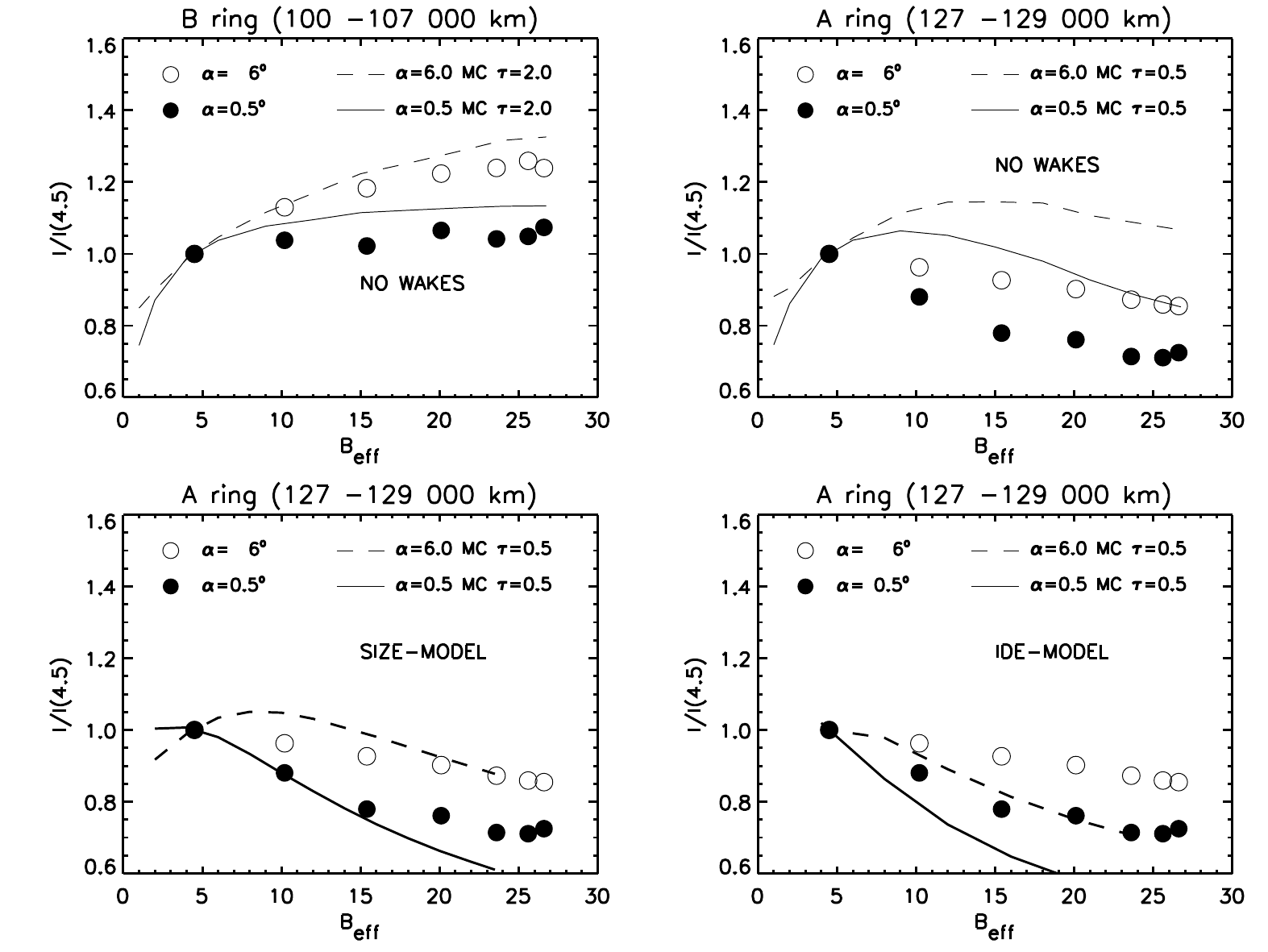}
  \caption{Comparison of the observed B and A ring tilt effects with photometric
    models, for $\alpha \sim 6^\circ$ and for $\alpha \sim
    0.5^\circ$. We have extracted the intensity at the ring ansa vs
    $\Beff$ at the indicated radial ranges.  The upper row displays
    observations for the B and A rings, together with results of
    photometric models, using non-gravitating particle simulations,
    with size distribution from 0.5 to 5.0 m, and with $\taud=2.0$
    and $\taud=0.5$.  In the lower row, the A ring observations are
    compared with the self-gravitating SIZE and IDE models explored in
    Salo et al.~ (2004) and French et al.~ (2007a). The $n_s=3.09$ power-law
    phase function with albedo A=0.5 is assumed.
    \label{fig:tilt_simu}}
\end{figure}

%%%%%%%%%%%%%%%%%%%%%%%%%%%%%%%%%%%%%%%%%%%%%%%%%%%%%%%%%%%%%%%%%%%%%%%%%%%%%%%%%%%%%%%%%%%%%%%%%%%%%%%%%
%%NEW
%%%%%%%%%%%%%%%%%%%%%%%%%%%%%%%%%%%%%%%%%%%%%%%%%%%%%%%%%%%%%%%%%%%%%%%%%%%%%%%%%%%%%%%%%%%%%%%%%%%%%%%%%

%%%%%%%%%%%%%%%%%%%%%%%%%%%%%%%%%%%%%%%%%%%%%%%%%%%%%%%%%%%%%%%%%%%%%%%%%%%%%%%%%%%%%%%%%%%%%%%%%%%%%%%%%
\begin{figure}[b]
\includegraphics[width=1\textwidth]{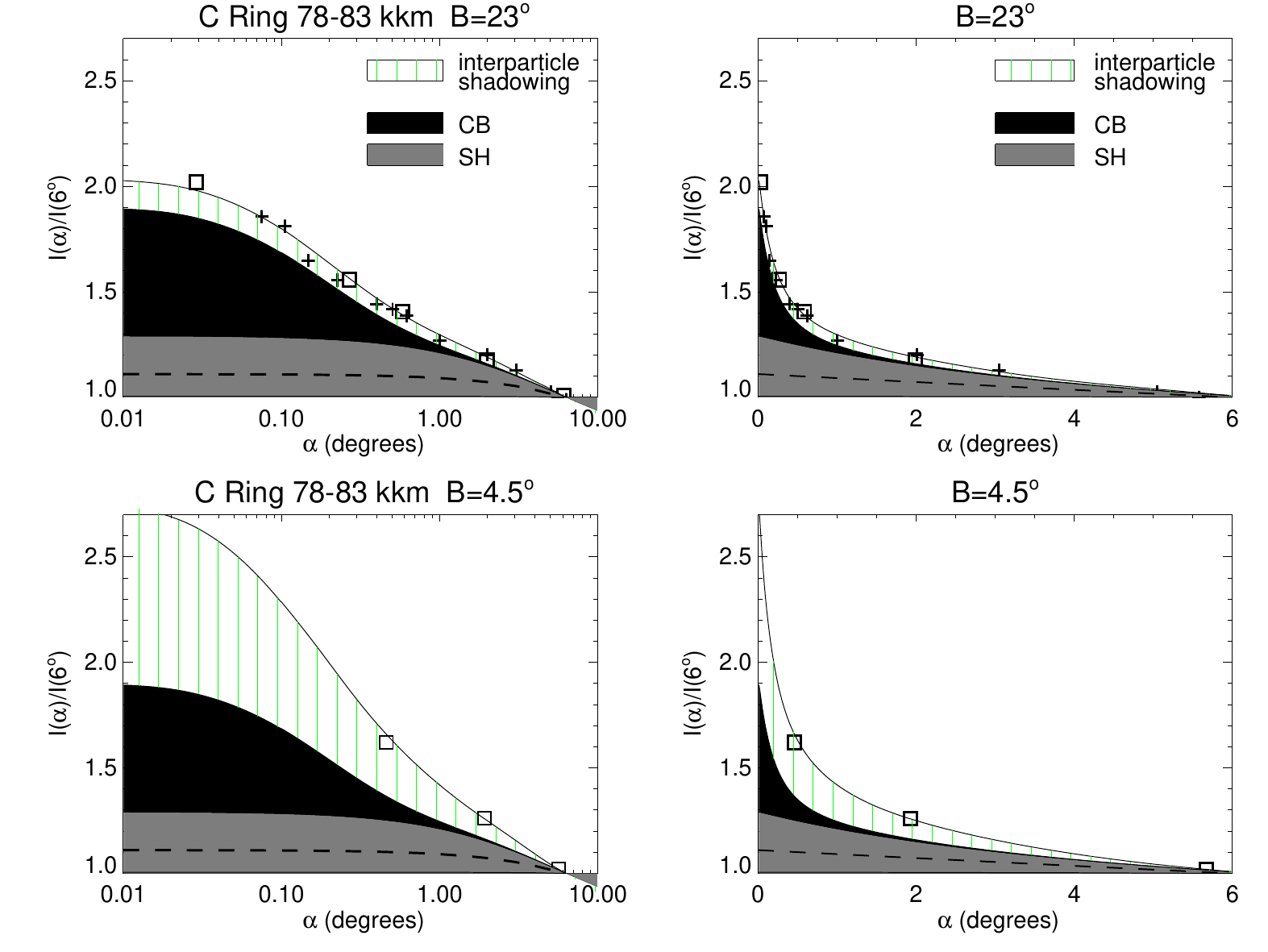}
  \caption{Modeled contributions to the C ring opposition effect,
    measured in terms of $I/I(6^\circ)$, shown for $B_{\rm
      eff}=23^\circ$ and $B_{\rm eff}=4.5^\circ$ (upper and lower
    frames, respectively), on logarithmic (left) and linear (right)
    scales. The models are for F336W filter: the vertically banded
    shaded regions indicate the SH (gray) and CB (black) contributions
    to the intrinsic component, obtained from a SH-CB fit to HST data
    from all elevations, after removal of the modeled interparticle
    contribution; they are identical for both $B_{\rm eff}$'s.  The
    shaded region indicates the interparticle contribution, which
    depends on $B_{\rm eff}$. The interparticle contribution is
    calculated for the best fitting model of Section 4 (dynamical
    optical depth $\taud=0.1$, $W_s=50$, Bridges et al.~(1984)
    coefficient of restitution formula; photometric calculations use a
    Lambert law with $A=0.5$).  Symbols indicate the HST observations
    at the F336W filter, for the indicated $B_{\rm eff}$. The dashed
    line indicates the relative change of an $n_s=3.07$ power law
    phase function.
    \label{fig:discussion_cb_sh_cring}}
\end{figure}

%%%%%%%%%%%%%%%%%%%%%%%%%%%%%%%%%%%%%%%%%%%%%%%%%%%%%%%%%%%%%%%%%%%%%%%%%%%%%%%%%%%%%%%%%%%%%%%%%%%%%%%%%
\begin{figure}[b]
\includegraphics[width=1\textwidth]{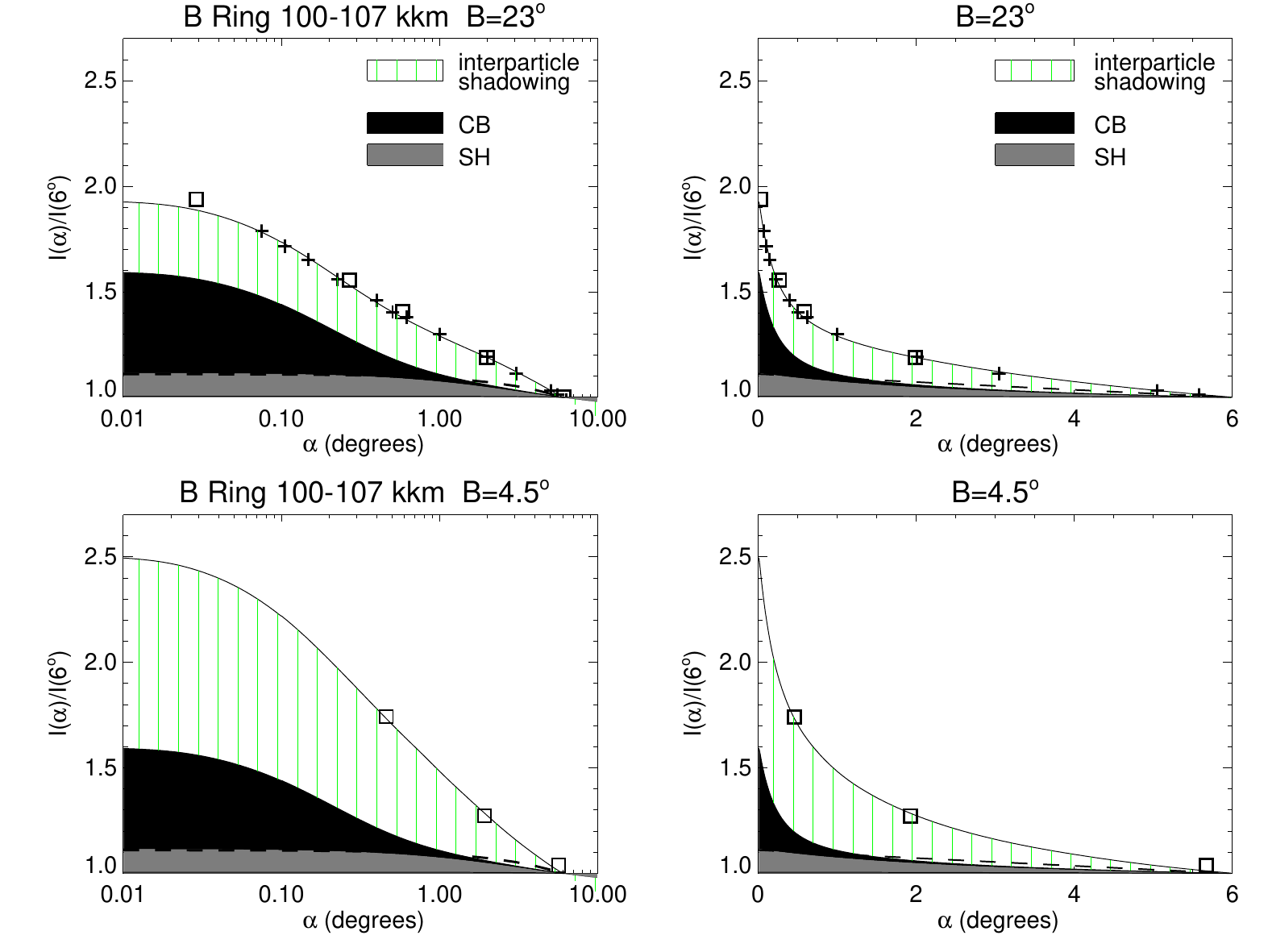}
  \caption{ Same as Fig. \ref{fig:discussion_cb_sh_cring}, except for
    the B ring region. The interparticle shadowing model is calculated
    for dynamical optical depth $\taud=2.0$,
    $W_s=5$, Bridges et al.~(1984) coefficient of restitution formula.
    \label{fig:discussion_cb_sh_bring}}
\end{figure}

%%%%%%%%%%%%%%%%%%%%%%%%%%%%%%%%%%%%%%%%%%%%%%%%%%%%%%%%%%%%%%%%%%%%%%%%%%%%%%%%%%%%%%%%%%%%%%%%%%%%%%%%%
\begin{figure}[b]
\includegraphics[width=1\textwidth]{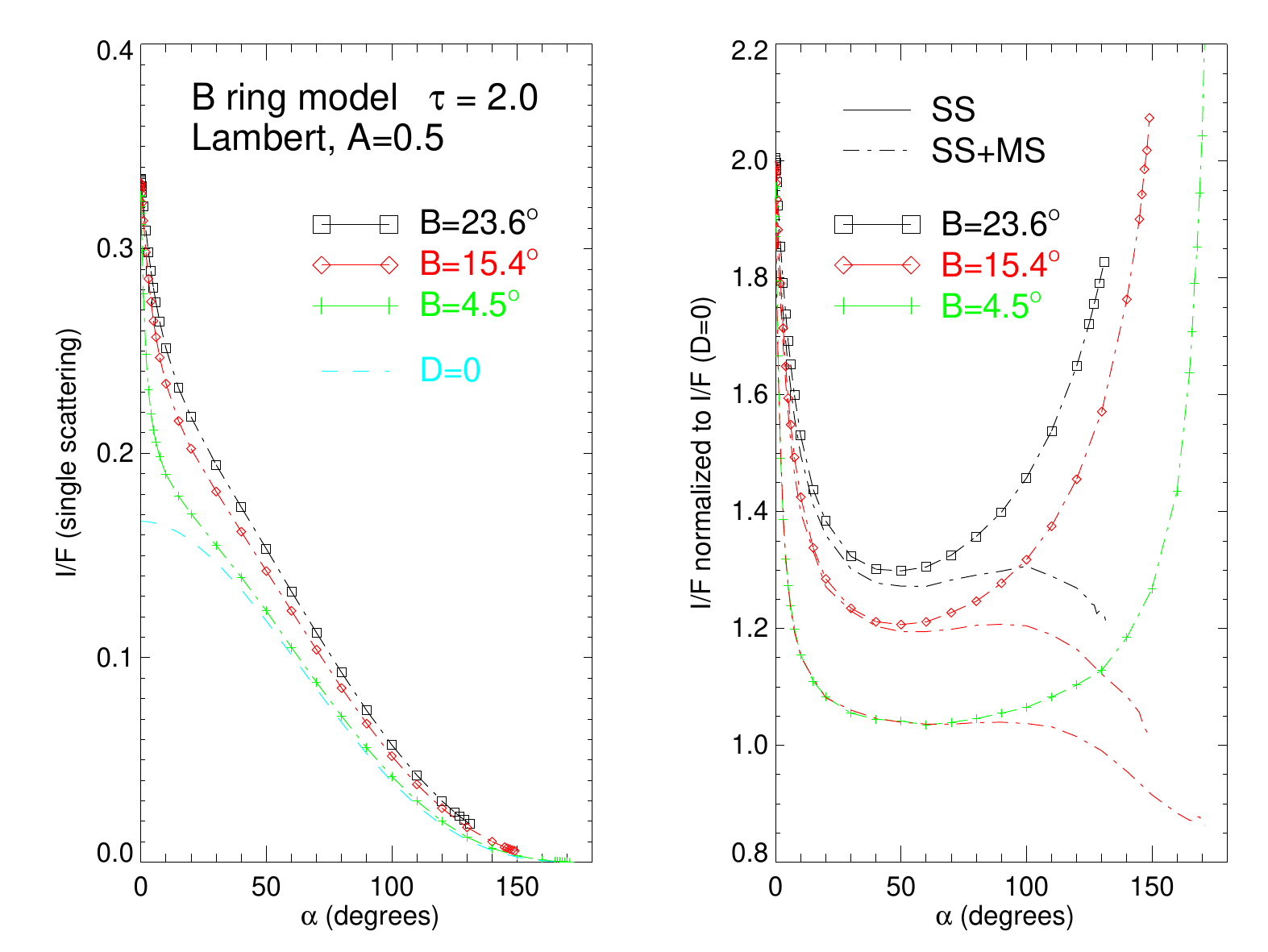}
  \caption{The behavior of the modeled B ring $I/F$ for a full range of
    phase angles (dynamical optical depth $\taud=2.0$, $W_s=5$,
    Bridges et al.~(1984) coefficient of restitution formula is
    assumed; together with Lambert phase function with Bond albedo
    $A=0.5$). At left, the single scattering contribution is shown,
    for three different $\beff$'s. For comparison, the classical $D=0$
    single scattering contribution is also shown by a dashed line.  At
    the right, the same single scattering model curves are shown,
    normalized to $D=0$ curve. Also shown by dashed lines are
    corresponding ratios when both single and multiple scattering are
    included.
    \label{fig:discussion_oppo_extended}}
\end{figure}

%%%%%%%%%%%%%%%%%%%%%%%%%%%%%%%%%%%%%%%%%%%%%%%%%%%%%%%%%%%%%%%%%%%%%%%%%%%%%%%%%%%%%%%%%%%%%%%%%%%%%%%%%
\begin{figure}[b]
\includegraphics[width=1\textwidth]{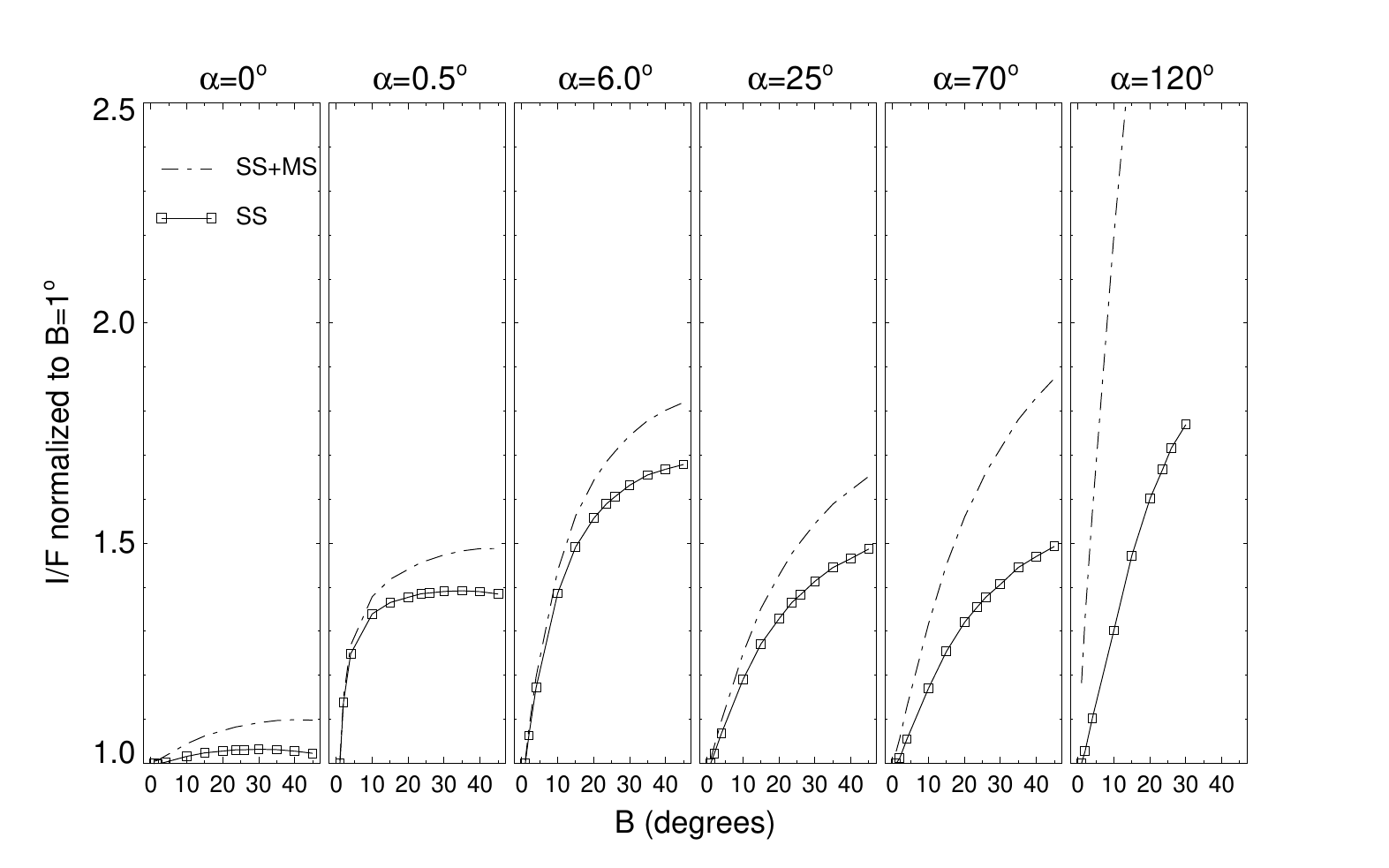}
  \caption{Comparison of the modeled tilt effect for a range of phase
    angles: the dynamical/photometric parameters are the same as in
    Fig. \ref{fig:discussion_oppo_extended}.
    \label{fig:discussion_tilt_extended}}
\end{figure}

\clearpage
%%%%%%%%%%%%%%%%%%%%%%%%%%%%%%%%%%%%%%%%%%%%%%%%%%%%%%%%%%%%%%%%%%%%%%%%%%%%%%%%%%%%%%%%%%%%%%%%%%%%%%%%%
\begin{deluxetable}{rlrlrlrlrl}

\tablewidth{-15pt}
\tabletypesize{\scriptsize}
\tablecaption{C Ring: 78,000-83,000 km}
\tablehead{
{\hfill F336W} & {}& {\hfill  F439W} & {}& {\hfill  F555W} & {}& {\hfill F675W} & {}& {\hfill F814W} & {} \\ \tableline\\
\colhead{$\alpha$} & \colhead{$I/F_{\rm corr}$} & 
\colhead{$\alpha$} & \colhead{$I/F_{\rm corr}$} & 
\colhead{$\alpha$} & \colhead{$I/F_{\rm corr}$} & 
\colhead{$\alpha$} & \colhead{$I/F_{\rm corr}$} & 
\colhead{$\alpha$} & \colhead{$I/F_{\rm corr}$}}
\startdata

%\cutinhead{\hskip0.4cm F439W \hskip1.6cm            F439W   \hskip1.6cm           F439W      \hskip1.6cm      F439W     \hskip1.6cm     F439W }

\sidehead{ Cycle 6 $|B_{\rm eff}| = 4.30^\circ - 4.66^\circ$}\\
 0.4605 &  0.1235 &  0.4607 &  0.1514 &  0.4609 &  0.1680 &  0.4611 &  0.1840 &  0.4613 &  0.1962 \\
 1.9249 &  0.0955 &  0.4617 &  0.1524 &  0.4619 &  0.1692 &  0.4621 &  0.1860 &  1.9257 &  0.1554 \\
 1.9315 &  0.0966 &  1.9251 &  0.1201 &  1.9253 &  0.1319 &  1.9255 &  0.1489 &  5.6708 &  0.1322 \\
 5.6705 &  0.0780 &  1.9318 &  0.1198 &  1.9320 &  0.1367 &  5.6709 &  0.1230 &   ...   &   ...   \\
 5.6711 &  0.0774 &  5.6704 &  0.0980 &  5.6704 &  0.1104 &   ...   &   ...   &   ...   &   ...   \\
  ...   &   ...   &  5.6710 &  0.0973 &  5.6709 &  0.1107 &   ...   &   ...   &   ...   &   ...   \\

\sidehead{ Cycle 7a $|B_{\rm eff}| = 10.01^\circ - 10.29^\circ$}\\
 0.2997 &  0.0880 &  0.2998 &  0.1108 &  0.2998 &  0.1204 &  0.2998 &  0.1351 &  0.2999 &  0.1377 \\
 0.3000 &  0.0879 &  0.3000 &  0.1109 &  0.3000 &  0.1209 &  0.3000 &  0.1311 &  0.3001 &  0.1385 \\
 0.4948 &  0.0818 &  0.4946 &  0.1027 &  0.4944 &  0.1115 &  0.4943 &  0.1210 &  0.4941 &  0.1274 \\
 0.4958 &  0.0811 &  0.4956 &  0.1016 &  0.4954 &  0.1094 &  0.4952 &  0.1209 &  0.4951 &  0.1260 \\
 0.9774 &  0.0733 &  0.4957 &  0.1014 &  0.9770 &  0.1014 &  0.9767 &  0.1099 &  0.9765 &  0.1169 \\
 0.9843 &  0.0717 &  0.9772 &  0.0930 &  0.9837 &  0.0985 &  0.9834 &  0.1087 &  0.9832 &  0.1132 \\
 1.9911 &  0.0650 &  0.9839 &  0.0911 &  1.9906 &  0.0905 &  1.9904 &  0.0999 &  1.9901 &  0.1043 \\
 1.9979 &  0.0645 &  0.9840 &  0.0910 &  1.9975 &  0.0893 &  1.9972 &  0.0992 &  1.9970 &  0.1027 \\
 6.0165 &  0.0554 &  1.9908 &  0.0831 &  6.0165 &  0.0790 &  6.0165 &  0.0858 &  6.0165 &  0.0903 \\
 6.0168 &  0.0546 &  1.9977 &  0.0818 &  6.0168 &  0.0779 &  6.0168 &  0.0862 &  6.0169 &  0.0893 \\
  ...   &   ...   &  6.0165 &  0.0717 &   ...   &   ...   &   ...   &   ...   &   ...   &   ...   \\
  ...   &   ...   &  6.0165 &  0.0714 &   ...   &   ...   &   ...   &   ...   &   ...   &   ...   \\
  ...   &   ...   &  6.0165 &  0.0716 &   ...   &   ...   &   ...   &   ...   &   ...   &   ...   \\
  ...   &   ...   &  6.0165 &  0.0714 &   ...   &   ...   &   ...   &   ...   &   ...   &   ...   \\
  ...   &   ...   &  6.0168 &  0.0706 &   ...   &   ...   &   ...   &   ...   &   ...   &   ...   \\

\sidehead{ Cycle 7b $|B_{\rm eff}| = 15.40^\circ - 15.49^\circ$}\\
 0.3156 &  0.0659 &  0.3156 &  0.0825 &  0.3158 &  0.0885 &  0.3158 &  0.0975 &  0.3159 &  0.1016 \\
 0.3160 &  0.0661 &  0.3157 &  0.0830 &  0.3162 &  0.0890 &  0.3163 &  0.0957 &  0.3164 &  0.1022 \\
 0.6866 &  0.0586 &  0.3161 &  0.0826 &  0.6861 &  0.0798 &  0.6859 &  0.0878 &  0.6857 &  0.0914 \\
 0.6927 &  0.0575 &  0.6864 &  0.0747 &  0.6923 &  0.0777 &  0.6921 &  0.0841 &  0.6918 &  0.0903 \\
 1.1982 &  0.0533 &  0.6925 &  0.0718 &  1.1978 &  0.0724 &  1.1975 &  0.0784 &  1.1973 &  0.0846 \\
 1.2042 &  0.0537 &  1.1980 &  0.0672 &  1.2034 &  0.0737 &  1.2031 &  0.0800 &  1.2029 &  0.0853 \\
 6.2567 &  0.0417 &  1.2036 &  0.0687 &  6.2567 &  0.0589 &  6.2566 &  0.0656 &  6.2566 &  0.0684 \\
 6.2568 &  0.0407 &  1.2038 &  0.0687 &  6.2567 &  0.0576 &  6.2567 &  0.0636 &  6.2567 &  0.0676 \\
  ...   &   ...   &  1.2039 &  0.0687 &   ...   &   ...   &   ...   &   ...   &   ...   &   ...   \\
  ...   &   ...   &  6.2567 &  0.0545 &   ...   &   ...   &   ...   &   ...   &   ...   &   ...   \\
  ...   &   ...   &  6.2567 &  0.0530 &   ...   &   ...   &   ...   &   ...   &   ...   &   ...   \\

\sidehead{ Cycle 8 $|B_{\rm eff}| = 20.05^\circ - 20.17^\circ$}\\
 0.2956 &  0.0542 &  0.2957 &  0.0682 &  0.2958 &  0.0718 &  0.2959 &  0.0790 &  0.2959 &  0.0826 \\
 0.2971 &  0.0544 &  0.2957 &  0.0680 &  0.2958 &  0.0727 &  0.2973 &  0.0795 &  0.2973 &  0.0848 \\
 0.4244 &  0.0523 &  0.2957 &  0.0681 &  0.2958 &  0.0725 &  0.4233 &  0.0771 &  0.4231 &  0.0806 \\
 0.4257 &  0.0506 &  0.2972 &  0.0693 &  0.2972 &  0.0738 &  0.4251 &  0.0741 &  0.4249 &  0.0776 \\
 6.1038 &  0.0346 &  0.4240 &  0.0664 &  0.4235 &  0.0702 &  6.1037 &  0.0534 &  6.1036 &  0.0580 \\
 6.1052 &  0.0340 &  0.4241 &  0.0662 &  0.4236 &  0.0707 &  6.1049 &  0.0527 &  6.1049 &  0.0562 \\
  ...   &   ...   &  0.4242 &  0.0666 &  0.4238 &  0.0700 &   ...   &   ...   &   ...   &   ...   \\
  ...   &   ...   &  0.4255 &  0.0643 &  0.4253 &  0.0675 &   ...   &   ...   &   ...   &   ...   \\
  ...   &   ...   &  6.1038 &  0.0455 &  6.1037 &  0.0494 &   ...   &   ...   &   ...   &   ...   \\
  ...   &   ...   &  6.1051 &  0.0444 &  6.1050 &  0.0474 &   ...   &   ...   &   ...   &   ...   \\
  ...   &   ...   &  6.1051 &  0.0442 &  6.1050 &  0.0475 &   ...   &   ...   &   ...   &   ...   \\
  ...   &   ...   &  6.1052 &  0.0443 &  6.1050 &  0.0471 &   ...   &   ...   &   ...   &   ...   \\

\sidehead{ Cycle 9 $|B_{\rm eff}| = 23.54^\circ - 23.69^\circ$}\\
 0.2681 &  0.0487 &  0.2682 &  0.0622 &  0.2683 &  0.0657 &  0.2684 &  0.0710 &  0.2685 &  0.0746 \\
 0.2702 &  0.0488 &  0.2682 &  0.0620 &  0.2684 &  0.0651 &  0.2705 &  0.0706 &  0.2706 &  0.0744 \\
 0.5852 &  0.0443 &  0.2703 &  0.0617 &  0.2704 &  0.0649 &  0.5862 &  0.0655 &  0.5864 &  0.0688 \\
 0.5870 &  0.0437 &  0.2703 &  0.0617 &  0.2704 &  0.0644 &  0.5880 &  0.0640 &  0.5882 &  0.0678 \\
 1.9914 &  0.0369 &  0.5854 &  0.0565 &  0.5858 &  0.0604 &  1.9925 &  0.0562 &  1.9929 &  0.0602 \\
 1.9915 &  0.0368 &  0.5856 &  0.0565 &  0.5860 &  0.0594 &  1.9927 &  0.0556 &  6.0951 &  0.0514 \\
 6.0947 &  0.0313 &  0.5873 &  0.0558 &  0.5876 &  0.0590 &  6.0950 &  0.0472 &  6.0954 &  0.0531 \\
 6.0952 &  0.0319 &  0.5874 &  0.0557 &  0.5878 &  0.0582 &  6.0954 &  0.0501 &   ...   &   ...   \\
  ...   &   ...   &  1.9918 &  0.0483 &  1.9922 &  0.0515 &   ...   &   ...   &   ...   &   ...   \\
  ...   &   ...   &  1.9919 &  0.0479 &  1.9923 &  0.0509 &   ...   &   ...   &   ...   &   ...   \\
  ...   &   ...   &  6.0948 &  0.0407 &  6.0949 &  0.0434 &   ...   &   ...   &   ...   &   ...   \\
  ...   &   ...   &  6.0948 &  0.0405 &  6.0949 &  0.0429 &   ...   &   ...   &   ...   &   ...   \\
  ...   &   ...   &  6.0948 &  0.0405 &  6.0950 &  0.0434 &   ...   &   ...   &   ...   &   ...   \\
  ...   &   ...   &  6.0953 &  0.0419 &  6.0953 &  0.0449 &   ...   &   ...   &   ...   &   ...   \\
\enddata
\label{table:cring}
\end{deluxetable}

\clearpage
%%%%%%%%%%%%%%%%%%%%%%%%%%%%%%%%%%%%%%%%%%%%%%%%%%%%%%%%%%%%%%%%%%%%%%%%%%%%%%%%%%%%%%%%%%%%%%%%%%%%%%%%%
\begin{deluxetable}{rlrlrlrlrl}

\tablewidth{-15pt}
\tabletypesize{\scriptsize}
\tablecaption{B Ring: 100,000-107,000 km}
\tablehead{
{\hfill F336W} & {}& {\hfill  F439W} & {}& {\hfill  F555W} & {}& {\hfill F675W} & {}& {\hfill F814W} & {} \\ \tableline\\
\colhead{$\alpha$} & \colhead{$I/F_{\rm corr}$} & 
\colhead{$\alpha$} & \colhead{$I/F_{\rm corr}$} & 
\colhead{$\alpha$} & \colhead{$I/F_{\rm corr}$} & 
\colhead{$\alpha$} & \colhead{$I/F_{\rm corr}$} & 
\colhead{$\alpha$} & \colhead{$I/F_{\rm corr}$}}
\startdata 

%\cutinhead{\hskip0.4cm F439W \hskip1.6cm            F439W   \hskip1.6cm           F439W      \hskip1.6cm      F439W     \hskip1.6cm     F439W }

\sidehead{ Cycle 6 $|B_{\rm eff}| = 4.30^\circ - 4.66^\circ$}\\
 0.4605 &  0.2681 &  0.4607 &  0.4421 &  0.4609 &  0.5588 &  0.4611 &  0.6222 &  0.4613 &  0.6515 \\
 1.9249 &  0.1875 &  0.4617 &  0.4432 &  0.4619 &  0.5592 &  0.4621 &  0.6207 &  1.9257 &  0.5055 \\
 1.9315 &  0.2043 &  1.9251 &  0.3298 &  1.9253 &  0.4261 &  1.9255 &  0.4948 &  5.6708 &  0.4406 \\
 5.6705 &  0.1595 &  1.9318 &  0.3595 &  1.9320 &  0.4640 &  5.6709 &  0.4211 &   ...   &   ...   \\
 5.6711 &  0.1601 &  5.6704 &  0.2875 &  5.6704 &  0.3737 &   ...   &   ...   &   ...   &   ...   \\
  ...   &   ...   &  5.6710 &  0.2862 &  5.6709 &  0.3740 &   ...   &   ...   &   ...   &   ...   \\

\sidehead{ Cycle 7a $|B_{\rm eff}| = 10.01^\circ - 10.29^\circ$}\\

 0.2997 &  0.2904 &  0.2998 &  0.4820 &  0.2998 &  0.6062 &  0.2998 &  0.6760 &  0.2999 &  0.7053 \\
 0.3000 &  0.2932 &  0.3000 &  0.4845 &  0.3000 &  0.6087 &  0.3000 &  0.6795 &  0.3001 &  0.7071 \\
 0.4948 &  0.2735 &  0.4946 &  0.4597 &  0.4944 &  0.5813 &  0.4943 &  0.6507 &  0.4941 &  0.6734 \\
 0.4958 &  0.2774 &  0.4956 &  0.4650 &  0.4954 &  0.5794 &  0.4952 &  0.6590 &  0.4951 &  0.6701 \\
 0.9774 &  0.2497 &  0.4957 &  0.4652 &  0.9770 &  0.5447 &  0.9767 &  0.6118 &  0.9765 &  0.6351 \\
 0.9843 &  0.2523 &  0.9772 &  0.4263 &  0.9837 &  0.5474 &  0.9834 &  0.6261 &  0.9832 &  0.6359 \\
 1.9911 &  0.2249 &  0.9839 &  0.4331 &  1.9906 &  0.5028 &  1.9904 &  0.5775 &  1.9901 &  0.5874 \\
 1.9979 &  0.2246 &  0.9840 &  0.4326 &  1.9975 &  0.5059 &  1.9972 &  0.5821 &  1.9970 &  0.5908 \\
 6.0165 &  0.1768 &  1.9908 &  0.3935 &  6.0165 &  0.4227 &  6.0165 &  0.4789 &  6.0165 &  0.4988 \\
 6.0168 &  0.1785 &  1.9977 &  0.3965 &  6.0168 &  0.4221 &  6.0168 &  0.4881 &  6.0169 &  0.4963 \\
  ...   &   ...   &  6.0165 &  0.3228 &   ...   &   ...   &   ...   &   ...   &   ...   &   ...   \\
  ...   &   ...   &  6.0165 &  0.3223 &   ...   &   ...   &   ...   &   ...   &   ...   &   ...   \\
  ...   &   ...   &  6.0165 &  0.3228 &   ...   &   ...   &   ...   &   ...   &   ...   &   ...   \\
  ...   &   ...   &  6.0165 &  0.3218 &   ...   &   ...   &   ...   &   ...   &   ...   &   ...   \\
  ...   &   ...   &  6.0168 &  0.3255 &   ...   &   ...   &   ...   &   ...   &   ...   &   ...   \\

\sidehead{ Cycle 7b $|B_{\rm eff}| = 15.40^\circ - 15.49^\circ$}\\
 0.3156 &  0.2935 &  0.3156 &  0.4891 &  0.3158 &  0.6099 &  0.3158 &  0.6928 &  0.3159 &  0.7053 \\
 0.3160 &  0.2947 &  0.3157 &  0.4890 &  0.3162 &  0.6153 &  0.3163 &  0.6863 &  0.3164 &  0.7118 \\
 0.6866 &  0.2661 &  0.3161 &  0.4895 &  0.6861 &  0.5673 &  0.6859 &  0.6475 &  0.6857 &  0.6582 \\
 0.6927 &  0.2672 &  0.6864 &  0.4508 &  0.6923 &  0.5756 &  0.6921 &  0.6427 &  0.6918 &  0.6672 \\
 1.1982 &  0.2469 &  0.6925 &  0.4545 &  1.1978 &  0.5495 &  1.1975 &  0.6175 &  1.1973 &  0.6417 \\
 1.2042 &  0.2426 &  1.1980 &  0.4288 &  1.2034 &  0.5440 &  1.2031 &  0.6129 &  1.2029 &  0.6368 \\
 6.2567 &  0.1867 &  1.2036 &  0.4223 &  6.2567 &  0.4405 &  6.2566 &  0.5088 &  6.2566 &  0.5189 \\
 6.2568 &  0.1862 &  1.2038 &  0.4217 &  6.2567 &  0.4441 &  6.2567 &  0.5139 &  6.2567 &  0.5231 \\
  ...   &   ...   &  1.2039 &  0.4216 &   ...   &   ...   &   ...   &   ...   &   ...   &   ...   \\
  ...   &   ...   &  6.2567 &  0.3381 &   ...   &   ...   &   ...   &   ...   &   ...   &   ...   \\
  ...   &   ...   &  6.2567 &  0.3419 &   ...   &   ...   &   ...   &   ...   &   ...   &   ...   \\

\sidehead{ Cycle 8 $|B_{\rm eff}| = 20.05^\circ - 20.17^\circ$}\\
 0.2956 &  0.2971 &  0.2957 &  0.4951 &  0.2958 &  0.6180 &  0.2959 &  0.7049 &  0.2959 &  0.7174 \\
 0.2971 &  0.2968 &  0.2957 &  0.4940 &  0.2958 &  0.6241 &  0.2973 &  0.6932 &  0.2973 &  0.7214 \\
 0.4244 &  0.2814 &  0.2957 &  0.4951 &  0.2958 &  0.6183 &  0.4233 &  0.6774 &  0.4231 &  0.6886 \\
 0.4257 &  0.2818 &  0.2972 &  0.4924 &  0.2972 &  0.6201 &  0.4251 &  0.6785 &  0.4249 &  0.6899 \\
 6.1038 &  0.1890 &  0.4240 &  0.4731 &  0.4235 &  0.5935 &  6.1037 &  0.5202 &  6.1036 &  0.5417 \\
 6.1052 &  0.1899 &  0.4241 &  0.4725 &  0.4236 &  0.5995 &  6.1049 &  0.5300 &  6.1049 &  0.5402 \\
  ...   &   ...   &  0.4242 &  0.4727 &  0.4238 &  0.5940 &   ...   &   ...   &   ...   &   ...   \\
  ...   &   ...   &  0.4255 &  0.4752 &  0.4253 &  0.5957 &   ...   &   ...   &   ...   &   ...   \\
  ...   &   ...   &  6.1038 &  0.3459 &  6.1037 &  0.4568 &   ...   &   ...   &   ...   &   ...   \\
  ...   &   ...   &  6.1051 &  0.3496 &  6.1050 &  0.4561 &   ...   &   ...   &   ...   &   ...   \\
  ...   &   ...   &  6.1051 &  0.3493 &  6.1050 &  0.4606 &   ...   &   ...   &   ...   &   ...   \\
  ...   &   ...   &  6.1052 &  0.3494 &  6.1050 &  0.4563 &   ...   &   ...   &   ...   &   ...   \\

\sidehead{ Cycle 9 $|B_{\rm eff}| = 23.54^\circ - 23.69^\circ$}\\

 0.2681 &  0.2992 &  0.2682 &  0.4980 &  0.2683 &  0.6286 &  0.2684 &  0.7115 &  0.2685 &  0.7241 \\
 0.2702 &  0.2991 &  0.2682 &  0.4974 &  0.2684 &  0.6223 &  0.2705 &  0.7090 &  0.2706 &  0.7229 \\
 0.5852 &  0.2706 &  0.2703 &  0.4970 &  0.2704 &  0.6261 &  0.5862 &  0.6644 &  0.5864 &  0.6751 \\
 0.5870 &  0.2702 &  0.2703 &  0.4961 &  0.2704 &  0.6212 &  0.5880 &  0.6665 &  0.5882 &  0.6779 \\
 1.9914 &  0.2286 &  0.5854 &  0.4588 &  0.5858 &  0.5847 &  1.9925 &  0.6038 &  1.9929 &  0.6219 \\
 1.9915 &  0.2288 &  0.5856 &  0.4583 &  0.5860 &  0.5781 &  1.9927 &  0.5970 &  6.0951 &  0.5538 \\
 6.0947 &  0.1929 &  0.5873 &  0.4590 &  0.5876 &  0.5868 &  6.0950 &  0.5318 &  6.0954 &  0.5438 \\
 6.0952 &  0.1928 &  0.5874 &  0.4591 &  0.5878 &  0.5804 &  6.0954 &  0.5341 &   ...   &   ...   \\
  ...   &   ...   &  1.9918 &  0.4068 &  1.9922 &  0.5277 &   ...   &   ...   &   ...   &   ...   \\
  ...   &   ...   &  1.9919 &  0.4061 &  1.9923 &  0.5229 &   ...   &   ...   &   ...   &   ...   \\
  ...   &   ...   &  6.0948 &  0.3531 &  6.0949 &  0.4669 &   ...   &   ...   &   ...   &   ...   \\
  ...   &   ...   &  6.0948 &  0.3538 &  6.0949 &  0.4622 &   ...   &   ...   &   ...   &   ...   \\
  ...   &   ...   &  6.0948 &  0.3531 &  6.0950 &  0.4664 &   ...   &   ...   &   ...   &   ...   \\
  ...   &   ...   &  6.0953 &  0.3493 &  6.0953 &  0.4578 &   ...   &   ...   &   ...   &   ...   \\
  
\enddata
\label{table:bring}
\end{deluxetable}
\clearpage

%%%%%%%%%%%%%%%%%%%%%%%%%%%%%%%%%%%%%%%%%%%%%%%%%%%%%%%%%%%%%%%%%%%%%%%%%%%%%%%%%%%%%%%%%%%%%%%%%%%%%%%%%
%-----------------

\begin{deluxetable}{rlrlrlrlrl}

\tablewidth{-15pt}
\tabletypesize{\scriptsize}
\tablecaption{A Ring: 127,000-129,000 km}
\tablehead{
{\hfill F336W} & {}& {\hfill  F439W} & {}& {\hfill  F555W} & {}& {\hfill F675W} & {}& {\hfill F814W} & {} \\ \tableline\\
\colhead{$\alpha$} & \colhead{$I/F_{\rm corr}$} & 
\colhead{$\alpha$} & \colhead{$I/F_{\rm corr}$} & 
\colhead{$\alpha$} & \colhead{$I/F_{\rm corr}$} & 
\colhead{$\alpha$} & \colhead{$I/F_{\rm corr}$} & 
\colhead{$\alpha$} & \colhead{$I/F_{\rm corr}$}}
\startdata 

%\cutinhead{\hskip0.4cm F439W \hskip1.6cm            F439W   \hskip1.6cm           F439W      \hskip1.6cm      F439W     \hskip1.6cm     F439W }

\sidehead{ Cycle 6 $|B_{\rm eff}| = 4.30^\circ - 4.66^\circ$}\\

 0.4605 &  0.2768 &  0.4607 &  0.4358 &  0.4609 &  0.5275 &  0.4611 &  0.5746 &  0.4613 &  0.6017 \\
 1.9249 &  0.2126 &  0.4617 &  0.4409 &  0.4619 &  0.5300 &  0.4621 &  0.5769 &  1.9257 &  0.4918 \\
 1.9315 &  0.2100 &  1.9251 &  0.3497 &  1.9253 &  0.4247 &  1.9255 &  0.4806 &  5.6708 &  0.3935 \\
 5.6705 &  0.1600 &  1.9318 &  0.3512 &  1.9320 &  0.4326 &  5.6709 &  0.3755 &   ...   &   ...   \\
 5.6711 &  0.1609 &  5.6704 &  0.2772 &  5.6704 &  0.3396 &   ...   &   ...   &   ...   &   ...   \\
  ...   &   ...   &  5.6710 &  0.2733 &  5.6709 &  0.3405 &   ...   &   ...   &   ...   &   ...   \\

\sidehead{ Cycle 7a $|B_{\rm eff}| = 10.01^\circ - 10.29^\circ$}\\

 0.2997 &  0.2594 &  0.2998 &  0.4071 &  0.2998 &  0.4943 &  0.2998 &  0.5417 &  0.2999 &  0.5658 \\
 0.3000 &  0.2592 &  0.3000 &  0.4057 &  0.3000 &  0.4891 &  0.3000 &  0.5375 &  0.3001 &  0.5601 \\
 0.4948 &  0.2436 &  0.4946 &  0.3832 &  0.4944 &  0.4652 &  0.4943 &  0.5116 &  0.4941 &  0.5339 \\
 0.4958 &  0.2446 &  0.4956 &  0.3883 &  0.4954 &  0.4676 &  0.4952 &  0.5218 &  0.4951 &  0.5301 \\
 0.9774 &  0.2202 &  0.4957 &  0.3883 &  0.9770 &  0.4366 &  0.9767 &  0.4796 &  0.9765 &  0.4993 \\
 0.9843 &  0.2247 &  0.9772 &  0.3544 &  0.9837 &  0.4375 &  0.9834 &  0.4886 &  0.9832 &  0.4998 \\
 1.9911 &  0.1976 &  0.9839 &  0.3603 &  1.9906 &  0.3993 &  1.9904 &  0.4490 &  1.9901 &  0.4596 \\
 1.9979 &  0.1994 &  0.9840 &  0.3600 &  1.9975 &  0.4033 &  1.9972 &  0.4533 &  1.9970 &  0.4636 \\
 6.0165 &  0.1516 &  1.9908 &  0.3239 &  6.0165 &  0.3278 &  6.0165 &  0.3622 &  6.0165 &  0.3789 \\
 6.0168 &  0.1530 &  1.9977 &  0.3303 &  6.0168 &  0.3271 &  6.0168 &  0.3705 &  6.0169 &  0.3790 \\
  ...   &   ...   &  6.0165 &  0.2590 &   ...   &   ...   &   ...   &   ...   &   ...   &   ...   \\
  ...   &   ...   &  6.0165 &  0.2590 &   ...   &   ...   &   ...   &   ...   &   ...   &   ...   \\
  ...   &   ...   &  6.0165 &  0.2594 &   ...   &   ...   &   ...   &   ...   &   ...   &   ...   \\
  ...   &   ...   &  6.0165 &  0.2582 &   ...   &   ...   &   ...   &   ...   &   ...   &   ...   \\
  ...   &   ...   &  6.0168 &  0.2631 &   ...   &   ...   &   ...   &   ...   &   ...   &   ...   \\

\sidehead{ Cycle 7b $|B_{\rm eff}| = 15.40^\circ - 15.49^\circ$}\\
 
 0.3156 &  0.2369 &  0.3156 &  0.3697 &  0.3158 &  0.4442 &  0.3158 &  0.4940 &  0.3159 &  0.5037 \\
 0.3160 &  0.2382 &  0.3157 &  0.3702 &  0.3162 &  0.4486 &  0.3163 &  0.4898 &  0.3164 &  0.5105 \\
 0.6866 &  0.2129 &  0.3161 &  0.3711 &  0.6861 &  0.4081 &  0.6859 &  0.4540 &  0.6857 &  0.4638 \\
 0.6927 &  0.2146 &  0.6864 &  0.3371 &  0.6923 &  0.4175 &  0.6921 &  0.4554 &  0.6918 &  0.4709 \\
 1.1982 &  0.1965 &  0.6925 &  0.3435 &  1.1978 &  0.3942 &  1.1975 &  0.4335 &  1.1973 &  0.4513 \\
 1.2042 &  0.1944 &  1.1980 &  0.3214 &  1.2034 &  0.3897 &  1.2031 &  0.4296 &  1.2029 &  0.4469 \\
 6.2567 &  0.1470 &  1.2036 &  0.3154 &  6.2567 &  0.3128 &  6.2566 &  0.3538 &  6.2566 &  0.3620 \\
 6.2568 &  0.1497 &  1.2038 &  0.3158 &  6.2567 &  0.3176 &  6.2567 &  0.3602 &  6.2567 &  0.3669 \\
  ...   &   ...   &  1.2039 &  0.3158 &   ...   &   ...   &   ...   &   ...   &   ...   &   ...   \\
  ...   &   ...   &  6.2567 &  0.2497 &   ...   &   ...   &   ...   &   ...   &   ...   &   ...   \\
  ...   &   ...   &  6.2567 &  0.2550 &   ...   &   ...   &   ...   &   ...   &   ...   &   ...   \\

\sidehead{ Cycle 8 $|B_{\rm eff}| = 20.05^\circ - 20.17^\circ$}\\
  0.2956 &  0.2253 &  0.2957 &  0.3505 &  0.2958 &  0.4209 &  0.2959 &  0.4713 &  0.2959 &  0.4798 \\
 0.2971 &  0.2233 &  0.2957 &  0.3506 &  0.2958 &  0.4249 &  0.2973 &  0.4600 &  0.2973 &  0.4795 \\
 0.4244 &  0.2125 &  0.2957 &  0.3504 &  0.2958 &  0.4213 &  0.4233 &  0.4481 &  0.4231 &  0.4558 \\
 0.4257 &  0.2132 &  0.2972 &  0.3491 &  0.2972 &  0.4203 &  0.4251 &  0.4533 &  0.4249 &  0.4598 \\
 6.1038 &  0.1402 &  0.4240 &  0.3335 &  0.4235 &  0.4015 &  6.1037 &  0.3382 &  6.1036 &  0.3518 \\
 6.1052 &  0.1437 &  0.4241 &  0.3333 &  0.4236 &  0.4043 &  6.1049 &  0.3480 &  6.1049 &  0.3567 \\
  ...   &   ...   &  0.4242 &  0.3331 &  0.4238 &  0.4014 &   ...   &   ...   &   ...   &   ...   \\
  ...   &   ...   &  0.4255 &  0.3375 &  0.4253 &  0.4054 &   ...   &   ...   &   ...   &   ...   \\
  ...   &   ...   &  6.1038 &  0.2409 &  6.1037 &  0.3041 &   ...   &   ...   &   ...   &   ...   \\
  ...   &   ...   &  6.1051 &  0.2466 &  6.1050 &  0.3068 &   ...   &   ...   &   ...   &   ...   \\
  ...   &   ...   &  6.1051 &  0.2455 &  6.1050 &  0.3097 &   ...   &   ...   &   ...   &   ...   \\
  ...   &   ...   &  6.1052 &  0.2461 &  6.1050 &  0.3069 &   ...   &   ...   &   ...   &   ...   \\

\sidehead{ Cycle 9 $|B_{\rm eff}| = 23.54^\circ - 23.69^\circ$}\\

  0.2681 &  0.2199 &  0.2682 &  0.3414 &  0.2683 &  0.4118 &  0.2684 &  0.4572 &  0.2685 &  0.4671 \\
 0.2702 &  0.2200 &  0.2682 &  0.3412 &  0.2684 &  0.4083 &  0.2705 &  0.4571 &  0.2706 &  0.4669 \\
 0.5852 &  0.1976 &  0.2703 &  0.3424 &  0.2704 &  0.4130 &  0.5862 &  0.4241 &  0.5864 &  0.4318 \\
 0.5870 &  0.1950 &  0.2703 &  0.3418 &  0.2704 &  0.4099 &  0.5880 &  0.4189 &  0.5882 &  0.4285 \\
 1.9914 &  0.1665 &  0.5854 &  0.3129 &  0.5858 &  0.3827 &  1.9925 &  0.3788 &  1.9929 &  0.3908 \\
 1.9915 &  0.1667 &  0.5856 &  0.3125 &  0.5860 &  0.3772 &  1.9927 &  0.3743 &  6.0951 &  0.3473 \\
 6.0947 &  0.1381 &  0.5873 &  0.3107 &  0.5876 &  0.3786 &  6.0950 &  0.3329 &  6.0954 &  0.3394 \\
 6.0952 &  0.1378 &  0.5874 &  0.3098 &  0.5878 &  0.3743 &  6.0954 &  0.3323 &   ...   &   ...   \\
  ...   &   ...   &  1.9918 &  0.2732 &  1.9922 &  0.3395 &   ...   &   ...   &   ...   &   ...   \\
  ...   &   ...   &  1.9919 &  0.2730 &  1.9923 &  0.3360 &   ...   &   ...   &   ...   &   ...   \\
  ...   &   ...   &  6.0948 &  0.2372 &  6.0949 &  0.2992 &   ...   &   ...   &   ...   &   ...   \\
  ...   &   ...   &  6.0948 &  0.2375 &  6.0949 &  0.2966 &   ...   &   ...   &   ...   &   ...   \\
  ...   &   ...   &  6.0948 &  0.2369 &  6.0950 &  0.2989 &   ...   &   ...   &   ...   &   ...   \\
  ...   &   ...   &  6.0953 &  0.2334 &  6.0953 &  0.2924 &   ...   &   ...   &   ...   &   ...   \\

\enddata
\label{table:aring}
\end{deluxetable}

\clearpage

%%%%%%%%%%%%%%%%%%%%%%%%%%%%%%%%%%%%%%%%%%%%%%%%%%%%%%%%%%%%%%%%%%%%%%%%%%%%%%%%%%%%%%%%%%%%%%%%%%%%%%%%%
%-----------------
%log-fits

\begin{deluxetable}{rrlrlrlrlrl}
\tablewidth{-15pt}
\tabletypesize{\scriptsize}
\tablecaption{Log-linear fits to normalized HST phase curves\tablenotemark{1}
}
\tablehead{
& \hspace{1cm}{\hfill F336W} & {}& {\hfill  F439W} & {}& {\hfill  F555W} & {}& {\hfill F675W} & {}& {\hfill F814W} & {} \\ \tableline\\
\colhead{$B_{\rm eff}$} &
\colhead{\hspace{1cm}$a$} & \colhead{$b$\hspace{.1cm}} & 
\colhead{ $a$} & \colhead{$b$\hspace{.1cm}  } & 
\colhead{$a$} & \colhead{$b$\hspace{.1cm} } & 
\colhead{ $a$} & \colhead{$b$\hspace{.1cm}  } & 
\colhead{ $a$} & \colhead{$b$\hspace{.1cm} }}
\startdata 
 \sidehead{C Ring 78,000 - 83.000 km}
  26.1 &-0.0048 & 0.0377 &-0.0060 & 0.0491 &-0.0061 & 0.0521 &-0.0063 & 0.0561 &-0.0067 & 0.0606  \\
  23.6 &-0.0055 & 0.0412 &-0.0067 & 0.0529 &-0.0068 & 0.0559 &-0.0071 & 0.0612 &-0.0071 & 0.0650  \\
  20.1 &-0.0066 & 0.0461 &-0.0079 & 0.0589 &-0.0082 & 0.0627 &-0.0086 & 0.0685 &-0.0086 & 0.0725  \\
  15.4 &-0.0082 & 0.0557 &-0.0097 & 0.0707 &-0.0100 & 0.0759 &-0.0105 & 0.0829 &-0.0112 & 0.0878  \\
  10.2 &-0.0110 & 0.0736 &-0.0127 & 0.0932 &-0.0139 & 0.1015 &-0.0153 & 0.1115 &-0.0159 & 0.1164  \\
   4.5 &-0.0182 & 0.1088 &-0.0217 & 0.1348 &-0.0232 & 0.1503 &-0.0248 & 0.1657 &-0.0257 & 0.1751  \\
 \sidehead{B Ring 100,000 - 107,000 km}
  26.1 &-0.0321 & 0.2488 &-0.0451 & 0.4337 &-0.0501 & 0.5594 &-0.0527 & 0.6317 &-0.0550 & 0.6564  \\
  23.6 &-0.0340 & 0.2533 &-0.0460 & 0.4359 &-0.0513 & 0.5568 &-0.0564 & 0.6365 &-0.0551 & 0.6505  \\
  20.1 &-0.0352 & 0.2529 &-0.0476 & 0.4345 &-0.0530 & 0.5531 &-0.0574 & 0.6290 &-0.0577 & 0.6448  \\
  15.4 &-0.0361 & 0.2524 &-0.0501 & 0.4325 &-0.0573 & 0.5503 &-0.0599 & 0.6227 &-0.0630 & 0.6406  \\
  10.2 &-0.0380 & 0.2484 &-0.0548 & 0.4247 &-0.0611 & 0.5389 &-0.0640 & 0.6101 &-0.0684 & 0.6276  \\
   4.5 &-0.0424 & 0.2299 &-0.0624 & 0.3917 &-0.0741 & 0.4993 &-0.0811 & 0.5568 &-0.0850 & 0.5784  \\
  \sidehead{A Ring 127,000 - 129,000 km}
  26.1 &-0.0243 & 0.1781 &-0.0325 & 0.2885 &-0.0356 & 0.3562 &-0.0363 & 0.3921 &-0.0379 & 0.4076  \\
  23.6 &-0.0260 & 0.1844 &-0.0334 & 0.2961 &-0.0361 & 0.3616 &-0.0394 & 0.4034 &-0.0388 & 0.4137  \\
  20.1 &-0.0270 & 0.1907 &-0.0344 & 0.3067 &-0.0373 & 0.3740 &-0.0405 & 0.4162 &-0.0405 & 0.4270  \\
  15.4 &-0.0299 & 0.2024 &-0.0396 & 0.3248 &-0.0440 & 0.3970 &-0.0450 & 0.4393 &-0.0474 & 0.4528  \\
  10.2 &-0.0355 & 0.2193 &-0.0499 & 0.3526 &-0.0539 & 0.4308 &-0.0568 & 0.4779 &-0.0600 & 0.4937  \\
   4.5 &-0.0464 & 0.2413 &-0.0648 & 0.3896 &-0.0749 & 0.4729 &-0.0779 & 0.5183 &-0.0827 & 0.5402  \\
 \enddata
\tablenotetext{1}{$a$ and $b$ are the fit parameters in $\frac{I(\alpha)}{F} = a \ln \alpha + b$ (Eq. \ref{eq:log}; phase angle $\alpha$ expressed in degrees, natural logarithm).}
\label{table:logfit_ab}
\end{deluxetable}

%-----------------
%%%%%%%%%%%%%%%%%%%%%%%%%%%%%%%%%%%%%%%%%%%%%%%%%%%%%%%%%%%%%%%%%%%%%%%%%%%%%%%%%%%%%%%%%%%%%%%%%%%%%%%%%
%-----------------
%log-fits: OE, I6

\begin{deluxetable}{rrlrlrlrlrl}

\tablewidth{-15pt}
\tabletypesize{\scriptsize}
\tablecaption{Log-linear fits to HST phase curves: the derived parameters\tablenotemark{1}}
\tablehead{
& {\hfill F336W} & {}& {\hfill  F439W} & {}& {\hfill  F555W} & {}& {\hfill F675W} & {}& {\hfill F814W} & {} \\ \tableline\\
\colhead{$B_{\rm eff}$} &
\colhead{\hspace{1cm}$OE$} & \colhead{$I/F~ (6^\circ)$} & 
\colhead{$OE$} & \colhead{$I/F ~(6^\circ)$} & 
\colhead{$OE$} & \colhead{$I/F~ (6^\circ)$} & 
\colhead{$OE$} & \colhead{$I/F~ (6^\circ)$} & 
\colhead{$OE$} & \colhead{$I/F ~(6^\circ)$}}
\startdata 
 \sidehead{C Ring 78,000 - 83,000 km}
 26.1 & 1.4125 & 0.0291 & 1.3869 & 0.0384 & 1.3696 & 0.0412 & 1.3467 & 0.0449 & 1.3436 & 0.0485   \\
  23.6 & 1.4386 & 0.0313 & 1.4070 & 0.0409 & 1.3862 & 0.0437 & 1.3640 & 0.0485 & 1.3359 & 0.0523  \\
  20.1 & 1.4741 & 0.0344 & 1.4390 & 0.0448 & 1.4238 & 0.0480 & 1.4018 & 0.0531 & 1.3732 & 0.0572  \\
  15.4 & 1.4945 & 0.0410 & 1.4495 & 0.0534 & 1.4298 & 0.0580 & 1.4084 & 0.0640 & 1.4092 & 0.0678  \\
  10.2 & 1.5104 & 0.0538 & 1.4465 & 0.0705 & 1.4534 & 0.0765 & 1.4541 & 0.0840 & 1.4511 & 0.0878  \\
   4.5 & 1.5932 & 0.0762 & 1.5602 & 0.0960 & 1.5296 & 0.1088 & 1.5078 & 0.1213 & 1.4940 & 0.1291  \\
 \sidehead{B Ring 100,000 - 107,000 km}
  26.1 & 1.4164 & 0.1913 & 1.3177 & 0.3528 & 1.2652 & 0.4696 & 1.2436 & 0.5373 & 1.2449 & 0.5579  \\
  23.6 & 1.4393 & 0.1924 & 1.3238 & 0.3534 & 1.2739 & 0.4650 & 1.2619 & 0.5354 & 1.2479 & 0.5519  \\
  20.1 & 1.4606 & 0.1899 & 1.3389 & 0.3492 & 1.2877 & 0.4580 & 1.2713 & 0.5261 & 1.2650 & 0.5414  \\
  15.4 & 1.4785 & 0.1877 & 1.3628 & 0.3428 & 1.3178 & 0.4477 & 1.2886 & 0.5154 & 1.2964 & 0.5278  \\
  10.2 & 1.5245 & 0.1802 & 1.4175 & 0.3264 & 1.3537 & 0.4294 & 1.3211 & 0.4954 & 1.3365 & 0.5051  \\
   4.5 & 1.6849 & 0.1539 & 1.5543 & 0.2799 & 1.5024 & 0.3665 & 1.4900 & 0.4114 & 1.4958 & 0.4261  \\
 \sidehead{A Ring 127,000 - 129,000 km}
  26.1 & 1.4494 & 0.1345 & 1.3504 & 0.2303 & 1.3025 & 0.2924 & 1.2760 & 0.3271 & 1.2775 & 0.3397  \\
  23.6 & 1.4680 & 0.1379 & 1.3515 & 0.2362 & 1.3018 & 0.2970 & 1.2946 & 0.3327 & 1.2800 & 0.3442  \\
  20.1 & 1.4711 & 0.1423 & 1.3483 & 0.2452 & 1.3016 & 0.3072 & 1.2925 & 0.3438 & 1.2836 & 0.3545  \\
  15.4 & 1.4980 & 0.1490 & 1.3874 & 0.2539 & 1.3433 & 0.3182 & 1.3115 & 0.3587 & 1.3206 & 0.3678  \\
  10.2 & 1.5663 & 0.1557 & 1.4712 & 0.2632 & 1.4010 & 0.3342 & 1.3750 & 0.3762 & 1.3860 & 0.3862  \\
   4.5 & 1.7297 & 0.1581 & 1.5885 & 0.2735 & 1.5498 & 0.3387 & 1.5114 & 0.3787 & 1.5240 & 0.3921  \\
\label{table:logfit_oe6}
\enddata
 \tablenotetext{1}{$OE \equiv I(\alpha=0.5^\circ)/I(\alpha=6.0^\circ)$}
\end{deluxetable}

%-----------------
\clearpage

%%%%%%%%%%%%%%%%%%%%%%%%%%%%%%%%%%%%%%%%%%%%%%%%%%%%%%%%%%%%%%%%%%%%%%%%%%%%%%%%%%%%%%%%%%%%%%%%%%%%%%%%%
\begin{deluxetable}{rlrlrlrrrl}

\tablewidth{-15pt}
\tabletypesize{\scriptsize}
\tablecaption{CB-SH (simplified Hapke) model parameters for the intrinsic opposition effect\tablenotemark{1}}
\tablehead{
\colhead{Ring} & \colhead{Filter \,\,\,\,\,} & \colhead{$A_i$}& \colhead{$B_{C0}$} & \colhead{$B_{S0}$} &\colhead{$h_{C}$} & \colhead{$h_{S}$} & \colhead{$SH(0)$} & \colhead{HWHM} &\colhead{RMS}}
\startdata
C ...... &   F336W & 0.725 & 0.516 & 0.781 &  0.0036 & 0.052 & 1.290 & 0.147 & 0.0183 \\ 
         &   F439W & 0.837 & 0.447 & 0.559 &  0.0030 & 0.029 & 1.305 & 0.125 & 0.0163 \\ 
         &   F555W & 0.841 & 0.452 & 0.526 &  0.0030 & 0.032 & 1.284 & 0.125 & 0.0193 \\ 
         &   F675W & 0.844 & 0.489 & 0.477 &  0.0032 & 0.036 & 1.247 & 0.131 & 0.0135 \\ 
         &   F814W & 0.852 & 0.467 & 0.469 &  0.0030 & 0.034 & 1.251 & 0.122 & 0.0155 \\ 
         &         &       &       &       &         &       &       &       &        \\       
 B ......&   F336W & 0.930 & 0.467 & 0.201 &  0.0039 & 0.031 & 1.116 & 0.162 & 0.0154 \\ 
         &   F439W & 0.975 & 0.426 & 0.084 &  0.0028 & 0.019 & 1.057 & 0.114 & 0.0147 \\ 
         &   F555W & 0.991 & 0.395 & 0.047 &  0.0022 & 0.006 & 1.037 & 0.091 & 0.0148 \\ 
         &   F675W & 0.993 & 0.431 & 0.000 &  0.0024 &-0.000 & 0.993 & 0.100 & 0.0164 \\ 
         &   F814W & 0.996 & 0.416 & 0.046 &  0.0024 & 0.000 & 1.041 & 0.099 & 0.0136 \\ 
         &         &       &       &       &         &       &       &       &        \\       
 A ......&   F336W & 0.693 & 0.559 & 0.578 &  0.0036 & 0.165 & 1.093 & 0.147 & 0.0218 \\ 
         &   F439W & 0.535 & 0.500 & 0.940 &  0.0027  & 0.591 & 1.039 & 0.113 & 0.0183 \\ 
         &   F555W & 0.524 & 0.461 & 0.928 &  0.0025  & 1.920 & 1.009 & 0.104 & 0.0191 \\ 
         &   F675W & 0.518 & 0.480 & 0.921 &  0.0024  & 43.44 & 0.994 & 0.098 & 0.0213 \\ 
         &   F814W & 0.518 & 0.481 & 0.924 &  0.0023  & ----- & 0.997 & 0.093 & 0.0166* \\ 
\enddata
\tablenotetext{1}{Intrinsic effect normalized to $\alpha=6^\circ$ (=$g_i(\alpha)$ defined by 
Eq.~(22)).
The $A_i,B_{C0}, B_{S0},h_{C}, h_{S}$ are the original parameters in the fits, while $SH(0) = A_i (1+B_{S0})$ and HWHM$=0.72 h_c$ expressed in degrees.}
\label{table:intrinsic_hapke}
\end{deluxetable}

\clearpage
%%%%%%%%%%%%%%%%%%%%%%%%%%%%%%%%%%%%%%%%%%%%%%%%%%%%%%%%%%%%%%%%%%%%%%%%%%%%%%%%%%%%%%%%%%%%%%%%%%%%%%%%%
\begin{deluxetable}{rlrlrlrrrl}

\tablewidth{-15pt} \tabletypesize{\scriptsize}
\tablecaption{Linear-exponential model parameters for
  the intrinsic opposition effect\tablenotemark{1}}
\tablehead{
\colhead{Ring} & \colhead{Filter \,\,\,\,\,} & \colhead{$a'$}& \colhead{$b'$} & \colhead{$d'$} &\colhead{$k'$} & \colhead{$a'/b'$} & \colhead{$k'/b'$} & \colhead{HWHM} &\colhead{RMS}}
\startdata
 C ......      &   F336W & 0.632 & 1.248 & 0.301 & -2.3252 & 0.507 &-0.033 & 0.208 & 0.0195 \\ 
         &   F439W & 0.603 & 1.223 & 0.283 & -2.0944 & 0.493 &-0.030 & 0.196 & 0.0202 \\ 
         &   F555W & 0.590 & 1.210 & 0.285 & -1.9575 & 0.487 &-0.028 & 0.198 & 0.0211 \\ 
         &   F675W & 0.579 & 1.193 & 0.292 & -1.7778 & 0.485 &-0.026 & 0.202 & 0.0159 \\ 
         &   F814W & 0.551 & 1.192 & 0.288 & -1.7842 & 0.462 &-0.026 & 0.199 & 0.0190 \\ 
         &         &       &       &       &         &       &       &       &        \\       
B ......       &   F336W & 0.492 & 1.095 & 0.307 & -0.9147 & 0.449 &-0.015 & 0.213 & 0.0150 \\ 
         &   F439W & 0.423 & 1.050 & 0.212 & -0.5067 & 0.403 &-0.008 & 0.147 & 0.0152 \\ 
         &   F555W & 0.389 & 1.025 & 0.170 & -0.2875 & 0.379 &-0.005 & 0.118 & 0.0153 \\ 
         &   F675W & 0.367 & 1.001 & 0.194 & -0.0456 & 0.367 &-0.001 & 0.134 & 0.0153 \\ 
         &   F814W & 0.372 & 1.004 & 0.184 & -0.0684 & 0.370 &-0.001 & 0.128 & 0.0152 \\ 
         &         &       &       &       &         &       &       &       &        \\       
A ......       &   F336W & 0.544 & 1.113 & 0.249 & -1.1088 & 0.489 &-0.017 & 0.173 & 0.0209 \\ 
         &   F439W & 0.469 & 1.056 & 0.192 & -0.5569 & 0.444 &-0.009 & 0.133 & 0.0188 \\ 
         &   F555W & 0.420 & 1.027 & 0.174 & -0.2891 & 0.409 &-0.005 & 0.120 & 0.0192 \\ 
         &   F675W & 0.416 & 1.011 & 0.173 & -0.1587 & 0.411 &-0.003 & 0.120 & 0.0209 \\ 
         &   F814W & 0.416 & 1.014 & 0.164 & -0.1740 & 0.411 &-0.003 & 0.114 & 0.0168 \\ 
\enddata
\label{table:intrinsic_linexp}
 \tablenotetext{1}{Intrinsic effect normalized to $\alpha=6^\circ$ (=$g_i(\alpha)$ defined by 
 Eq.~(22)). The $a', b', d', k'$ are the original fit parameters, $a/b', k'/b'$ indicate the normalized amplitude of the exponential part, and the normalized linear slope, the HWHM$= d' \ln 2$ expressed in degrees.}
\end{deluxetable}
\clearpage

%%%%%%%%%%%%%%%%%%%%%%%%%%%%%%%%%%%%%%%%%%%%%%%%%%%%%%%%%%%%%%%%%%%%%%%%%%%%%%%%%%%%%%%%%%%%%%%%%%%%%%%%%
%---------------- oe_i oe_e

\begin{deluxetable}{lcccrrrr}

\tablewidth{-15pt} \tabletypesize{\scriptsize} \tablecaption{Modeled
  intrinsic and interparticle opposition effects at $B_{\rm
    eff}=23^\circ$\tablenotemark{1}}
\tablehead{
\colhead{Filter \,\,\,\,\,} & \colhead{$OE_i(0.5^\circ)$}& \colhead{$OE_e(0.5^\circ)$} & \colhead{$OE(0.5^\circ)$} &\colhead{$OE_i(0.0^\circ)$} &\colhead{$OE_e(0.0^\circ)$} & \colhead{$OE(0.0^\circ)$} & \colhead{$OE_{obs}(0^\circ)$}}
\startdata
\sidehead{C Ring 78,000 - 83,000 km, model: $\taud=0.1$, $0.1 -5.0$ m}
   F336W  &  1.35 &  1.05 &  1.42 &  1.86 &  1.07 &  1.99 &  2.02 \\
  F439W  &  1.32 &  1.05 &  1.39 &  1.79 &  1.07 &  1.92 &  1.96 \\
  F555W  &  1.30 &  1.05 &  1.37 &  1.76 &  1.07 &  1.89 &  1.94 \\
  F675W  &  1.28 &  1.05 &  1.35 &  1.76 &  1.07 &  1.88 &  1.91 \\
  F814W  &  1.28 &  1.05 &  1.34 &  1.73 &  1.07 &  1.86 &  1.88 \\
\sidehead{B Ring 100,000 - 107,000 km, model: $\taud=2$, $1 -5.0$ m}
  F336W  &  1.18 &  1.20 &  1.42 &  1.57 &  1.22 &  1.91 &  1.94 \\
  F439W  &  1.09 &  1.20 &  1.31 &  1.42 &  1.22 &  1.74 &  1.77 \\
  F555W  &  1.05 &  1.20 &  1.26 &  1.35 &  1.22 &  1.66 &  1.70 \\
  F675W  &  1.03 &  1.20 &  1.24 &  1.34 &  1.22 &  1.63 &  1.66 \\
  F814W  &  1.03 &  1.20 &  1.24 &  1.34 &  1.22 &  1.63 &  1.67 \\
\sidehead{A Ring 127,000 - 129,000 km, model: $\taud=1$, $1 -5.0$ m}
  F336W  &  1.18 &  1.21 &  1.43 &  1.62 &  1.24 &  2.00 &  2.02 \\
  F439W  &  1.09 &  1.21 &  1.32 &  1.46 &  1.24 &  1.81 &  1.85 \\
  F555W  &  1.05 &  1.21 &  1.27 &  1.38 &  1.24 &  1.71 &  1.75 \\
  F675W  &  1.04 &  1.21 &  1.25 &  1.37 &  1.24 &  1.70 &  1.73 \\
  F814W  &  1.04 &  1.21 &  1.25 &  1.37 &  1.24 &  1.70 &  1.73 \\
\enddata
 \tablenotetext{1}{The symbols $OE(0.5^\circ) \equiv
  I(0.5^\circ)/I(6^\circ)$, and $OE(0^\circ) \equiv
  I(0^\circ)/I(6^\circ)$. The total opposition effect is the product of the
  intrinsic and external (interparticle) contributions; $OE(0^\circ) =
  OE_i(0^\circ)  \times OE_e(0^\circ)$, $OE(0.5^\circ) = OE_i(0.5^\circ) \times OE_e(0.5^\circ)$.}
\label{table:intrinsic_external_23}
\end{deluxetable}

%%%%%%%%%%%%%%%%%%%%%%%%%%%%%%%%%%%%%%%%%%%%%%%%%%%%%%%%%%%%%%%%%%%%%%%%%%%%%%%%%%%%%%%%%%%%%%%%%%%%%%%%%
\begin{deluxetable}{lcccrrrr}

\tablewidth{-15pt} \tabletypesize{\scriptsize} \tablecaption{Modeled
  intrinsic and interparticle opposition effects at
  $B_{\rm eff}=4.5^\circ$\tablenotemark{1}}
\tablehead{
\colhead{Filter \,\,\,\,\,} & \colhead{$OE_i(0.5^\circ)$}& \colhead{$OE_e(0.5^\circ)$} & \colhead{$OE(0.5)$} &\colhead{$OE_i(0.0^\circ)$} &\colhead{$OE_e(0.0^\circ)$} & \colhead{$OE(0.0^\circ)$}}
\startdata
\sidehead{C Ring 78,000 - 83,000 km, model: $\taud=0.1$, $0.1 -5.0$ m}
  F336W  &  1.35 &  1.21 &  1.63 &  1.86 &  1.46 &  2.72  \\
  F439W  &  1.32 &  1.21 &  1.59 &  1.79 &  1.46 &  2.62  \\
  F555W  &  1.30 &  1.21 &  1.57 &  1.76 &  1.46 &  2.58  \\
  F675W  &  1.28 &  1.21 &  1.55 &  1.76 &  1.46 &  2.57  \\
  F814W  &  1.28 &  1.21 &  1.54 &  1.73 &  1.46 &  2.54  \\ 
\sidehead{B Ring 100,000 - 107,000 km, model: $\taud=2$, $1 -5.0$ m}
  F336W  &  1.18 &  1.45 &  1.72 &  1.57 &  1.58 &  2.48  \\
  F439W  &  1.09 &  1.45 &  1.58 &  1.42 &  1.58 &  2.25  \\
  F555W  &  1.05 &  1.45 &  1.52 &  1.35 &  1.58 &  2.14  \\
  F675W  &  1.03 &  1.45 &  1.50 &  1.34 &  1.58 &  2.11  \\
  F814W  &  1.03 &  1.45 &  1.50 &  1.34 &  1.58 &  2.12  \\
\sidehead{A Ring 127,000 - 129,000 km, model: $\taud=1$, $1 -5.0$ m}
  F336W  &  1.18 &  1.48 &  1.74 &  1.62 &  1.63 &  2.64  \\
  F439W  &  1.09 &  1.48 &  1.61 &  1.46 &  1.63 &  2.39  \\
  F555W  &  1.05 &  1.48 &  1.56 &  1.38 &  1.63 &  2.26  \\
  F675W  &  1.04 &  1.48 &  1.53 &  1.37 &  1.63 &  2.24  \\
  F814W  &  1.04 &  1.48 &  1.53 &  1.37 &  1.63 &  2.24  \\
\enddata
 \tablenotetext{1}{The symbols $OE(0.5^\circ) \equiv
  I(0.5^\circ)/I(6^\circ)$, and $OE(0^\circ) \equiv
  I(0^\circ)/I(6^\circ)$. The total opposition effect is the product of the
  intrinsic and external (interparticle) contributions; $OE(0^\circ) =
  OE_i(0^\circ) \times OE_e(0^\circ)$, $OE(0.5^\circ) = OE_i(0.5^\circ) \times OE_e(0.5^\circ)$.
 The last column lists the observed $I(0.5^\circ)/I(6^\circ)$.
}
\label{table:intrinsic_external_4}
\end{deluxetable}
%%%%%%%%%%%%%%%%%%%%%%%%%%%%%%%%%%%%%%%%%%%%%%%%%%%%%%%%%%%%%%%%%%%%%%%%%%%%%%%%%%%%%%%%%%%%%%%%%%%%%%%%%


\begin{thebibliography}{}
%
\bibitem[{Akkermans} et~al., 1988]{akkermans1988}
{Akkermans}, E., {Wolf}, P., {Maynard}, R., and {Maret}, G., 1988.
\newblock {Theoretical study of the coherent backscattering of light by
  disordered media}.
%\newblock { J.~Phys., vol.~49, p.~77-98 (1988).}, 49:77--98.
\newblock {J.~Phys., 49,77--98}.

\bibitem[Altobelli et~al., 2009]{altobelli2009} Altobelli, N.,
  Spilker, L., Leyrat, C., Pilorz, S., Edgington, S., Flanders,
  A., 2009.  \newblock {Thermal phase curves observed in Saturn's
    main rings by Cassini-CIRS: detection of an opposition effect?}.
  \newblock { Geophys. Res. Lett}, 36, L10105.


\bibitem[{Araki} and {Tremaine}, 1986]{araki1986}
{Araki}, S. and {Tremaine}, S., 1986.
\newblock {The dynamics of dense particle disks.}
\newblock { Icarus}, 65,83--109.


\bibitem[{Bobrov}, 1970]{bobrov1970}
{Bobrov}, M.~S., 1970.
\newblock { {The Rings of Saturn}}.
\newblock NASA TT F-701 (Translation of {Kol'tsa Saturna}, Nauka, Moscow).


\bibitem[{{Bridges} et~al.(1984){Bridges}, {Hatzes}, and
    {Lin}}]{bridges1984} {Bridges}, F.~G., {Hatzes}, A., {Lin},
  D. N.~C., 1984. {Structure, stability and evolution of Saturn's
    rings}. { Nature} 309, 333--335.

\bibitem[{Camichel}, 1958]{camichel1958}
{Camichel}, H, 1958.
\newblock {Mesures photom{\'e}triques de Saturne et de son anneau}.
\newblock { Annales d'Astrophysique}, 21, 231--242.

\bibitem[{{Colombo} et~al.(1976){Colombo}, {Goldreich}, and
  {Harris}}]{colombo1976}
{Colombo}, G., {Goldreich}, P., {Harris}, A.~W., 1976. {Spiral structure as an
  explanation for the asymmetric brightness of Saturn's A ring}. Nature 264,
  344--345.


\bibitem[{Colwell} et~al., 2006a]{colwell2006}
{Colwell}, J.~E., {Esposito}, L.~W., and {Srem{\v c}evi{\'c}}, M., 2006.
\newblock {Self-gravity wakes in Saturn's A ring measured by stellar
  occultations from Cassini}.
\newblock { Geophys. Res. Lett.}, 33, 7201.

\bibitem[{Colwell} et~al., 2007]{Colwell07b}
{Colwell}, J.~E., {Esposito}, L.~W., {Srem{\v c}evi{\'c}}, M., {Stewart},
  G.~R., and {McClintock}, W.~E., 2007.
\newblock {Self-gravity wakes and radial structure of Saturn's B ring}.
\newblock { Icarus}, 190, 127--144.


\bibitem[{Cook} et~al., 1973]{cook1973}
{Cook}, A.~F., {Franklin}, F.~A., and {Palluconi}, F.~D., 1973.
\newblock {Saturn's Rings - A survey}.
\newblock { Icarus}, 18, 317--337.

%\bibitem
%[Cuzzi(1984)]{cuzzi84}
%{{Cuzzi, J.~N., Lissauer, J.~J., Esposito, L.~W., Holberg, J.~B.,
%        Marouf, E.~A., Tyler, G.~L., Boishchot, A.} 1984.
%  {Saturn's rings -- Properties and processes.} In {\it Planetary
%  Rings} {(R.~Greenberg and A.~Brahic, Eds.) pp.~73--199. Univ. of 
%  Arizona Press, Tucson.}}


\bibitem[{{Cuzzi} et~al.(2002){Cuzzi}, {French}, and {Dones}}]{cuzzi2002}
{Cuzzi}, J.~N., {French}, R.~G., {Dones}, L., 2002. {HST multicolor (255 - 1042
  nm) photometry of Saturn's main rings. I. Radial profiles, phase and opening
  angle variations, and regional spectra}. Icarus 158, 199--223.


\bibitem[{D\'eau} et al.  (2009)]{deau2009} D\'eau, E., Charnoz, S,
  Dones, L, Brahic, A, Porco, C. {The opposition effect in Saturn's
    rings seen by Cassini/ISS: I. Morphology of phase
    curves}. Submitted to Icarus.

 \bibitem[{Dollfus}, 1996]{Dollfus1996}
{Dollfus}, A., 1996.
\newblock {Saturn's rings: Optical reflectance polarimetry}.
\newblock { Icarus}, 124, 237--261.


\bibitem[{{Dones} et~al.(1993){Dones}, {Cuzzi}, and {Showalter}}]{dones1993}
{Dones}, L., {Cuzzi}, J.~N., {Showalter}, M.~R., 1993. {Voyager photometry of
  Saturn's A ring}. Icarus 105, 184--215.

\bibitem[Drossart (1993)]{drossart1993}
Drossart, P., 1993. {Optics on a fractal surface and the photometry of the regoliths}. Planet. Space Sci. 41, 381-393.



 
%\bibitem
%[DoyleDonesCuzzi(1989)]{doyle1989}
%{Doyle, L. R., Dones, L., Cuzzi, J. N., 1989.}
%        {Radiative transfer modeling of Saturn's outer B ring}
%  	{Icarus} {80}, {104--135}.



\bibitem[{Esposito} and {Lumme}, 1977]{esposito1977}
{Esposito}, L.~W. and {Lumme}, K., 1977.
\newblock {The tilt effect for Saturn's rings}.
\newblock { Icarus}, 31, 157--167.

\bibitem[{Ferrari} and {Leyrat}, 2006]{ferrari2006} {Ferrari}, C.,
  {Leyrat}, C., 2006.  
\newblock {Thermal emission of spherical
    spinning ring particles. The standard model}.  
\newblock {
    Astronomy and Astrophysics} 447, 745--760.

\bibitem[{Ferrari} et~al., 2009]{ferrari2009}
{Ferrari}, C., {Brooks}, S., {Edgington}, S., {Leyrat}, C., {Pilorz}, S., and
  {Spilker}, L., 2009.
\newblock {Structure of self-gravity wakes in Saturn's A ring as measured by
  Cassini CIRS}.
\newblock { Icarus}, 199, 145--153.

\bibitem[{Franklin} et~al., 1987]{franklin1987}
{Franklin}, F.~A., {Cook}, A.~F., {Barrey}, R.~T.~F., {Roff}, C.~A., {Hunt},
  G.~E., and {de Rueda}, H.~B., 1987.
\newblock {Voyager observations of the azimuthal brightness variations in
  Saturn's rings}.
\newblock { Icarus}, 69, 280--296.

\bibitem[{{French} and {Nicholson}(2000)}]{french2000}
{French}, R.~G., {Nicholson}, P.~D., 2000. {Saturn's Rings II. Particle sizes
  inferred from stellar occultation data}. Icarus 145, 502--523.


%\bibitem[French et al.(2006)]{french2006} French, R.~G., McGhee, 
%C.~A., Frey, M., Hock, R., Rounds, S., Jacobson, R., \& Verbiscer, A., 2006,
%\newblock {Astrometry of Saturn's satellites from the Hubble Space Telescope WFPC2}
%\newblock \pasp, 118, 246--259.

%\bibitem[French et al.(2003)]{Fre03} French, R.~G., McGhee, 
%C.~A., Dones, L., \& Lissauer, J.~J., 2003, 
%\newblock {Saturn's wayward shepherds: the peregrinations of Prometheus and Pandora}
%\newblock Icarus, 162, 143

\bibitem[{French} et~al., 2007a]{french2007a}
{French}, R.~G., {Salo}, H., {McGhee}, C.~A., and {Dones}, L., 2007a.
\newblock {HST observations of azimuthal asymmetry in Saturn's rings}.
\newblock { Icarus}, 189, 493--522.

\bibitem[{French} et~al., 2007b]{french2007b}
{French}, R.~G., {Verbiscer}, A, {Salo}, H., {McGhee}, C.~A., and {Dones}, L., 2007b.
\newblock {Saturn's rings at true opposition}.
\newblock { \pasp}, 119, 623-6422.

\bibitem[{Hapke}, 1986]{hapke1986} {Hapke}, B.,  1986.  
{Bidirectional reflectance spectroscopy. IV - The extinction
coefficient and the opposition effect}.  {Icarus}, 67,
264--280.

\bibitem[Hapke(1990)]{hapke1990} {Hapke}, B., 1990. { Coherent
backscatter and the radar characteristics of outer planet satellites.}
Icarus 88, 407--417.

\bibitem[{Hapke}, 2002]{hapke2002} {Hapke}, B.,  2002.  
{Bidirectional reflectance spectroscopy. V - The coherent backscatter opposition effect and anisotropic scattering}.  {Icarus}, 157,
523-534.


\bibitem[{Hapke} et~al., 2006]{hapke2006} {Hapke}, B.~W., and 29
  colleagues, 2006.  \newblock {Cassini Observations of the Opposition
    Effect of Saturn's Rings 2.  Interpretation: Plaster of Paris as
    an Analog of Ring Particles}.  Lunar
  Planet. Inst. Conf. Abstr. 37, 1466-1467.



\bibitem[{Hapke} et~al., 2009]{hapke2009}
{Hapke}, B.~W., {Shepard}, M.~K., {Nelson}, R.~M., {Smythe}, W.~D., and
  {Piatek}, J.~L., 2009.
\newblock {A quantitative test of the ability of models based on the equation
  of radiative transfer to predict the bidirectional reflectance of a
  well-characterized medium}.
\newblock { Icarus}, 199, 210--218.



\bibitem[{Hedman} et~al., 2007]{Hedman07a}
{Hedman}, M.~M., {Nicholson}, P.~D., {Salo}, H., {Wallis}, B.~D., {Buratti},
  B.~J., {Baines}, K.~H., {Brown}, R.~H., and {Clark}, R.~N., 2007.
\newblock {Self-gravity wake structures in Saturn's A ring revealed by Cassini
  VIMS}.
\newblock { Astronomical Journal}, 133, 2624--2629.


\bibitem[{H\"ameen-Anttila} and {Pyykk\"o}, 1972]{hameen-anttila1972}
{H\"ameen-Anttila}, K.~A. and {Pyykko}, S., 1972.
\newblock {Photometric behaviour of Saturn's rings as a function of the
  Saturnocentric latitudes of the Earth and the Sun}.
\newblock { Astronomy and Astrophysics}, 19, 235--247.


\bibitem[{{H\"ameen-Anttila} and {Vaaraniemi}(1975)}]{hameen-anttila1975}
{H\"ameen-Anttila}, K.~A., {Vaaraniemi}, P., 1975. {A theoretical photometric
  function of Saturn's rings}. Icarus 25, 470--478.

\bibitem[{Irvine}, 1966]{irvine1966} {Irvine}, W.~M., 1966.  \newblock
{The shadowing effect in diffuse reflection}.  \newblock
{J. Geophys. Res.}, 71, 2931--2937.

%\bibitem[{{Julian} and {Toomre}(1966)}]{julian1966}
%{Julian}, W.~H., {Toomre}, A., 1966. {Non-axisymmetric responses of
%  differentially rotating disks of stars}. Astrophys. J. 146, 810--830.
%  

\bibitem[{Irvine} et~al., 1988]{irvine1988}
{Irvine}, W.~M., {Muinonen}, K., and {Lumme}, K., 1988.
\newblock {Is the mutual shadowing explanation for the opposition effect of
  Saturn's rings still valid?}
\newblock Bull. Am. Astron. Soc. 20, 853.


\bibitem[{Johnson} et~al., 1980]{johnson_pe1980}
{Johnson}, P.~E., {Kemp}, J.~C., {King}, R., {Parker}, T.~E., and {Barbour},
  M.~S. (1980).
\newblock {New results from optical polarimetry of Saturn's rings}.
\newblock { Science}, 283, 146--149.

\bibitem[Kaasalainen et al.(2003)]{2003Icar..161...34K} Kaasalainen, S., 
Piironen, J., Kaasalainen, M., Harris, A.~W., Muinonen, K., 
\& Cellino, A., 2003, Icarus, 161, 34--46

%OK
\bibitem[{{Lumme}(1970)}]{lumme1970}
{Lumme}, K., 1970. {On photometric properties of Saturn's
  rings}. {Astrophys. Space Sci.} 8, 90--101.

\bibitem[{Lumme} and {Bowell}, 1981]{lumme1981}
{Lumme}, K. and {Bowell}, E., 1981.
\newblock {Radiative transfer in the surfaces of atmosphereless bodies. I -
  Theory.}
\newblock { Astron. J}, 86, 1694--1704.


\bibitem[{Lumme} and {Irvine}, 1976a]{lumme1976a}
{Lumme}, K. and {Irvine}, W.~M., 1976a.
\newblock {Azimuthal brightness variations of Saturn's rings}.
\newblock { Astrophys. J}, 204, L55--L57.

%OK
\bibitem[{Lumme} and {Irvine}, 1976b]{lumme1976b}
{Lumme}, K., {Irvine}, W.~M., 1976b. 
\newblock {Photometry of Saturn's  rings}. Astron. J. 81, 865--893.




%OK
\bibitem[{{Lumme} et~al.(1983){Lumme}, {Irvine}, and {Esposito}}]{lumme1983}
{Lumme}, K., {Irvine}, W.~M., {Esposito}, L.~W., 1983. {Theoretical
  interpretation of the ground-based photometry of Saturn's B ring}. Icarus 53,
  174--184.

\bibitem[{Lyot}, 1929]{lyot1929}
{Lyot}, B., 1929.
\newblock {Recherce sur la polarization de la lumi\`ere des plan\`etes et de
  quelques substances terrestres}.
\newblock { Ann. Obs. Meudon VIII, Fasc 1. (NASA-TTF-187, 1964)}.


\bibitem[{{Marouf} et~al.(1983){Marouf}, {Tyler}, {Zebker}, {Simpson}, and
  {Eshleman}}]{marouf1983}
{Marouf}, E.~A., {Tyler}, G.~L., {Zebker}, H.~A., {Simpson}, R.~A., {Eshleman},
  V.~R., 1983. {Particle size distributions in Saturn's rings from Voyager 1
  radio occultation}. Icarus 54, 189--211.


\bibitem[McGhee et al.(2005)]{mcghee2005} McGhee, C.~A., French,
  R.~G., Dones,L., Cuzzi, J.~N.,Salo, H.~J.,\& Danos, R., 2005.
{HST observations of spokes in Saturn's B ring}.
  Icarus, 173, 508-521

\bibitem[{Mishchenko}, 1992]{mishchenko1992}
{Mishchenko}, M.~I., 1992.
\newblock {The angular width of the coherent back-scatter opposition effect -
  an application to icy outer planet satellites}.
\newblock { Aastrophys. Space Sci.}, 194, 327--333.


\bibitem[{Mishchenko}, 1993]{mishchenko1993}
{Mishchenko}, M.~I., 1993.
\newblock {On the nature of the polarization opposition effect exhibited by
  Saturn's rings}.
\newblock {Astrophys. J.}, 411, 351--361.

\bibitem[{Mishchenko} and {Dlugach}, 1992]{mishchenko1992b}
{Mishchenko}, M.~I. and {Dlugach}, Z.~M., 1992.
\newblock {Can weak localization of photons explain the opposition effect of
  Saturn's rings?}.
\newblock {MNRAS}, 254, 15P--18P.


\bibitem[Morishima et~al., 2009]{morishima2009}
Morishima, R., Salo, H., and Ohtsuki, K., 2009.
\newblock {A multi-layer model for thermal infrared emission of Saturn's rings:
  basic formulation and implications for Earth-based observations}.
\newblock { Icarus}, 201, 634-654.


\bibitem[{Muinonen} et~al., 1991]{muinonen2001}
{Muinonen}, K.~O., {Sihvola}, A.~H., {Lindell}, I.~V., and {Lumme}, K.~A., 1991.
\newblock {Scattering by a small object close to an interface. II. Study of
  backscattering}.
\newblock { Journal of the Optical Society of America A}, 8, 477--482.

%\bibitem[{Muinonen} et~al., 2002]{muinonen2002}
%{Muinonen}, K., {Piironen}, J., {Kaasalainen}, S., and {Cellino}, A., 2002.
%\newblock {Asteroid photometric and polarimetric phase curves: empirical
%  modeling}.
%\newblock { Memorie della Societa Astronomica Italiana}, 73, 716--+.

\bibitem[{Nelson} et~al., 2000]{nelson2000}
{Nelson}, R.~M., {Hapke}, B.~W., {Smythe}, W.~D., and {Spilker}, L.~J., 2000.
\newblock {The opposition effect in simulated planetary regoliths. Reflectance
  and circular polarization ratio change at small phase angle}.
\newblock { Icarus}, 147, 545--558.

\bibitem[{Nelson} et~al., 2002]{nelson2002}
{Nelson}, R.~M., {Smythe}, W.~D., {Hapke}, B.~W., and {Hale}, A.~S., 2002.
\newblock {Low phase angle laboratory studies of the opposition effect: search
  for wavelength dependence}.
\newblock { Planetary Space Sci.}, 50, 849--856.

\bibitem[{Nelson}, 2008]{nelson2008}
{Nelson}, R.~M., 2008.
\newblock {Laboratory Investigations Relevant to Cassini VIMS Reports of Coherent Constructive Interference in Saturn's Rings}.
\newblock {37th COSPAR Scientific Assembly, Plenary meeting, July 3-20, 2008, Montr\'eal, Canada, Paper number: B08-0009-08}



\bibitem[{{Nicholson} et~al.(2005){Nicholson}, {French}, {Campbell}, {Margot}, {Nolan}, {Black} and {Salo}}]{nicholson2005}
{Nicholson}, P., {French}, R.~G., {Campbell}, D.~B., {Margot},J.-L., {Nolan}, M.~C., {Black}, G.~J., and {Salo}, H., 2005.
{Radar imaging of Saturn's rings}. Icarus, 177, 32-62


\bibitem[{{Peltoniemi} and {Lumme}(1992)}]{peltoniemi1992}
{Peltoniemi}, J.~I., {Lumme}, K., 1992. Light scattering by closely packed
  particulate media. J.\ Opt.\ Soc.\ Am.\ A 9, 1320--1326.


\bibitem[{Plass} and {Kattawar}, 1968]{plass1968}
{Plass}, G.~N. and {Kattawar}, G.~W., 1968.
\newblock {Monte Carlo calculations of light scattering from clouds}.
\newblock { Applied Optics} 7, 415--419.


\bibitem[{Porco} et~al., 2008]{porco2008}
{Porco}, C.~C., {Weiss}, J.~W., {Richardson}, D.~C., {Dones}, L., {Quinn}, T.,
  and {Throop}, H., 2008.
\newblock {Simulations of the dynamical and light-scattering behavior of
  Saturn's rings and the derivation of ring particle and disk properties}.
\newblock { Astron. J.}, 136, 2172--2200.


\bibitem[{Poulet} et~al. (2002)]{poulet2002}
{Poulet, F., J. N. Cuzzi, R.~G.~French, and Dones, L., 2002}. 
{A study of Saturn's ring phase curves from HST observations.}
{Icarus} {158,} {224--248.}

%ok
\bibitem[{{Price}(1973)}]{price1973}
{Price}, M.~J., 1973.  {Optical scattering Properties of Saturn's
  Rings}. {Astron. J.} 78, 113-120.

\bibitem[{Rosenbush} et~al., 1997]{rosenbush1997}
{Rosenbush}, V., {Avramchuk}, V., Rosenbush, A., and Mishchenko, M., 1997.
\newblock {Polarization properties of the Galilean satellites of Jupiter:
observations and preliminary analysis}.
\newblock { Astrophys. J.}
  487, 402-412.

\bibitem[{Salo}, 1988]{salo1988}
{Salo}, H., 1988.
\newblock {Monte Carlo modeling of the net effects of coma scattering and
  thermal reradiation on the energy input to cometary nucleus}.
\newblock { Icarus} 76, 253--269.

\bibitem[{Salo}, 1992a]{salo1992a}
{Salo}, H., 1992a.
\newblock {Numerical simulations of dense collisional systems. II - Extended
  distribution of particle sizes}.
\newblock { Icarus}, 96 85--106.


\bibitem[{{Salo}(1992b)}]{salo1992b}
{Salo}, H., 1992b. {Gravitational wakes in Saturn's rings}. Nature 359,
  619--621.

\bibitem[{{Salo}(1995)}]{salo1995}
{Salo}, H., 1995. {Simulations of dense planetary rings. III. Self-gravitating
  identical particles.} Icarus 117, 287--312.

\bibitem[{Salo} et~al., 2001]{salo2001b}
{Salo}, H., {Schmidt}, J., and {Spahn}, F., 2001.
\newblock {Viscous overstability in Saturn's B-ring: I. Direct simulations and
  measurement of transport coefficients}.
\newblock { Icarus}, 153, 295--315.


\bibitem[{{Salo} and {Karjalainen}(2003)}]{salo2003}
{Salo}, H., {Karjalainen}, R., 2003. {Photometric modeling of Saturn's rings:
  I. Monte Carlo method and the effect of nonzero volume filling factor}.
  Icarus 164, 428--460 (SK2003).

\bibitem[Salo et~al.(2004)]{salo2004} Salo, H., Karjalainen, R.,
French, R.~G., 2004. {Photometric modeling of Saturn's rings
II. Azimuthal asymmetry in reflected and transmitted light}.\ Icarus
170, 70--90.

\bibitem[{{Salo} and {Schmidt}(2010)}]{salo2010}
{Salo}, H., {Schmidt}, J., 2010. {N-body simulations of viscous instability of planetary rings}.
  Icarus 206, 390--409.

\bibitem[{Schmidt} et~al. (2009)]{schmidt2009} Schmidt, J., Ohtsuki,
    K., Rappaport, N., Salo, H., Spahn, F., 2009.  {Dynamics of
      Saturn's dense rings}.  In: Brown, R.H., Dougherty, M. (Eds.),
     Saturn after Cassini-Huygens. Kluwer, pp. 413-458.


\bibitem[Shepard and Helfenstein (2007)]{shepard2007} {Shepard}, M.~K. and {Helfenstein}, P., 2007. {A test of the Hapke photometric model}. J. Geophys. Res 112,
EO3001, 1-17.


\bibitem[{Shkuratov}, 1988]{shkuratov1988}
{Shkuratov}, I.~G., 1988.
\newblock {A diffraction mechanism for the formation of the opposition effect
  of the brightness of surfaces having a complex structure}.
\newblock { Kinematika i Fizika Nebesnykh Tel}, 4, 33--39.

\bibitem[{Shkuratov} et~al., 1999]{shkuratov1999}
{Shkuratov}, Y.~G., {Kreslavsky}, M.~A., {Ovcharenko}, A.~A., {Stankevich},
  D.~G., {Zubko}, E.~S., {Pieters}, C., and {Arnold}, G., 1999.
\newblock {Opposition effect from Clementine data and mechanisms of
  backscatter}.
\newblock { Icarus}, 141,132--155.

\bibitem[{Shkuratov} et~al., 2007]{shkuratov2007}
{Shkuratov}, Y., {Bondarenko}, S., {Kaydash}, V., {Videen}, G., {Mu{\~n}oz},
  O., and {Volten}, H., 2007.
\newblock {Photometry and polarimetry of particulate surfaces and aerosol
  particles over a wide range of phase angles}.
\newblock { Journal of Quantitative Spectroscopy and Radiative Transfer}
  106, 487--508.


\bibitem[Showalter et al.(1996)]{showalter1996}
{Showalter, M. R., Bollinger, K. J., Nicholson, P. D., and Cuzzi, J. N., 1996.}
        {The Rings Node for the Planetary Data System.}
	{Plan. Space Sci.}
        {44}, {33--45}. 


\bibitem[{{Thompson} et~al.(1981){Thompson}, {Lumme}, {Irvine}, {Baum}, and
  {Esposito}}]{thompson1981}
{Thompson}, W.~T., {Lumme}, K., {Irvine}, W.~M., {Baum}, W.~A., {Esposito},
  L.~W., 1981. Saturn's rings - Azimuthal variations, phase curves, and radial
  profiles in four colors. Icarus 46, 187--200.


\bibitem[{{Toomre} and {Kalnajs}(1991)}]{toomre1991}
{Toomre}, A., {Kalnajs}, A.~J., 1991. Spiral chaos in an orbital patch. In:
  {Sundelius, B. (Ed.), Dynamics of Disc Galaxies. Almquist-Wiksell, pp.
  341--358}.

\bibitem[{{van de Hulst}, H.~C.}]{vandehulst}
   {van de Hulst, H.~C.}, 1980. Multiple Light Scattering. Tables, Formulas and Applications.,
 Academic Press, New York


%\bibitem[Verbiscer et al.(2005)]{Ver05} Verbiscer, A.~J., 
%French, R.~G., \& McGhee, C.~A.\ 2005, Icarus, 173, 66 

\bibitem[{{Wisdom} and {Tremaine}(1988)}]{wisdom1988}
{Wisdom}, J., {Tremaine}, S., 1988. Local simulations of planetary rings.
  Astron. J. 95, 925--940.

\end{thebibliography}
\end{document}